\documentclass[a4paper,fleqn,usenatbib]{mnras}

\usepackage{newtxtext,newtxmath}
\usepackage[T1]{fontenc}
\usepackage{ae,aecompl,eurosym}

\usepackage{graphicx}	% Including figure files
\usepackage{amsmath}	% Advanced maths commands

\usepackage{amssymb}	% Extra maths symbols

\def\teff{\hbox{$T_{\rm eff}$}}
\def\logg{\hbox{$\log g$}}

\def\ms{\hbox{m\,s$^{-1}$}}
\def\dms{\hbox{dm\,s$^{-1}$}}
\def\cms{\hbox{cm\,s$^{-1}$}}
\def\phpspA{\hbox{ph\,s$^{-1}$\,\AA$^{-1}$}}

\def\kms{\hbox{km\,s$^{-1}$}}

\def\mic{\hbox{$\mu$m}}

\def\emr{}
\def\emq{}

\def\arcsec{\hbox{$^{\prime\prime}$}}
\def\degr{\hbox{$^\circ$}}
\def\degC{\hbox{$^\circ$C}}

\newcommand{\hei}{\hbox{He$\;${\sc i}}}

\newcommand{\pbe}{\hbox{Pa${\beta}$}}
\newcommand{\pga}{\hbox{Pa${\gamma}$}}
\newcommand{\bga}{\hbox{Br${\gamma}$}}
\newcommand{\hdo}{\hbox{H$_2$O}}

\newcommand{\cod}{\hbox{CO$_2$}}
\newcommand{\chq}{\hbox{CH$_4$}}
\newcommand{\od}{\hbox{O$_2$}}
\newcommand{\oh}{\hbox{OH}}

\title[SPIRou: nIR velocimetry \& spectropolarimetry at the CFHT]{SPIRou: nIR velocimetry \& spectropolarimetry at the CFHT} 
\author[J.-F.~Donati et al.]{J.-F.~Donati$^1$\thanks{E-mail: jean-francois.donati@irap.omp.eu},
           D.~Kouach$^2$, C.~Moutou$^1$, R.~Doyon$^3$, X.~Delfosse$^4$, E.Artigau$^3$, 
\newauthor S.~Baratchart$^2$, M.~Lacombe$^2$, G.~Barrick$^5$, G.~H\'ebrard$^6$, F.~Bouchy$^{7,8}$, L.~Saddlemyer$^9$,
\newauthor L.~Par\`es$^1$, P.~Rabou$^4$, Y.~Micheau$^2$, F.~Dolon$^{10}$, V.~Reshetov$^9$, Z.~Challita$^2$, 
\newauthor A.~Carmona$^{1,4}$, N.~Striebig$^2$, S.~Thibault$^{11}$, E.~Martioli$^{12,6}$, N.~Cook$^3$, P.~Fouqu\'e$^{5,1}$, 
\newauthor T.~Vermeulen$^5$, S.Y.~Wang$^{13}$, L.~Arnold$^{5,10}$, F.~Pepe$^7$, I.~Boisse$^8$, P.~Figueira$^{14,15}$, 
\newauthor J.~Bouvier$^4$, T.P.~Ray$^{16}$, C.~Feugeade$^1$, J.~Morin$^{17}$, S.~Alencar$^{18}$, M.~Hobson$^8$,
\newauthor B.~Castilho$^{12}$, S.~Udry$^7$, N.C.~Santos$^{15}$, O.~Hernandez$^{3,19}$, T.~Benedict$^5$, P.~Vall\'ee$^3$, 
\newauthor G.~Gallou$^1$, M.~Dupieux$^1$, M.~Larrieu$^1$, S.~Perruchot$^{10}$, R.~Sottile$^{10}$, F.~Moreau$^{10}$, 
\newauthor C.~Usher$^5$, M.~Baril$^5$, F.~Wildi$^7$, B.~Chazelas$^7$, L.~Malo$^3$, X.~Bonfils$^4$,  
\newauthor D.~Loop$^9$, D.~Kerley$^9$, I.~Wevers$^9$, J.~Dunn$^9$, J.~Pazder$^9$, S.~Macdonald$^9$, 
\newauthor B.~Dubois$^2$, E.~Carri\'e$^2$, H.~Valentin$^1$, F.~Henault$^4$, C.H.~Yan$^{13}$, T.~Steinmetz$^{20}$
\newauthor {\it\small  Affiliations are listed at the end of the paper}
} 

\date{Accepted 2020 August 19. Received 2020 August 18; in original form 2020 May 12} 

\pubyear{2020}

\volume{{\rm in press}}
\begin{document}

\label{firstpage}
\pagerange{\pageref{firstpage}--\pageref{lastpage}}
\maketitle

\begin{abstract}
This paper presents an overview of SPIRou, the new-generation near-infrared spectropolarimeter / precision velocimeter recently installed on 
the 3.6-m Canada-France-Hawaii Telescope (CFHT).  Starting from the two main science goals, namely the quest for planetary systems around nearby 
M dwarfs and the study of magnetized star / planet formation, we outline the instrument concept that was designed to efficiently address these 
forefront topics, and detail the in-lab and on-sky instrument performances measured throughout the intensive testing phase that SPIRou was 
submitted to before passing the final acceptance review in early 2019 and initiating science observations.  With a central position among the
newly started programmes, the SPIRou Legacy Survey (SLS) Large Programme was allocated 300 CFHT nights until at least mid 2022.  

We also briefly describe a few of the first results obtained in the various science topics that SPIRou started investigating, focusing in 
particular on planetary systems of nearby M dwarfs, transiting exoplanets and their atmospheres, magnetic fields of young stars, but also on 
alternate science goals like the atmospheres of M dwarfs and the Earth's atmosphere.  We finally conclude on the essential role that 
SPIRou and the CFHT can play in coordination with forthcoming major facilities like the JWST, the ELTs, PLATO and ARIEL over the decade.  
\end{abstract}

\begin{keywords}
stars: planetary systems --
stars: formation --
stars: magnetic fields --
techniques: radial velocities -- 
instrumentation: polarimeters -- 
instrumentation: spectrographs
\end{keywords}

%%%%%%%%%%%%%%%%%%%%%%%%%%%%%%%%%%%%%%%%%%%%%%%%%%

%%%%%%%%%%%%%%%%% BODY OF PAPER %%%%%%%%%%%%%%%%%%

\section{Introduction}
\label{sec:intr}

Over the last two decades, high-resolution spectrographs working at visible and optical wavelengths on 4m-class telescopes have been quite
successful at obtaining forefront results in various research fields relevant to the physics of stars and planets.  In particular, HARPS on the 
3.6-m European Southern Observatory (ESO) telescope \citep{Mayor03} with its exquisite sub-\ms\ velocimetric precision has been instrumental 
in unveiling and characterizing hundreds of planetary systems orbiting stars other than the Sun \citep[e.g.,][]{Mayor19};  similarly, ESPaDOnS on the 
3.6-m Canada-France-Hawaii Telescope \citep[CFHT,][]{Donati03} revolutionized the field of stellar magnetometry thanks to its polarimetric 
capabilities \citep{Donati09}, by revealing and mapping the surface magnetic fields in stars of all types, including newborn  
stars where magnetic fields are known to play a key role \citep[e.g.,][]{Bouvier17, Pudritz19}.  More recently, new-generation velocimeters 
\citep[like ESPRESSO on the ESO 4$\times$8-m Very Large Telescope / VLT,][]{Pepe13} are pushing further into the \dms\ precision regime to 
{\emr unveil Earth-like planets located in the habitable zone (HZ) of their host stars, i.e., in the adequate distance range from the star for 
liquid water to be able to pool on the planet surface. } 

To deepen this exploration further and reach yet new classes of stars, in particular very-low-mass dwarfs and very young class-I protostars currently 
out of reach of existing facilities given their intrinsic faintness at visible wavelengths, new instruments working in the near infrared (nIR) were 
designed and constructed.  These include {\emr GIANO on the 3.6-m Telescopio Nazionale Galileo \citep{Oliva18}}, Carmenes on the 3.5-m Telescope at Calar 
Alto \citep{Quirrenbach12}, iSHELL on the 3.2-m NASA InfraRed Telescope Facility \citep{Rayner16}, HPF on the 10-m Hobby-Ebberly Telescope 
\citep{Mahadevan14}, IRD on the 8-m Subaru Telescope \citep{Kotani14, Kotani18} {\emr and soon NIRPS on the 3.6-m ESO telescope \citep{Bouchy19}}. 
On-sky operations have started in the last few years for most of these instruments, often aiming at extensive velocimetric surveys of nearby M dwarfs.  
This exploration represents a critical step in our long-term quest to understand the emergence of life, by paving the ground for future scrutinizing 
studies of the atmospheres of the best Earth-like planet candidates with large facilities such as the 6-m James Webb Space Telescope (JWST), the 
Extremely Large Telescopes \citep[ELTs\footnote{Including the 39-m ESO's ELT, the Thirty Meter Telescope / TMT, and the 25-m Giant Magellan Telescope / 
GMT}, e.g.,][]{Cirasuolo19}, and later-on the ARIEL space probe \citep{Tinetti17}.  

SPIRou \citep[standing for SpectroPolarim\`etre InfraRouge, i.e., Infrared SpectroPolarimeter,][]{Donati18} is one of these new instruments, 
designed and constructed for the CFHT, and inspired from both HARPS and ESPaDOnS.  Covering the whole YJHK bands (0.95-2.5~\mic) in a single exposure, 
SPIRou includes an achromatic polarimeter fiber-feeding a cryogenic high-resolution \'echelle spectrograph capable of both high-precision velocimetry 
and spectropolarimetry of stars.  The two main science goals that motivated the construction of SPIRou are the quest for planetary systems around 
nearby M dwarfs on the one hand, and the study of magnetized star / planet formation on the other;  this core programme, 
called the SPIRou Legacy Survey (SLS), was allocated 300 CFHT nights from 2019 until at least mid 2022.  Whereas precision velocimetry 
is key for the first goal, spectropolarimetry is required for the other and can yield at the same time optimal proxies for monitoring 
activity in the surveyed M dwarfs \citep{Hebrard16, Haywood16}.  Beyond these two main goals, 
SPIRou can efficiently explore a wide variety of science topics, from planetary atmospheres (both for Solar-System planets and exoplanets) to 
weather patterns of brown dwarfs, stellar dynamos, stellar archaeology and seasonal variations in the Earth's atmosphere.  

In this paper, we start by detailing the science goals of SPIRou (in Sec.~\ref{sec:scig}), describe its technical characteristics 
(in Sec.~\ref{sec:inst}), its measured performances (in Sec.~\ref{sec:perf}), and present a short overview of the first 
results so far, mostly within the SLS (in Sec.~\ref{sec:pano}).  We finally conclude by outlining how the potential of SPIRou can be 
best exploited in the coming decade and beyond, in particular to prepare and contribute to future explorations to be carried out with planned major 
facilities from both ground and space, including the JWST, the ELTs, PLATO and ARIEL (in Sec.~\ref{sec:conc}).

\section{SPIRou science : main goals \& complementary objectives} 
\label{sec:scig}

We start this overall description of SPIRou by outlining the two main science goals for which SPIRou was designed and on which the SLS 
is focused, as well as the complementary objectives that SPIRou can tackle.  We also present the international science consortium 
involved in the funding and construction of SPIRou.

\subsection{Planetary systems of M dwarfs}

In the quest for worlds other than the Solar System, in particular those harbouring Earth-like planets located in the HZ of their host stars, 
M dwarfs have attracted tremendous interest over the last few years 
\citep[e.g.,][]{Bonfils13, Muirhead15}.  In addition to being the most populated stellar class in the solar neighbourhood, M dwarfs 
were shown to be planet rich (especially in low-mass planets), hosting on average $>$2 planets per star 
\citep[e.g.,][]{Dressing15, Gaidos16, Gillon17} with a large fraction of them located in the HZ \citep{Bonfils13}.  
In this context, studying planetary system architectures of M dwarfs and unveiling detailed statistics on the planet occurrence rates come as 
obvious goals for investigating the extent to which the masses of the host stars (and thereby the properties of the parent protoplanetary disc) 
impact star / planet formation.  Moreover, future atmospheric studies with, e.g, the JWST, the ELTs or ARIEL, of all HZ Earth-like planets 
to be detected and characterized through transit photometry and velocimetry will only be possible for those orbiting the nearest stars, 
which happen to be M dwarfs.  

As of now, the number of known planets and systems in the solar neighbourhood (within, say, 10~pc) is quite limited, of order 10\%\ of the 
actual planet population predicted from what we know of planetary systems of early-M dwarfs;  for instance, only about 35 planets 
around 19 stars\footnote{Namely Gl~15A, GJ~54.1=YZ~Cet, GJ~1061, GJ~191=Kapteyn's star, GJ~3323, Gl~229, GJ~273, GJ~338B, Gl~411, Gl~447=Ross~128, 
GJ~551=Proxima~Cen, Gl~581, Gl~625, Gl~628, Gl~674, Gl~687, Gl~752, Gl~832 and GJ~876 according to the NASA exoplanet catalog 
at {\tt https://exoplanetarchive.ipac.caltech.edu.}} have yet been identified among the 
140 M dwarfs within 6.6~pc of the Sun, which should together host a total of 310 planets \citep[assuming an average of 2.2~planets per M 
dwarf,][]{Gaidos16}.  One of the reasons for this relative ignorance is that few high-precision velocimeters were, until recently, capable of 
detecting and characterizing planetary systems around M dwarfs, especially for late-type ones, thereby significantly limiting the accessible sample;  
being carried out with visible spectrographs, surveys concentrating on M dwarfs have indeed always been photon-noise limited at a radial velocity (RV) 
precision worse than 1~\ms\ \citep[e.g.,][]{Bonfils13}, and as such severely limited our ability to detect planets.  Another reason is that a 
proper characterization of the planet masses in multi-planet systems require a large number of visits, typically a few hundreds, in order to disentangle 
the contributions of all planets \citep[e.g.,][]{Udry19}, even when the planet periods are known from transit photometry \citep[e.g.,][]{Cloutier19}.  
Last but not least, M dwarfs are on average more active than the Sun, making it harder to accurately determine the planet masses if stellar activity is 
not faithfully modeled, which requires a dense and regular monitoring of the host star \citep[e.g.,][]{Klein19}.  

{\emr SPIRou and the SLS tackle this goal in two complementary ways, through (i)~a systematic RV monitoring of about 70 nearby M dwarfs 
(with an average of about 150 visits per star and a special emphasis on planets lying within HZs), called the SLS Planet Search (SLS-PS), and 
(ii)~a RV follow-up of about 20 of the most interesting transiting planet candidates to be uncovered by photometric surveys (with an average of 
about 40 visits per star), called the SLS Transit Follow-up (SLS-TF).  Each visit provides a spectrum with a signal to noise ratio (SNR) of 250 to 
150 per pixel (for early-M to late-M dwarfs respectively), expected to yield a photon-noise RV precision of $\simeq$1~\ms, with spectropolarimetry 
being used in most cases to simultaneously monitor magnetic activity.  With average H magnitudes of 7 and 9 respectively, SLS-PS and SLS-TF targets  
require typical exposure times of about 6~min and 40~min per visit.} 

The main goal of the SLS-PS is to unveil and characterize (i.e., measure the orbital parameters and derive lower limits on the masses of) a 
large fraction of the still unknown planets in the immediate neighborhood of the Solar System, in order to achieve a large census and 
statistical study of the planet population around nearby M dwarfs (especially the less-explored lowest-mass stars) and thereby 
make a major step forward in our understanding of how systems like ours form and evolve into maturity.  
SLS-PS observations will yield an accurate measurement of the occurrence of Earth-like planets orbiting mid- to late-M dwarfs, especially 
those located within the HZ, to be compared with prior estimates from HARPS and Kepler data in the specific case of early-M dwarfs.  
{\emr From Monte-Carlo simulations, we estimate that the SLS-PS, totalling 150 CFHT nights (50\%\ of the SLS allocation) 
should detect $\simeq$60 new exoplanets {\emr \citep[yielding an average planet detection efficiency of $\simeq$40\%\ assuming 2.2 
planets per star, in line with][]{Cloutier18}}, including $\simeq$25 Earth-mass planets ($\simeq$6 of which located in the HZ of their host stars).  } 
The SLS-PS will also serve as a pioneering exploration to pin down the closest and very best 
planet candidates for future detailed characterizations of their atmospheres with the JWST and eventually the ELTs \citep[including for 
non-transiting planets, by combining high-dispersion spectroscopy with high-contrast imaging,][]{Snellen15}.  The SLS-PS input catalog, 
based on preparatory observations with ESPaDOnS at the CFHT, includes a list of all potential targets ranked according to a merit function 
taking into account several parameters including brightness and rotation rate, or equivalently activity level \citep{Moutou17, Fouque18}.  

In the case of the SLS-TF to which 75 CFHT nights (25\%\ of the SLS allocation) are dedicated, the main goals are (i)~to constrain the 
mass-radius diagram of low mass planets that will revolutionize our understanding of their internal structure and bulk composition and 
(ii)~to provide key mass measurements of the best planets amenable to atmospheric characterization with SPIRou through high-resolution 
transit spectroscopy \citep{Brogi18, Brogi19} and later-on with the JWST, the ELTs and ARIEL.  The bulk of the SLS-TF targets to be 
monitored is provided by TESS \citep{Ricker16}.  Knowing in advance both orbital periods and transit times from photometry partly 
compensates for SLS-TF targets being in average 2 magnitudes fainter than SLS-PS ones and ensures that the velocimetric 
signal can be detected with only a few tens of visits for simple systems, even when activity dominates the RV signal 
\citep[][Klein et al.\ 2020a, submitted]{Klein20}.  
Among the $\simeq$500 planets orbiting M dwarfs that TESS expects to unveil, we will monitor about 20 of the most 
interesting ones around the brightest stars visible from CFHT, in particular those located in the HZ of their host stars 
\citep[$\simeq$10 for mid M dwarfs with K<10,][]{Barclay18}.  The SLS-TF also includes continuous transit monitoring for 
a few known transiting planets (e.g., HD~189733~b, AU~Mic~b) to model the Rossiter-McLaughlin effect \citep{Moutou20, Martioli20} 
and characterize the properties of the planet atmospheres (Darveau-Bernier et al.\ 2020; Boucher et al.\ 2020; 
Klein et al.\ 2020b, in prep).  

As most SLS observations are using spectropolarimetry, SPIRou data can investigate at the same time the small- 
and large-scale magnetic properties and the associated activity of the host stars.  By modeling the magnetic activity of the sample stars, 
we not only expect to improve the precision at which RV curves are secured \citep[thanks to filtering techniques using optimal proxies, 
e.g.,][]{Hebrard16, Haywood16} and thereby enhance the sensitivity to low-mass planets, but also to provide opportunities for characterizing
magnetic fields and dynamo processes of largely- and fully-convective M dwarfs whose observational properties are still enigmatic 
\citep{Morin08b, Morin10, Morin11, Yadav15, Shulyak17}.  Last but not least, this approach will also allow us to study the extent to which magnetic 
fields and winds of active M dwarfs can impact the habitability of their HZ planets \citep{Vidotto13, Gudel14, Strugarek15, Vidotto19}.

\subsection{Star/planet formation}

Studying how low-mass stars and their planets form and migrate comes as a logical counterpart to the observation of mature planetary 
systems.  More specifically, we aim at investigating the key role that magnetic fields play at all phases of the star/planet formation 
process \citep[e.g.,][]{Konigl91, Andre09, Bouvier17, Pudritz19, Hennebelle20}.  
By controlling accretion, triggering outflows and jets, and producing intense X-rays, magnetic fields indeed critically impact the physics
of pre-main sequence (PMS) stars \citep{Baraffe10, Feiden16} and of their accretion discs \citep{Shu07}, and largely dictate their angular
momentum evolution \citep{Bouvier07}.  In particular, magnetic fields are thought to couple accreting PMS stars with their discs.
At some stage in the process, fields carve magnetospheric gaps in the central disc regions and trigger funneled inflows \& outflows from 
the inner discs, forcing the host stars to spin down \citep{Romanova04, Romanova11, Zanni13, Davies14}.  Magnetic fields presumably affect
planet formation as well \citep{Johansen09}, can stop or even reverse planet migration \citep{Baruteau14}, and may prevent close-in
planets, including hot-Jupiters \citep[hJs,][]{Lin96, Romanova06}, from falling into their host stars.

With the Magnetic PMS star/planet survey (SLS-MP), the SLS amplifies the effort initiated on this topic 15 years ago 
through spectropolarimetric observations of T~Tauri stars (TTSs) and inner protostellar accretion discs of the FUOr type, with ESPaDOnS at the 
CFHT in particular \citep[e.g.,][]{Donati05, Donati12, Donati20, Yu17, Hill19}.  This exploratory work led to the characterization of the large-scale 
magnetic fields of TTSs, to novel constraints about the fossil or dynamo origin of these fields, to the discovery that the magnetic topologies 
largely reflect the internal structure of PMS stars \citep{Gregory12}, and to the study of how such fields impact star/planet formation.  
Along with contemporary photometric data, these observations also revealed that close-in giant planets can indeed already be present around 
PMS stars at as early an age as a few Myr \citep{David16, Donati17, Yu17, David19};  these results provide evidence that planet-disc interactions 
are instrumental in planet formation, likely shaping the early architecture of planetary systems, and thereby yield strong observational constraints 
on the timescale of planet formation.  

To expand this exploration further, the SLS-MP carries out a spectropolarimetric monitoring of about {\emr 55 PMS stars}  in nearby star forming 
regions, including a handful of very young partly-embedded class-I protostars in which magnetic fields are still mostly unknown 
\citep[e.g.,][]{Flores19}, with the goal of characterizing these fields, of unveiling the potential presence of close-in giant planets, 
and ultimately, of investigating in a more systematic fashion the impact of magnetic fields on star/planet formation.  
As the missing link between the youngest class-0 protostars whose fields are surveyed with ALMA / NOEMA at mm wavelengths \citep[e.g.,][]{Maury10} 
and the older TTSs observed with optical instruments like ESPaDOnS, class-I protostars are key for fingerprinting the impact of magnetic fields 
on star/planet formation, by revealing the topologies of their fields, by telling us how dynamos and magnetospheric accretion behave when 
accretion is much stronger and more stochastic than in TTSs, by finding out how large a magnetospheric gap these fields are able to carve at 
the centre of the disc and how stars react to this process.  Given their intrinsic faintness at optical wavelengths (as a result of extinction 
by the dust cocoon in which they are still embedded), class-I protostars can only be studied at nIR wavelengths, hence the pioneering role that  
SPIRou and the SLS-MP are expected to play in this task, without suffering from too small a wavelength range in a single exposure like previous 
unsuccessful studies on such objects \citep[e.g., with CRIRES on the VLT,][]{Viana12}.  

Besides class-I protostars, the SLS-MP sample features about {\emr 50 TTSs}, either still surrounded by and actively accreting from their disc (called 
classical TTSs / cTTSs), or having already dissipated their inner discs (called weak-line TTSs / wTTSs).  The SLS-MP includes in particular 
cTTSs with medium to strong accretion rates and very-low-mass PMS stars for which only few spectropolarimetric observations exist so
far, so as to sample the whole range of masses and accretion patterns \citep{Cody14, Sousa16}.  Regularly monitoring such 
cTTSs for several rotation cycles will shed light on their variability, and especially on the dynamics of the star-disk interaction process 
\citep[e.g.,][]{Alencar18}.  Similarly, by monitoring a larger sample of wTTSs than the one already explored \citep[e.g.,][]{Yu19}, the SLS-MP 
ambitions to characterize the population of newborn close-in giant planets at early evolutionary stages, estimate the occurrence frequency of such 
planets around PMS stars and compare it with that around mature Sun-like stars;  ultimately, one may even detect and estimate the magnetic fields 
of these hJs using radio observations \citep{Vidotto17}.  
For all stars, Zeeman-Doppler Imaging \citep[ZDI, e.g.,][and references therein]{Donati20} will be used to infer brightness 
distributions and magnetic topologies at the surfaces of stars, and maps of accretion spots in the accreting PMS stars;  by extracting the main 
properties of the reconstructed large-scale fields, one can then investigate how dynamo processes respond to the evolutionary changes in the 
structure of PMS stars as they evolve along Hayashi tracks \citep{Gregory12, Emeriau17, Villebrun19}. 

The SLS-MP aims at securing {\emr 20 to 35 visits} per star, 
each visit yielding an intensity and a circularly polarized spectrum with typical SNRs of about 200 per pixel, for a total allocation of 
75 CFHT nights (25\%\ of the SLS allocation).  {\emr The typical exposure time is $\simeq$20~min, corresponding to an average H magnitude of 8.3. }

\subsection{Complementary science objectives}

SPIRou can also tackle a large number of additional science programmes, both within the SLS through its Legacy dimension and beyond the two main 
goals on which the SLS focuses.

By providing a wide and homogeneous set of nIR spectra for a large sample of M dwarfs and PMS stars, the SLS programmes offer the opportunity to 
assess theoretical atmospheric models of cool and very cool stars in much more detail than currently possible, with the goal of obtaining more 
precise determinations of the spectroscopic parameters characterizing the observed targets (in particular \teff, \logg, metallicity, also essential 
for planet characterization).  
It makes it possible to further constrain the key physical ingredients (including magnetic fields) playing a role in the atmospheres 
of very-cool stars and affecting their thermal and convection patterns;  such studies will lead to new sets of synthetic spectra that better match 
nIR observations \citep{Rajpurohit13, Allard13, Rajpurohit18a, Rajpurohit18b}, whereas the use of multi-component model atmospheres will 
improve the spectroscopic characterization of very active PMS stars \citep{Gully17}.  

Studying weather patterns in the atmospheres of brown dwarfs is another particularly exciting option.  These objects are known to exhibit 
photometric variations on short timescales \citep{Artigau09}, attributed to the presence of atmospheric clouds rotating in and out of view, and 
subject to temporal variability on timescales of only a few rotation cycles.  Using tomographic imaging applied to time-series of high-resolution
spectra, one can recover surface maps of the cloud patterns \citep{Crossfield14} and potentially their temporal evolution as well for the brightest 
targets.  With its high sensitivity and large spectral domain that can partly compensate for the intrinsic faintness of these very cool objects, 
SPIRou is a promising instrument for carrying out such studies.  

SPIRou is equally well suited for investigating the dynamics and chemistry of planetary atmospheres in our Solar System \citep{Machado14,Machado17},
and potentially of giant close-in exoplanet atmospheres as well, even when not transiting \citep{Snellen10, Brogi12}.  Last but not
least, SPIRou will also offer the opportunity of studying at high spectral resolution extremely metal-poor stars as relics of
the early universe, providing us with precious clues about the chemical evolution and formation of the Milky Way \citep{Reggiani16}.

Independently of the science programmes they are collected for, SPIRou data can also be used to characterise the chemical content of the Earth's 
atmosphere, and in particular of the main molecular species such as \hdo, \cod, \chq\ and \od, through fitting the very rich telluric spectrum 
fingerprinted in each single SPIRou observations, using tools such as TAPAS, TelFit or Molecfit \citep{Bertaux14, Gullikson14, Smette15, Ulmer-Moll19}.  
The extensive collection of spectra that SPIRou will collect over the coming decade provides the community with a novel way of quantitatively 
investigating the chemical content, as well as the seasonal and yearly evolution, of the Earth's atmosphere over the best astronomical site in the 
Northern Hemisphere.  SPIRou spectra can also be exploited to study \oh\ and \od\ terrestrial night-glow emission lines from the sky, as previously 
done from optical spectra \citep{Cosby06}.

\subsection{The SPIRou / SLS science consortium}

The SLS, whose main science topics are outlined above, was allocated 300 observing nights at the CFHT from 2019 to at least mid 2022 (and beyond, until 
a completion rate of at least 80\%\ is achieved if not by mid-2022), with the ambitious goal of addressing the following 6 key questions through its 
multi-faceted inter-connected components (SLS-PS, SLS-TF and SLS-MP): 
\begin{itemize}
\item How diverse are planetary systems of nearby M dwarfs?  How frequently do these stars harbor close-in or HZ Earth-mass planets?  
How does this diversity evolve with the host-star mass down to the brown-dwarf limit, a mostly unexplored mass range so far?
\item How does the density, structure and bulk composition of low-mass transiting planets change with planet mass and stellar irradiation, 
in particular for the HZ ones for which very few data exist?
\item What are the as-of-yet undetected systems, that are best suited for characterizations, of planetary atmospheres in particular, with 
future facilities like the JWST and the ELTs?
\item What kind of magnetic fields do largely- and fully-convective M dwarfs generate as they spin down to low activity levels for which 
almost no data are available? How do their fields and associated activity impact planet detectability and habitability?
\item What types of magnetic fields do low-mass PMS stars (and their accretion discs when relevant) host, in particular the youngest and 
strongly-accreting stars, for which very few magnetic observations exist? How do these fields change as stars and planets evolve?
\item How do such fields impact accretion/ejection processes and star/planet formation, and what is their role in the formation and 
survival of close-in planets, especially hJs known for their huge dynamic impact on the early architecture of planetary systems?
\end{itemize}

The SPIRou / SLS science  consortium involves about 150 scientists worldwide and includes in particular the core SPIRou science team who proposed the 
instrument, secured its funding, closely followed its design, construction and validation, and developed optimal tools for data reduction and 
velocimetric / spectropolarimetric analyses. Deeply rooted in the Franco-Canadian collaboration, born and grown strong and fruitful with the 
construction and operation of the CFHT,  the SLS / SPIRou consortium also gathers collaborators from other CFHT partners (Taiwan, Hawaii) as well 
as institutes involved in the construction / funding of SPIRou (in Brazil, Switzerland, Portugal and Ireland) from outside the CFHT community.

\section{Instrument characteristics} 
\label{sec:inst}

\begin{figure}
\includegraphics[scale=0.70,bb=0 20 434 160]{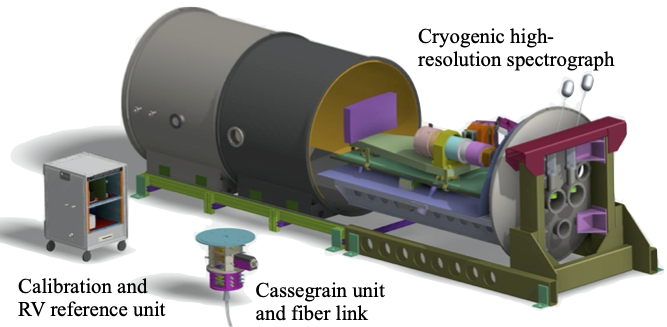}
\caption{CAD view of the main SPIRou units}
\label{fig:spirou}
\end{figure}

SPIRou is inspired from previous instruments, namely HARPS regarding precision velocimetry and ESPaDOnS regarding spectropolarimetry.  
It includes a cryogenic high-resolution spectrograph adapted from the vacuum-operated spectrograph of HARPS, a Cassegrain 
unit with an achromatic polarimeter derived from that of ESPaDOnS, a fiber-feed evolved from those of both ESPaDOnS 
and HARPS, and a calibration / RV reference unit copied from that of HARPS (see Fig.~\ref{fig:spirou} for a CAD view 
and {\tt http://spirou.irap.omp.eu/Gallery2/Photos} for pictures).  

We describe these units below, and outline the installation and operation of SPIRou at the CFHT.  We also briefly describe the 
SPIRou project team, managed by both Observatoire Midi Pyr\'en\'ees and the Institut de Recherche en Astrophysique et Plan\'etologie (OMP/IRAP) 
in Toulouse (France), and the worldwide institutes involved in the design, construction, tests and operation of SPIRou.

\subsection{Cassegrain unit}
\label{sec:cas}

\begin{figure}
\includegraphics[scale=0.45,bb=20 0 300 420]{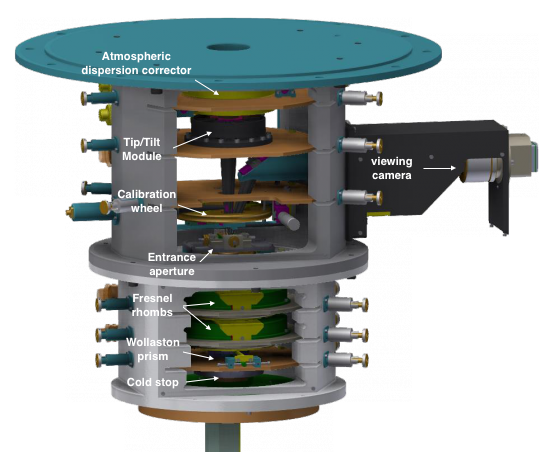}
\caption{CAD view of the middle and lower structures of the SPIRou Cassegrain unit.  The cold stop at the bottom of the lower structure, 
yielding no improvement in thermal background, was finally not used.  }
\label{fig:cas}
\end{figure}

The SPIRou Cassegrain unit consists of 3 structures mounted on top of each other and fixed at the Cassegrain focus of the telescope, 
for a total height, width and weight of 800~mm, 770~mm and 120~kg.  A CAD view of the lower two modules is shown in Fig.~\ref{fig:cas}.

The upper Cassegrain structure essentially serves as a mechanical interface with the telescope, ensuring that the instrument aperture
is ideally placed with respect to the telescope focus.  It also includes an option for feeding the instrument with a fully polarized beam,
allowing one to achieve a complementary polarimetric diagnostic of all optical components above the polarimeter (hence its designation as 
the ``end-to-end calibration module'') 

The lower Cassegrain structure first includes the instrument aperture, which, instead of being a genuine hole in the middle of a tilted 
mirror (as for ESPaDOnS and HARPS), consists of a circular region (of diameter 1.29\arcsec, i.e., 180~\mic) located at the centre of a silica plate (called 
the field mirror) whose lower surface features a reflective coating outside of the central region.  As a result, all incoming light falling 
outside the entrance aperture, as well as a few \% of the light falling inside, is reflected off by the field mirror towards the viewing camera 
(see below).  This structure also includes a focal reducer turning the f/8 incoming beam entering the aperture, into a f/4 beam with which the 
science optical fibers are fed with minimum Focal Ratio Degradation (FRD, see Sec.~\ref{sec:fib}).  This focal reducer involves a doublet and 
a triplet working at infinite conjugate ratio, both optimized {\emr regarding throughput and image quality over the spectral range} of SPIRou.  
Last but not least, this structure features an achromatic polarimeter, consisting of two 3/4-wave ZnSe dual Fresnel rhombs coupled to a 
Wollaston prism, splitting the incoming beam into 2 orthogonally polarized beams and feeding the twin science optical fibers (see Sec.~\ref{sec:fib}).  
By tuning the orientation 
of the rhombs, one can measure the amount of either circular or linear polarization in the incoming stellar light;  by coating one of the 
internal reflection surfaces of each rhomb, we can ensure that the polarimetric analysis is achromatic to better than a few degrees.  
ZnSe rhombs turned out to be quite tricky to assemble, with molecular adherence between the rhombs being very fragile and requiring 
the rhomb surface micro-roughness to be better than a few nm not to break under vibrations or stress (e.g., induced by temperature 
variations and gradients);  the stress-free mounting of the rhombs into their mechanical barrels also proved complex to master, as was 
the need to constrain the beam deviation to no more than a few \arcsec\ to minimize the potential RV impact of rotating the rhombs.  
After several attempts, fabrication challenges were successfully overcome by WinLight Optics (France);  {\emr the corresponding rhombs, made 
of ZnSe blanks from a different provider and with higher throughput in the blue, are progressively reaching CFHT, with one already 
mounted in SPIRou in June 2020 and a few more expected soon. } 

A cold stop was initially implemented to mask the central pupil obscuration by the telescope secondary mirror and the associated thermal 
emission, but turned out to bring no significant reduction in the level of thermal background while decreasing the throughput by $\simeq$10\%\ 
and was finally not used.  Baffles 
with gold coating are also inserted between the polarimeter optical elements for the same purpose, and in particular between the field mirror 
and the entrance doublet, and between the triplet and the science fiber head.  The thermal background that the polarimeter generates, scaling 
up with temperature, is about 15~\phpspA\ per science fiber {\emr at 2.35~\mic}, for a polarimeter temperature of 2\degC.  

The middle Cassegrain structure features several ancillary facilities used for both calibration and observing purposes.  It first includes
an atmospheric dispersion corrector (ADC) made of two motorized twin-prism pairs cancelling the impact of atmospheric refraction on the incoming 
beam up to airmasses of 2.5;  it ensures that the position of the star on the entrance aperture is achromatic over the full spectral range to a 
precision better than 0.03\arcsec.  
% The ADC prisms rotate at all times during observations, but can also rotate between exposures only if needed.  
The middle structure also includes a tip-tilt module (TTM) stabilizing the image of the star at the 
instrument aperture to ensure that the average image over a typical exposure time (of at least 10~sec) is stable with respect to the 
entrance aperture to better than 0.01\arcsec\ RMS up to H$\simeq$10.  This device works in conjunction with a nIR viewing camera (model 
Owl~640 from Raptor Photonics) coupled to a JH 
filter, looking at the instrument entrance aperture and sending back information to the TTM at a frequency of up to 50~Hz.  The main goal of 
the TTM is to minimise shifts of the observed star with respect to the entrance aperture, and thereby the systematic RV errors that 
may result from these shifts when scrambling of the near-field image by the optical fibers is not perfect (see Sec.~\ref{sec:fib});  at the 
same time, the TTM maximizes the stellar flux injected into the instrument.  
The algorithm implemented to measure the shift of the star with respect to the aperture (based on a least-squares Gaussian fit incorporating 
a low-reflectivity central circular aperture) and correct it through the TTM proved to be working well, both in lab tests and on-sky 
observations \citep{Barrick18}.  The final element of this module is a calibration wheel allowing one to inject light from the calibration unit 
(see Sec.~\ref{sec:cal}) into the instrument, with linear polarization along various directions if need be.  

The Cassegrain unit also includes the possibility of carrying an `internal alignment verification' (IAV) by sending light backwards from 
a small 50~\mic\ fiber inserted in the fiber connector beside the main science fibers (see Sec.~\ref{sec:fib}) into a second aperture 1~mm off 
the main instrument aperture in the field mirror;  with a dedicated prism installed in the calibration wheel, one can monitor the position of the 
fiber image with respect to the second aperture using the viewing camera, and detect potential shifts (to a precision better than 1~\mic) 
that may result from mechanical flexures or optical instabilities occurring between the main entrance aperture and the science fibers.  
Observed motions are of order 1~\mic\ RMS during observations across all telescope positions.  

Altogether, the throughput of the Cassegrain unit is about 50\%\ in the K band, falling down to $\simeq$30\% in the Y band as the main result 
of ZnSe absorption from the rhombs.  
The Cassegrain structure was designed, assembled and tested in a dedicated clean room by the SPIRou team 
members at OMP/IRAP \citep{Pares12}.

\begin{figure}
\includegraphics[scale=0.21,bb=0 0 800 660]{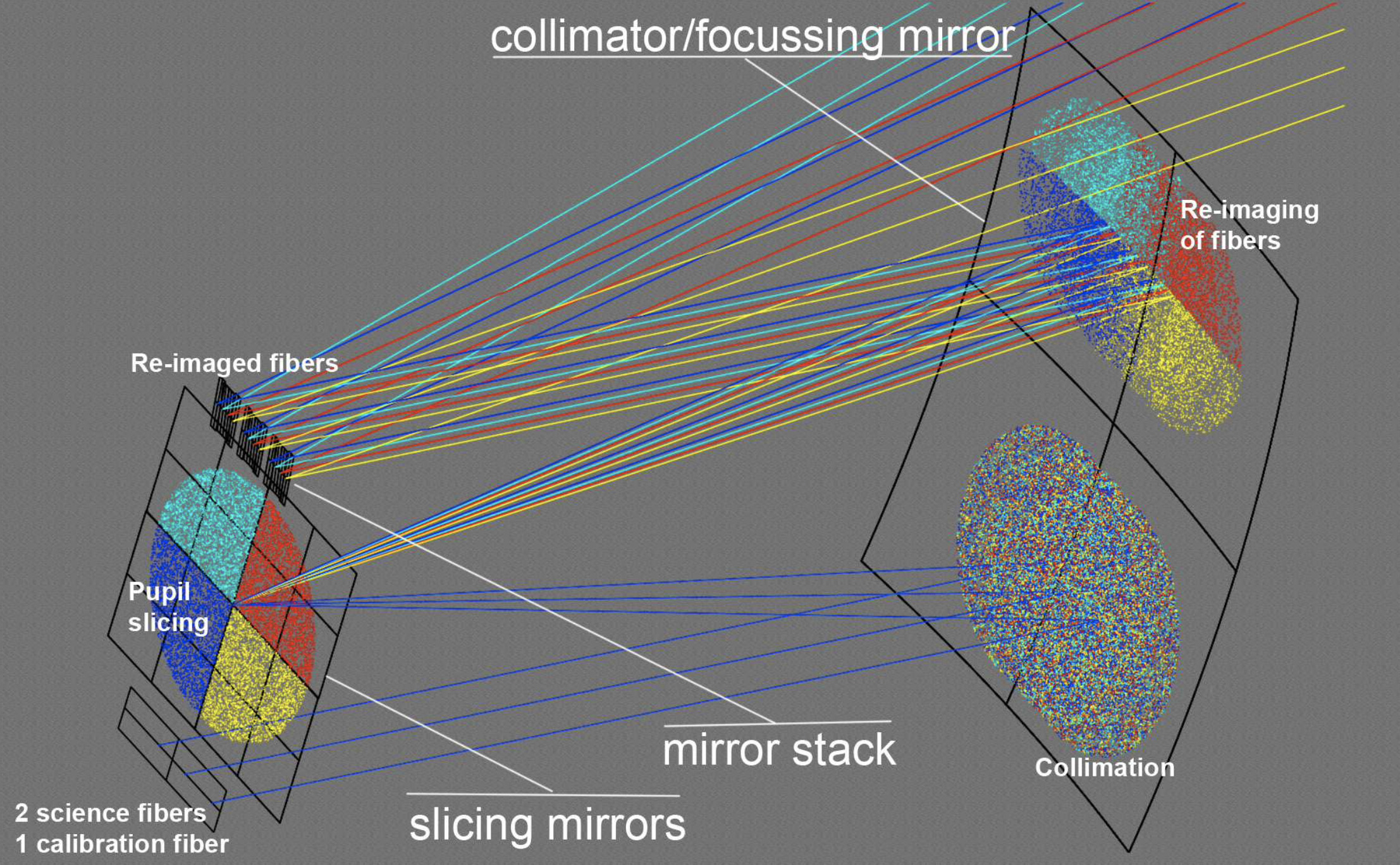} \\ 
\includegraphics[scale=0.20,bb=0 15 800 260]{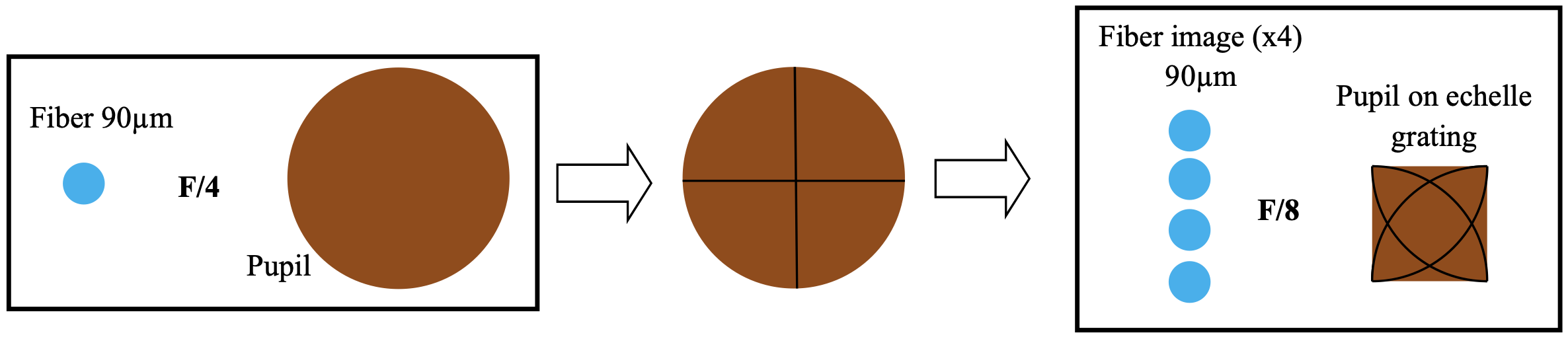} 
\caption{Optical layout of the SPIRou pupil slicer (top), with a schematic view of the pupil slicing principle for a single fibre (bottom)} 
\label{fig:slicer}      
\end{figure}

\subsection{The fiber link \& pupil slicer}
\label{sec:fib}

The fiber link features two 35-m long science fibers conveying the light from the twin orthogonally polarized beams coming out of the Cassegrain
polarimeter to the spectrograph, plus a 10-m long reference fiber bringing light from the calibration unit to the {\emq spectrograph} 
(see Sec.~\ref{sec:cal}).  All 3 fibers are circular fluoride fibers with a 90-\mic\ diameter, engineered by Le~Verre~Fluor\'e (LVF) in 
France from purified ZBLAN glass and ensuring a throughput of at least 90\%\ over the entire spectral range of SPIRou.  
{\emq The fiber link also includes a 35-m long ZBLAN fiber (of diameter 300~\mic) conveying the light of calibration lamps to the Cassegrain 
unit when needed, so that both science channels and the reference channel can be calibrated simultaneously in a homogeneous way.}  

On the Cassegrain side, the fiber link includes a fiber box that links the dual fiber head collecting light from the twin stellar 
images formed by the Wollaston prism, to a pair of single connectors, one for each of the two science fibers.  Although increasing the 
number of fibers connections, this setup makes it far easier, safer and more repeatable to de-connect and re-connect the science fibers to the 
Cassegrain unit each time SPIRou is removed from the Cassegrain focus of the CFHT.  On the spectrograph side, the fiber link features a triple 
hermetic feedthrough used to inject light from the three fibers within the cryostat (see Sec.~\ref{sec:spc}).  Following the feedthroughs, 
the fiber link includes a pupil slicer fed through three 1.4-m segments of 90-$\mu$m-diameter octagonal fiber (also engineered by LVF), 
each connected to a feedthrough.  The pupil slicer per se 
consists of a double-pass collimator and a pupil-slicing mosaic of 4 flat mirrors located at the focus of the collimator and slicing the pupil into 
4 equal 90\degr\ sectors; the twelve individual images (4 images for each of the three fibers) formed after a second pass through the collimator are focused 
on a stack of twelve small flat mirrors ensuring that the pupils of all individual beams overlap into a square pupil once imaged onto the spectrograph 
grating (see Fig.~\ref{fig:slicer}).  In addition of being extremely compact (only a few cm$^3$), this device has the key advantage of 
ensuring a high throughput with no loss of resolving power, and of delivering sliced images with identical shapes and flux 
distributions (as opposed to more conventional image slicers for which slices have different shapes), a prime asset for precision 
velocimetry where image stability is critical.  

Including absorption, FRD, connector and imaging losses, the fiber link provides an average throughput over the spectral range of $\simeq$60\%.
The combination of the circular and octagonal fibers, although intended to ensure a scrambling of the near-field image between polarimeter output 
and spectrograph input of order 1000, was however found to be much less efficient, with a scrambling performance about 10$\times$ lower than 
expected;  this degradation was recently showed to derive from excess stress in the way the pupil slicer holds the octagonal fibers.  
A hardware upgrade of the pupil slicer {\emr (with a stress-free mount of the octagonal fibers)}  is planned to improve near-field scrambling from 
the fiber link.  Regarding thermal background, the 
contribution from the fibers is moderate, with only about 5~\phpspA\ per fiber, whereas that from the hermetic feedthroughs, depending on the room 
temperature, is found to be about 25~\phpspA\ at 2.35~\mic\ for a temperature of 15\degC.  The fiber link, designed, assembled and tested by the 
SPIRou team members at OMP/IRAP \citep{Micheau12, Micheau15, Micheau18}.

\subsection{Cryogenic high-resolution spectrograph \& detector}
\label{sec:spc}

\begin{figure}
%\hspace{-2mm}\includegraphics[scale=0.16,bb=0 0 434 930]{fig/spirou_spec.png}
\hspace{-2mm}\includegraphics[scale=0.22,bb=0 10 434 650]{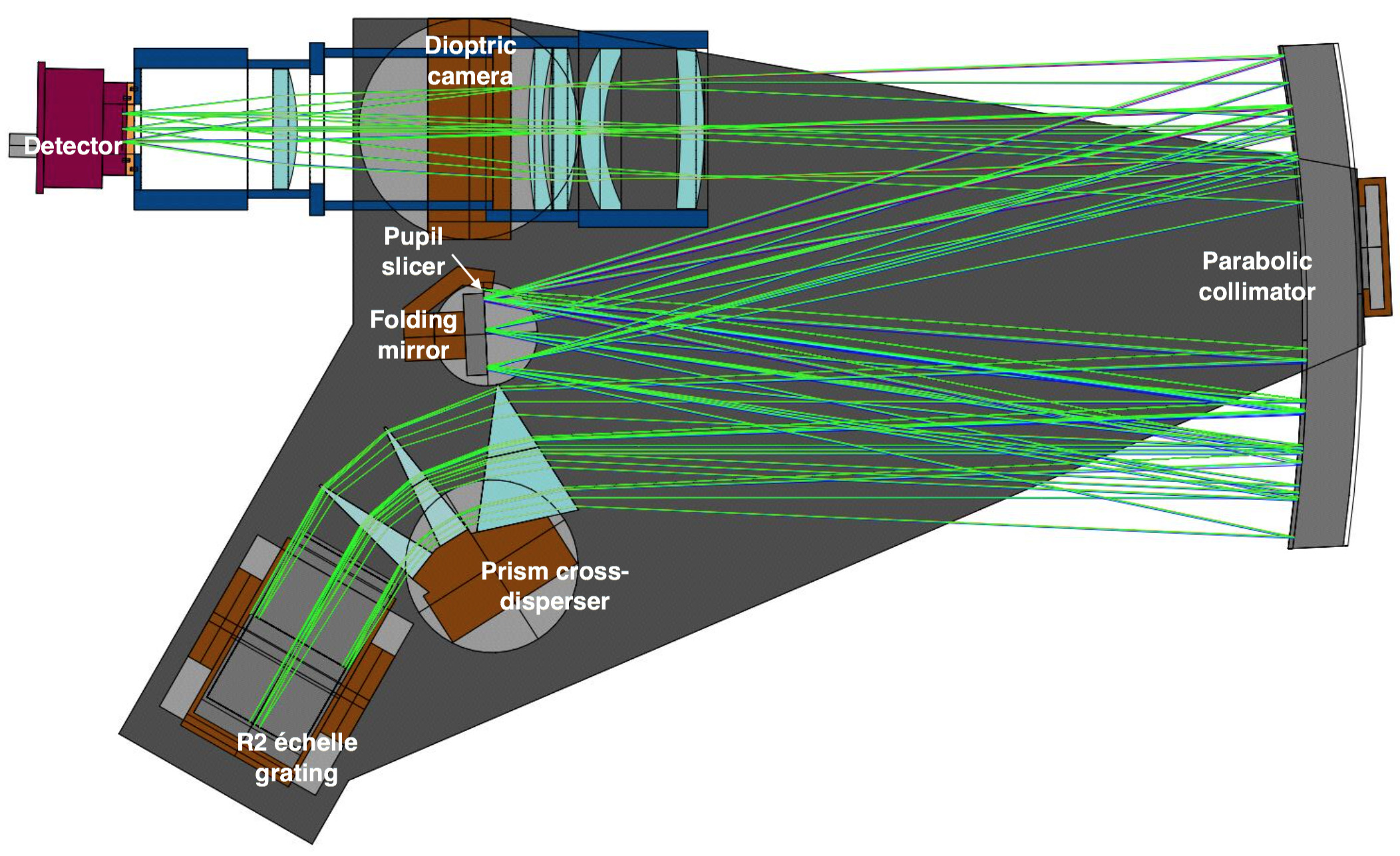}
\caption{Optical layout of the SPIRou spectrograph}
\label{fig:spectro}
\end{figure}

The SPIRou spectrograph is a bench-mounted high-resolution \'echelle spectrograph in a dual-pupil optical layout similar to that 
of ESPaDOnS and HARPS (see Fig.~\ref{fig:spectro}), the main difference being that the prism-train cross-disperser is used in double pass\footnote{The 
blue and red pupils on the grating are thus shifted by a few mm with respect to the pupil at mid wavelength, as visible on Fig~\ref{fig:spectro}.}.  
Light is injected into the spectrograph by the pupil slicer fed with 3 separate fibers, each one entering the spectrograph through a dedicated hermetic 
feedthrough (see Sec.~\ref{sec:fib}).  The pupil slicer generates a telecentric f/8 beam and a slit-like image featuring a triple set of 4 circular 
slices of overall dimension 90$\times$1530~\mic$^2$ (one set of 4 slices for each of the 3 incoming fibers).  This f/8 beam first meets the main 
collimator (an off-axis parabola with 1200-mm focal length), projecting a 150-mm square pupil onto the dispersing 
optical components, i.e., a double-pass triple-prism train cross-disperser (made of two ZnSe and one Infrasil prisms, of clear aperture 
190$\times$206~mm) and a R2 \'echelle grating (of clear aperture $154\times306$~mm$^2$, w/ 23.2~gr/mm).  Following a second pass on the 
off-axis parabola, the converging f/8 beam is reflected off a flat mirror near which an intermediate image of the full spectrum 
forms.  This spectrum is then re-injected into a fully-dioptric 5-lens camera (500-mm focal length, 220-mm clear aperture) 
after a third pass on the main collimator, forming a converging f/3.33 beam and the final scaled-down spectrum on a 4k$\times$4k H4RG detector 
(with 15-$\mu$m square pixels) located at the focus of the camera.  

\begin{figure*}
\hbox{\hspace{2mm}\includegraphics[scale=0.198,bb=0 20 1000 500]{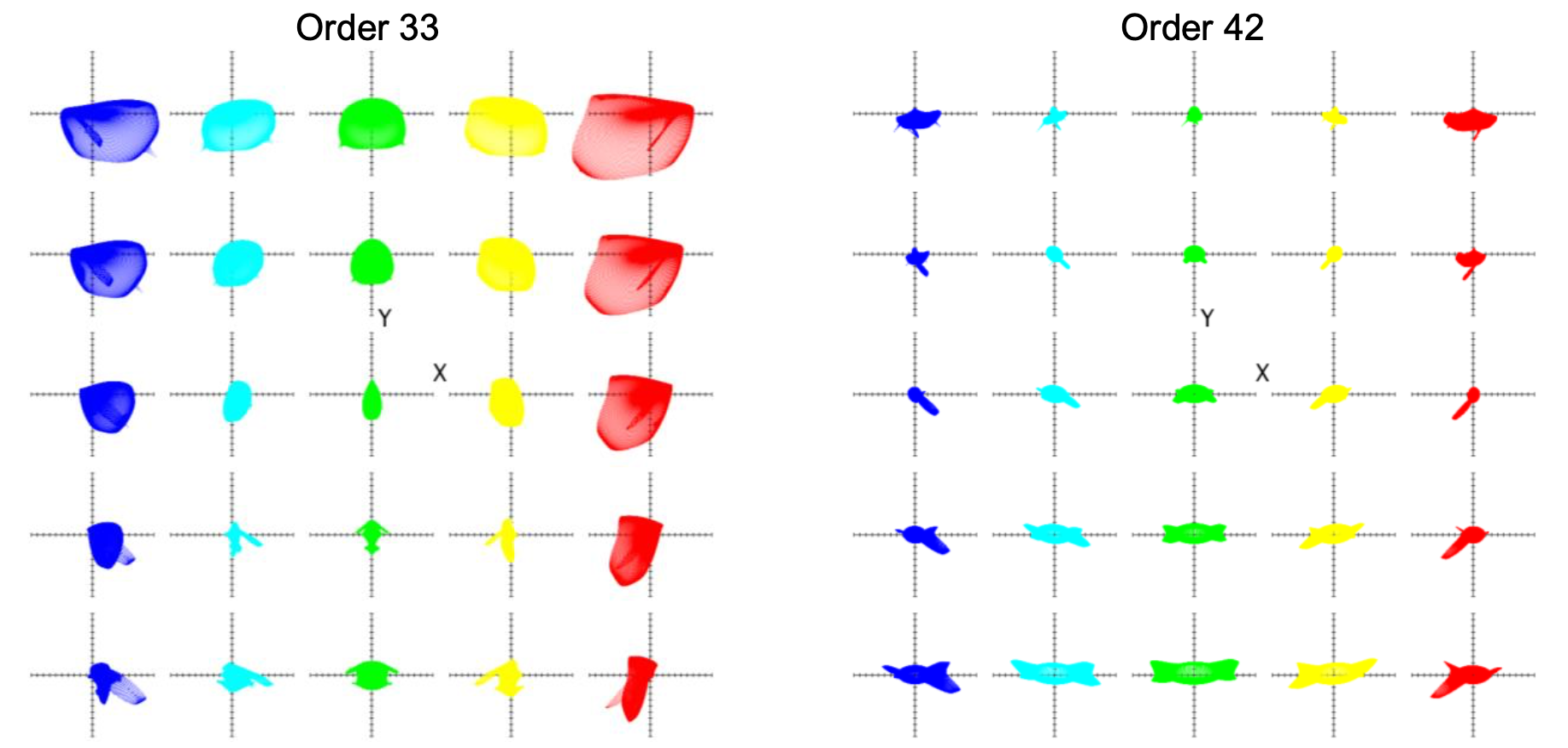}\includegraphics[scale=0.20,bb=0 20 500 500]{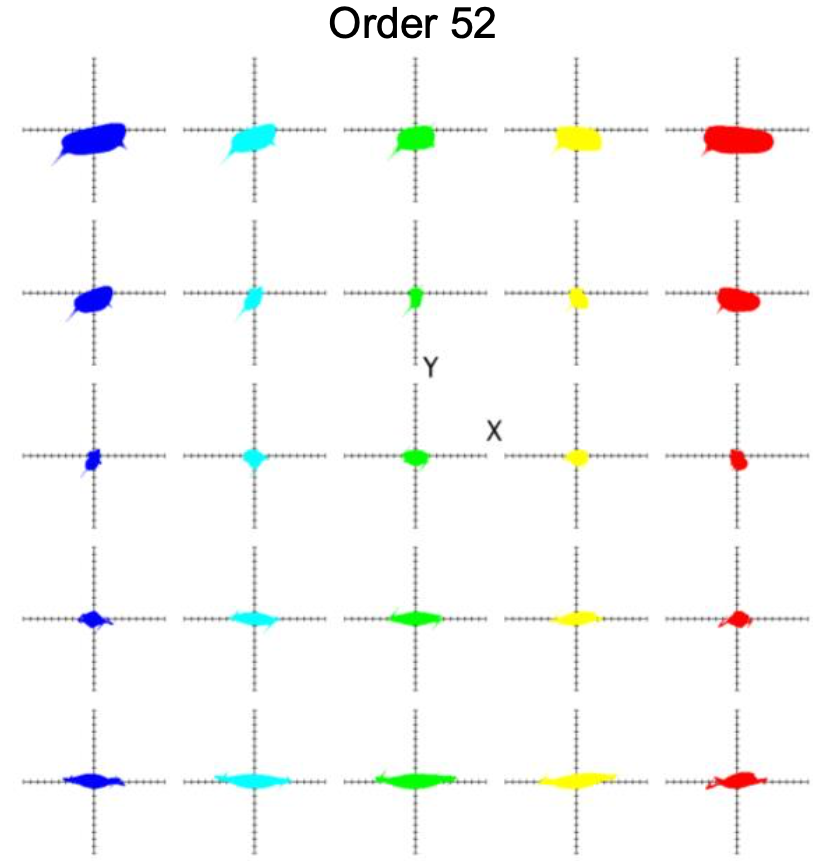}\includegraphics[scale=0.2,bb=0 20 1000 500]{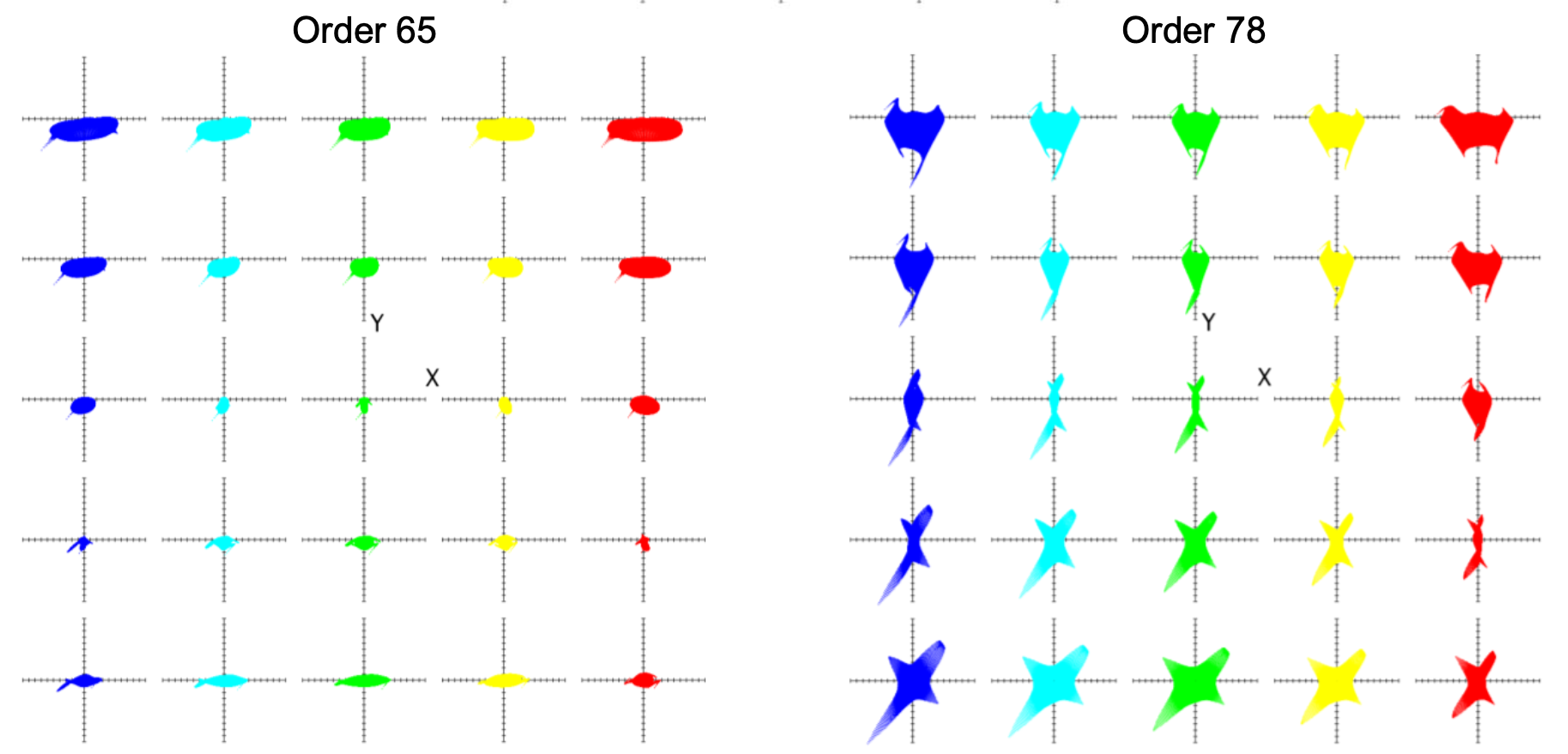}} 
\caption{Spot diagrams of the SPIRou spectrograph across the full spectral range (for orders 33, 42, 52, 65 and 78, corresponding to wavelengths 
2.40, 1.83, 1.48, 1.18 and 0.99~\mic\ respectively), for 5 wavelengths across each order (X direction) and for 
5 positions across the slit (Y direction).  The cross represents the size of one H4RG detector pixel (15~\mic). }
\label{fig:psf}
\end{figure*}

With this design, the spectrograph is able to fit the entire nominal spectral range of SPIRou (0.98--2.35~$\mu$m) on the H4RG detector (as 
46 orders numbered \#78 to \#33) and even fit 2 additional orders on both sides (\#80 to \#31) to further extend the usable spectral domain 
(0.95-2.50~\mic) albeit with a small 2-nm gap between the two reddest orders.  
Each order is about 48-pixel wide, with the distance between consecutive orders varying from 53 to 96~pixels.  
The spot diagrams (see Fig.~\ref{fig:psf}) feature a 
full-width at half-maximum (FWHM) across the slit smaller than half a pixel in most orders and everywhere smaller than 0.65~pixel, ensuring that 
the spectrograph is diffraction limited over the full spectral domain.  The spectral pixel size is equal to 2.28~\kms\ within 15\%\ throughout 
the whole domain, staying virtually constant at the centre of all orders and varying across orders.  The FWHM of the instrument response 
is dominated by the (non Gaussian) slicer profile ($\simeq$4~\kms), with minor contributions from the detector pixels (1.8~\kms) and from the 
optical point-spread-function (1.5~\kms);  as a result of the non-Gaussian shape of the instrument profile, the resolving power as derived from 
narrow calibration lines ($\simeq$64k) is predicted to be slightly smaller than that estimated from its broadening impact on the (wider) spectral 
lines of SPIRou stellar targets ($\simeq$70k).  {\emr Thanks to the tilted slit with respect to the detector pixels ensuring that the spectrum 
is sampled on a different pixel grid across the slit, it is possible to extract a spectrum with adequate sampling at full resolution, despite 
the detector pixel size (of 2.28~\kms) being 6\% larger than half the resolution element (of 4.29~\kms).}   

The SPIRou H4RG detector is a science-grade device engineered by Teledyne Imaging Sensors (TIS, USA), then mounted on its Focal Plane Array and 
extensively tested by the SPIRou team members at Universit\'e de Montr\'eal \citep[UdeM, Canada,][]{Artigau18};  at $\simeq$80~K, it features a quantum 
efficiency ranging from 80 to 95\%\ over the spectral domain, a cutoff wavelength of $\simeq$2.4~\mic, a readout noise of 12~e$^-$ in Correlated Double 
Sampling (CDS) readout mode, a median dark current of 0.007~e$^-$/s, and an average crosstalk to neighbouring pixels of 0.8\%.  The detector flatness 
is better than 10~\mic\ peak-to-valley (PtV), with a RMS under 5~\mic.  

The camera, assembled at Universit\'e Laval in Quebec City (UL, Canada) and tested at cryogenic temperatures at UdeM, follows a concept similar to 
that of WIRCAM, CFHT's 
wide-field infrared camera \citep{Thibault03}, with nylon pads holding the lenses to ensure minimal stress at operating temperatures.  Its throughput 
ranges from 95\%\ in the blue down to 90\%\ in the K band (reflecting the reduced transmission of the fourth camera lens made of S-FTM16).  
The parabola, engineered by SESO (France), features a protected silver coating ensuring a reflectivity of 98\%\ over the full domain, with 
wave-front errors smaller than $\lambda/2$ PtV.  The flat mirror and the prisms were engineered by Optical Surfaces (UK);  the prism-train 
transmission peaks at 90\%\ in the K band, falling down to 70\%\ towards the blue as a result of the higher ZnSe absorption.  Finally, the R2 
grating, fabricated by Richardson Lab (USA), exhibits an average transmission of 70\%\ throughout the domain, with wave-front errors smaller than $\lambda/2$ 
PtV.  As a whole, the spectrograph exhibits a total throughput of $\simeq$50\%, falling down to $\simeq$30\%\ on the blue side as a main result of 
ZnSe absorption.  
The spectrograph design was achieved by the SPIRou team members at Institut de Plan\'etologie et d'Astrophysique de Grenoble (IPAG) in France, with 
contributions from team members at UL \citep{Thibault12}.

\begin{figure}
%%% \center{\includegraphics[angle=-90,scale=0.34]{fig/spirou_foc.ps}} 
\center{\includegraphics[scale=0.34,bb=158 40 611 476]{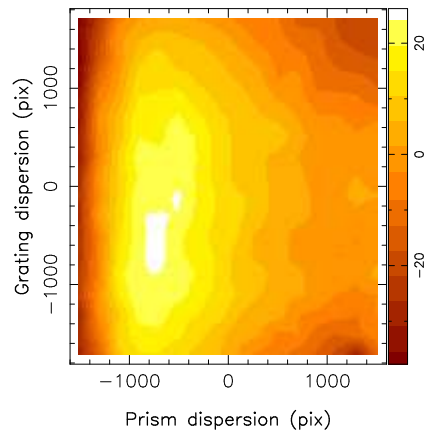}} 
\caption{Distance to the optimal focus (in \mic) over a 3200$\times$3600-pixel region covering most of the SPIRou spectrum (red is on the left 
and orders run vertically).  The focus variations mostly reflect the longitudinal chromatism of the optical design, with a 9.5~\mic\ shift with 
respect to the optimal focus. } 
\label{fig:foc}
\end{figure}

The alignment of the optical components, achieved at room temperature, consists of 5 main steps.  We start by installing four mechanical masks featuring 
pinholes at predefined locations, defining the reference path of the optical beam, and accurately positioned on the optical bench (in front of the 
parabola and of the folding flat mirror) and on the parabola;  with the dispersing elements (prisms and grating) replaced by a flat mirror located 
at the correct distance and angle, the parabola and the fold mirror are then tuned in position and orientation with respect to the reference masks 
on the optical bench until a collimated red laser beam passes through all pinholes.  Using several paper targets, the dispersing components and the 
camera are then installed on the bench so that the spot from the collimated red laser and/or from a fibred nIR laser beam matches the expected position 
on each optical surface.  Focusing the slicer and the folding mirror is achieved thanks to 
a large autocollimator (1-m focal length and 10-cm pupil), down to a precision of about 50~\mic.  
The fourth step consists in tuning the detector focus and tilt by introducing calibrated shims in the detector mount until the proper setting is 
achieved, down to a precision of about 10~\mic;  this step is the most delicate one, requiring several cooling cycles to converge.  
In the last step, we move the parabola to its {\emr nominal `cold position at room temperature', i.e., the position that it should have at room temperature 
to ensure perfect optical alignment and focus once the whole spectrograph reaches its operating temperature of 73.5~K.}   
Once all components on the spectrograph bench reached operating temperature, we use a Hartmann mask in front of the camera (alternatively hiding the top 
and bottom halves of the beam) to estimate the distance to the optimal focus for all regions of the detector;  {\emr the detector is positioned with respect 
to the overall optimal focus to a precision better than 10~\mic\ (see Fig.~\ref{fig:foc}). } 
The spectrograph optical alignment was devised and optimized by the SPIRou team members at OMP/IRAP, to ensure a straightforward reassembly and 
realignment at the CFHT \citep{Challita18}.  

The whole spectrograph with its optical bench (of total weight 500~kg) is enclosed in a cryogenic dewar (of external diameter 1.73~m and length 2.87~m) 
and mounted on a table supported at three points by an hexapod system from an internal support frame at room temperature (for a total weight of 3.5~tons).  
The spectrograph and optical bench are evacuated (down to a typical pressure of $10^{-6}$~mbar) and cooled down to 73.5~K via two cryocoolers acting on 
a cold bus that spreads cold to the whole optical bench.  The bench and spectrograph optics are enclosed in a multiple layer of active and passive 
thermal shields to further stabilize the temperature of all optical components to better than 1~mK.  
This high-level of thermal stability is required to ensure that the science channel does not drift in velocity with respect to 
the reference channel by more than a fraction of a \ms\ in a timescale of one night.  

\begin{figure}
\includegraphics[scale=0.2,bb=0 0 700 420]{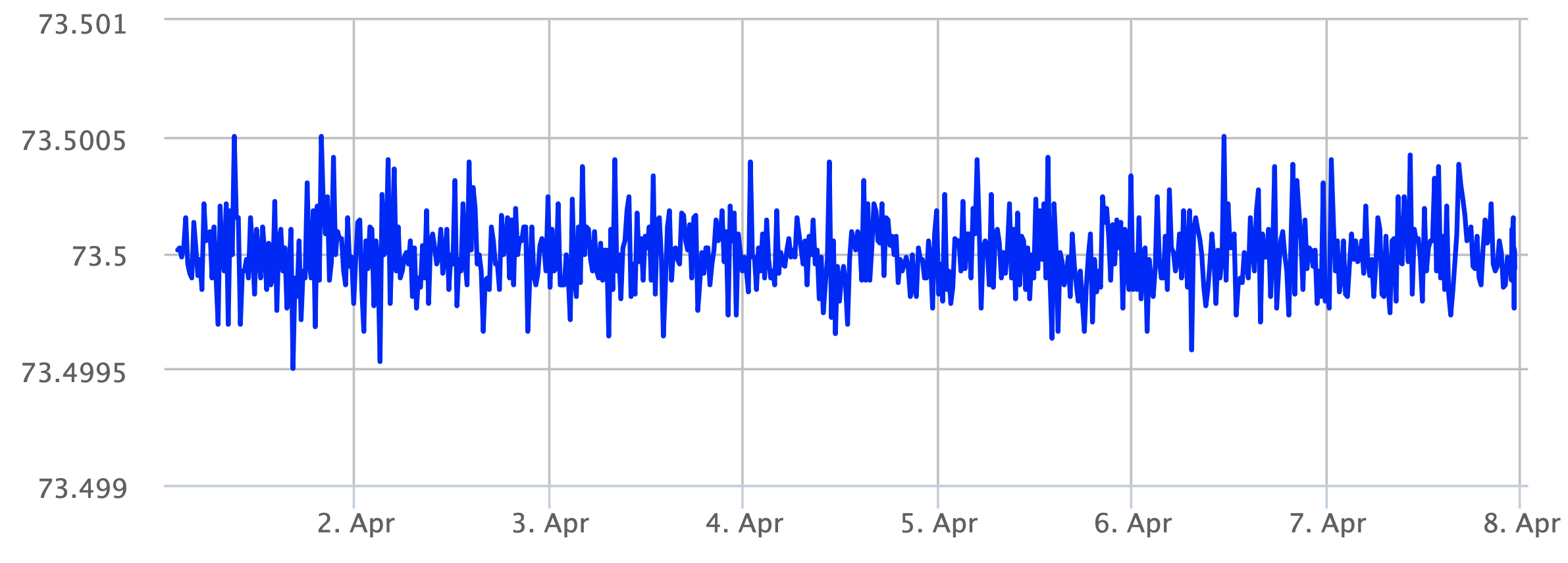}
\includegraphics[scale=0.2,bb=0 0 700 420]{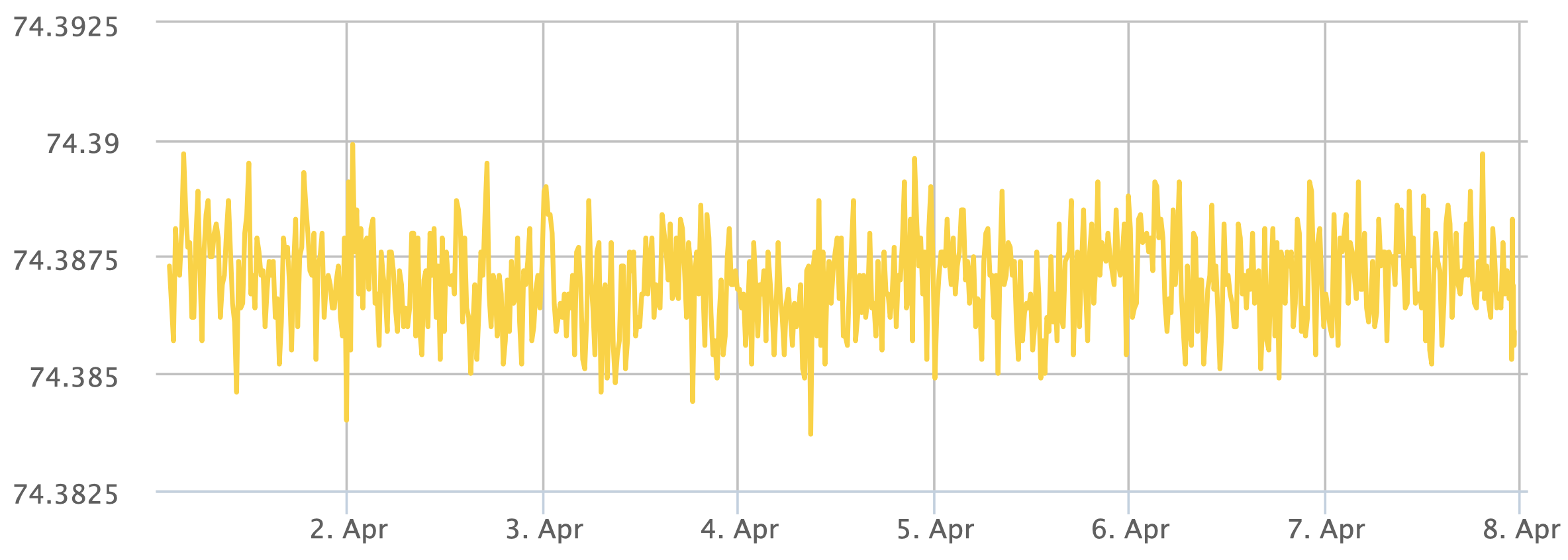}
\caption{Temperature of one of the 3 control points of the SPIRou spectrograph bench (top panel) and of the (uncontrolled) SPIRou grating over a timescale 
of a week in early April 2020, respectively showing thermal stabilities of 0.2 and 1.0~mK RMS over 24 hr periods. } 
\label{fig:temp}
\end{figure}

The temperature of the optical bench is stabilized at three specific points (located below the grating, the camera and the parabola respectively) 
through 3 control loops equipped with Isotech MicroK-500 / Cernox CX-1080 sensors, ensuring a typical RMS stability of 0.2~mK on timescales of 24~hr 
(see Fig.~\ref{fig:temp} top panel);  the temperature of the active shield is controlled with more standard components at a typical precision 
of a few mK RMS.  The (uncontrolled) temperature of optical components and of various bench locations are constantly monitored every few seconds, 
and are all found to be stable to within 1~mK RMS (see, e.g., Fig.~\ref{fig:temp} bottom panel showing the SPIRou grating temperature over a 
timescale of a week), demonstrating that the technical requirement on the thermal stability of the spectrograph is met. 
The typical cool-down time of the whole instrument from ambient temperature to 73.5~K is about 9~d, whereas the additional time for all optical 
components to stabilize at a level of 1~mK RMS is another 11~d.  With a warm up time of about 6~d, the duration of a full thermal cycle is close to a 
month.  

The design and construction of the cryogenic dewar were achieved by the SPIRou team members at the Hertzberg Institute of the National Research Council 
of Canada \citep[NRC-H, Victoria, Canada,][]{Reshetov12}, then delivered to OMP/IRAP where SPIRou was integrated and extensively tested in a clean room.  

\subsection{Calibration \& RV reference unit}
\label{sec:cal}

\begin{figure*}
\vspace{-5mm}
%%% \mbox{\includegraphics[scale=0.48,bb=15 30 550 375]{fig/spirou_fptot.ps}\includegraphics[scale=0.48,bb=15 30 550 375]{fig/spirou_hc.ps}} 
%%% \mbox{\includegraphics[scale=0.48,bb=15 30 550 375]{fig/spirou_lfc.ps}\includegraphics[scale=0.48,bb=15 30 550 375]{fig/spirou_adleo.ps}} 
\mbox{\includegraphics[scale=0.48,bb=45 260 570 600]{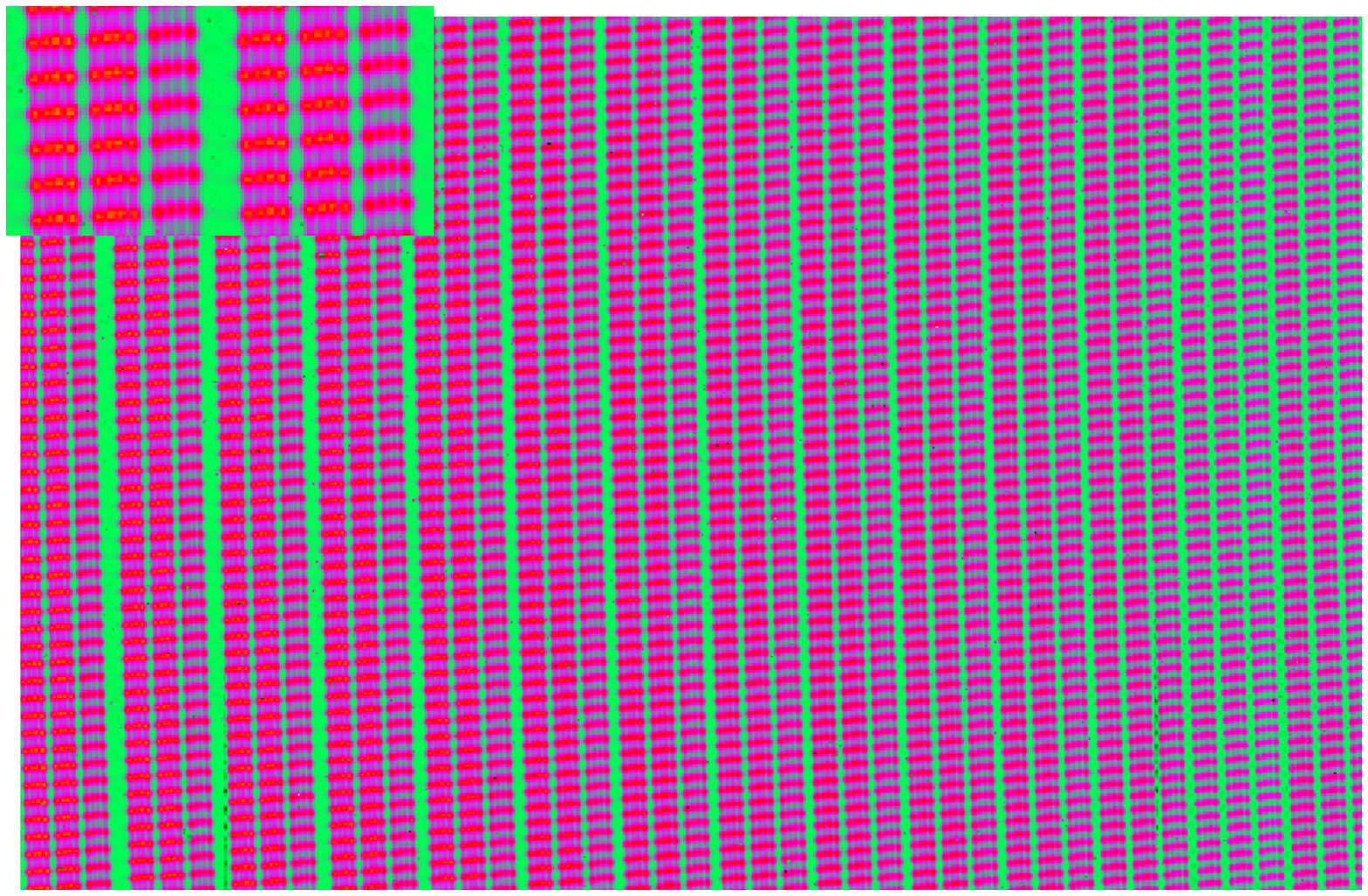}\includegraphics[scale=0.48,bb=45 260 570 600]{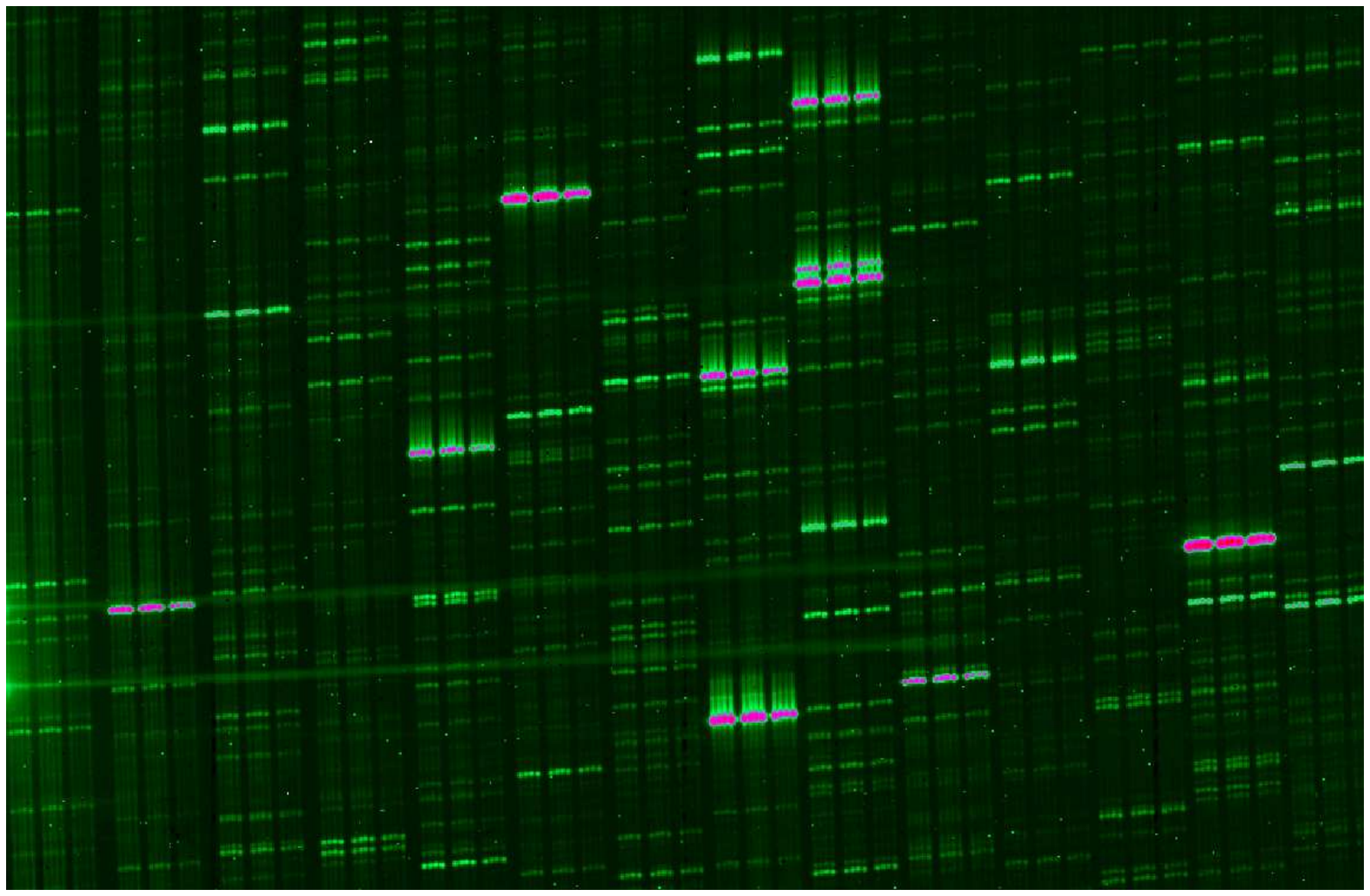}} 
\mbox{\includegraphics[scale=0.48,bb=45 260 570 600]{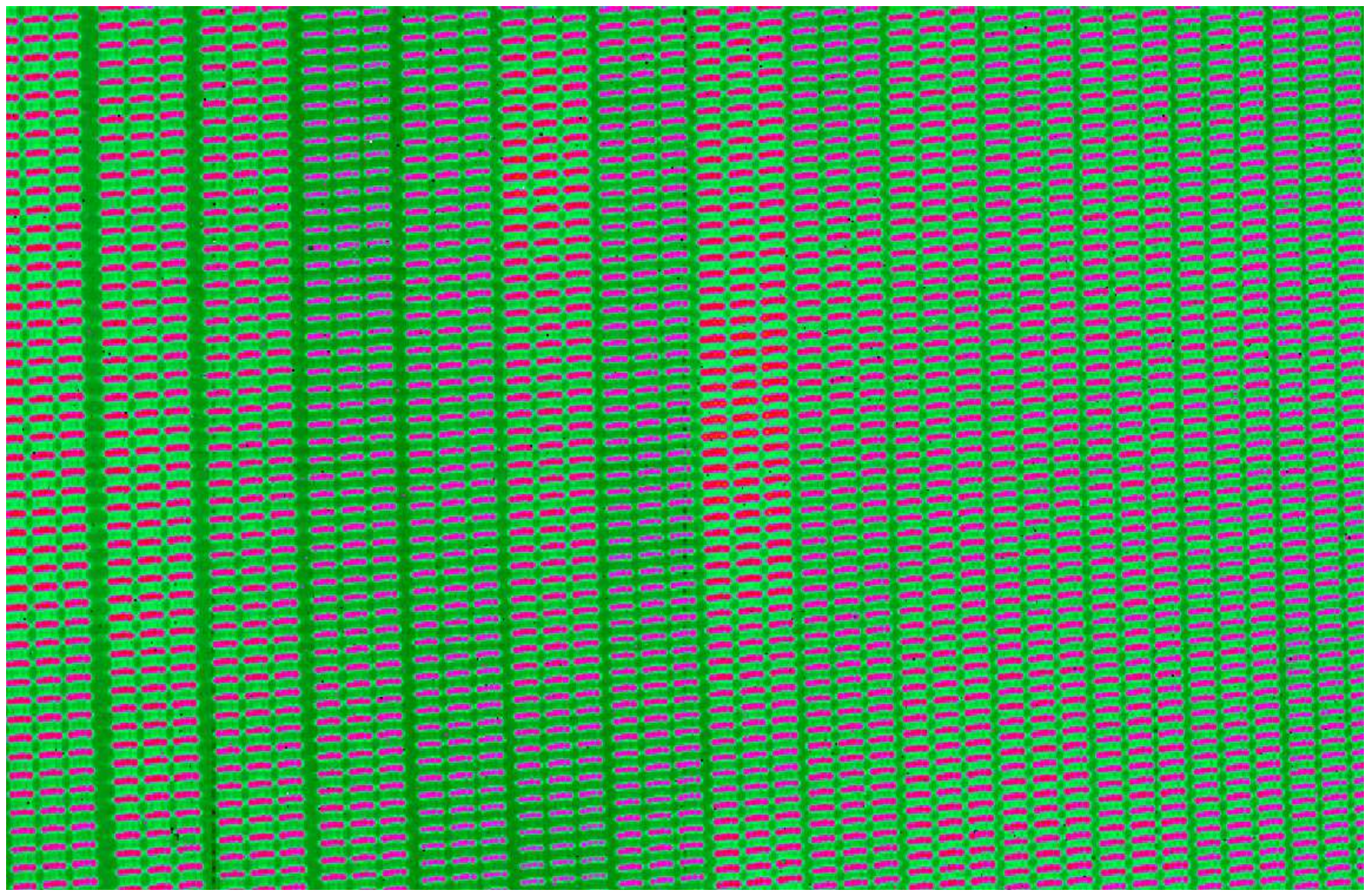}\includegraphics[scale=0.48,bb=45 260 570 600]{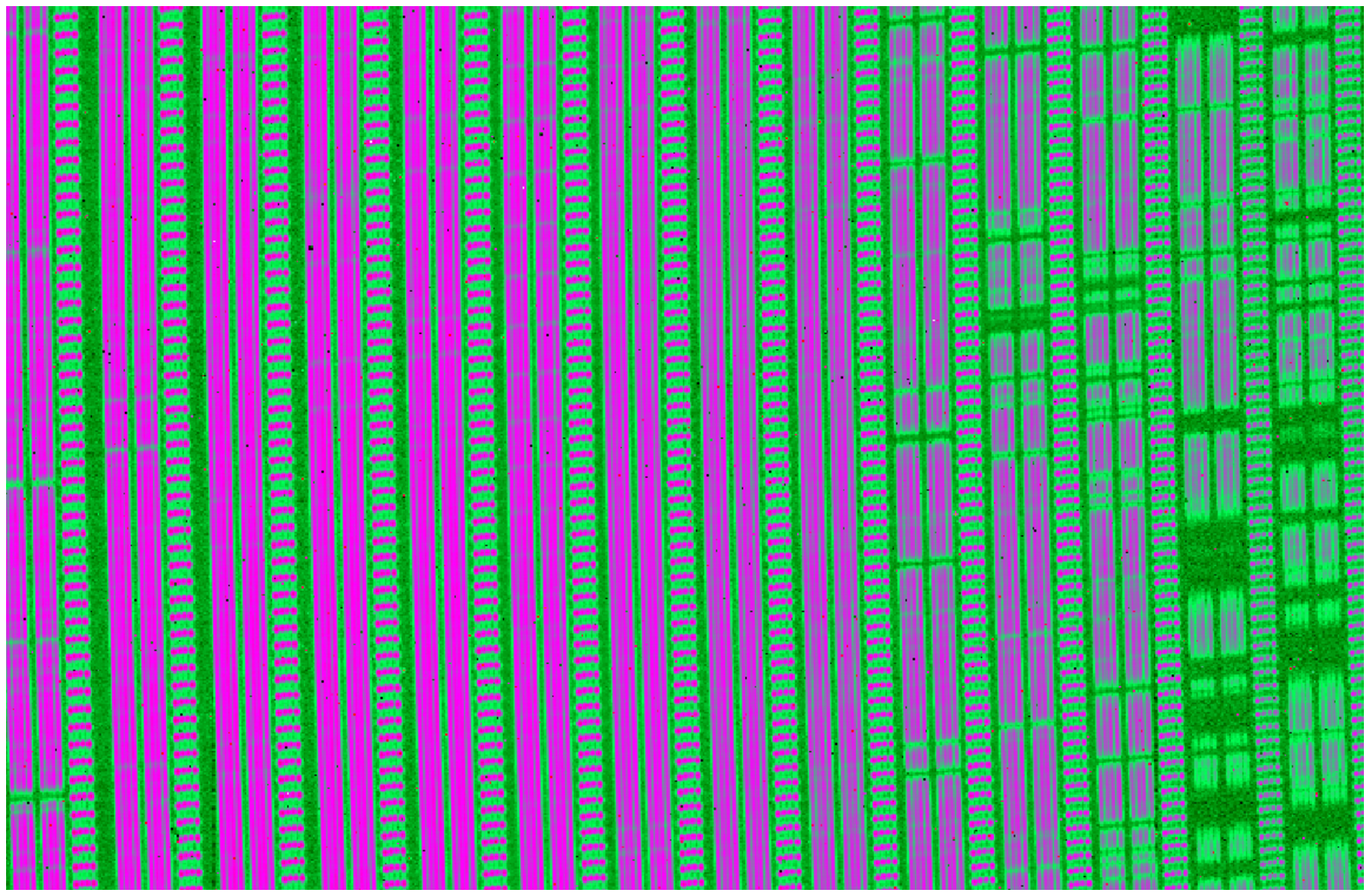}} 
\caption{Examples of SPIRou calibration and science frames collected on 2020 March 11, in the case of the FP (top left), HC (top right), LFC 
(bottom left) and science+FP 
exposures (for the active M dwarf AD~Leo).  The image only shows a portion at the centre of the detector, with 14 (out of the 50) recorded orders 
(running vertically, from order \#43 on the left to \#56 on the right, corresponding to mid wavelengths of 1787 to 1373~nm).  Each order 
displays a restricted spectral interval about flux maximum, and features 3 spectra, the left and middle ones corresponding to the science 
channels, and the right one associated with the reference channel.  In the FP, HC, and LFC frames, all 3 channels feature calibration 
spectra, whereas the science frame exhibits the spectrum of AD~Leo (with both stellar and telluric absorption lines clearly visible) 
accompanied by a FP spectrum on the reference channel.  The insert in the top left image is a zoom on the 3 channels of two consecutive orders, 
showing the quadruple-slice pattern of each channel.}  
\label{fig:cal}
\end{figure*}

The purpose of the calibration unit, located close to the spectrograph cryostat, is to provide light from the various lamps needed to accurately 
calibrate observed spectra.  It includes: 
\begin{itemize} 
\item a halogen lamp (with a radiation temperature of 3000~K) to collect the flat field (FF) exposures needed to 
      locate orders on the detector and to correct for the spectral response of the instrument; 
\item a U/Ne hollow-cathode (HC) lamp featuring about 2k lines over the spectral domain, to secure a precise absolute pixel-to-wavelength calibration 
      (a Th/Ar HC lamp is also available, with less lines over the domain and thus not used in operations); 
\item an evacuated ($\simeq$0.1~mbar) and temperature-stabilised ($\simeq$1~mK RMS on 24~hr slots) Fabry-Perot (FP) etalon coupled to a halogen lamp, 
      featuring about 15k lines across the spectral range, to track the detailed shape of the slit-like image that the pupil slicer generates at the 
      spectrograph entrance, and to monitor spectrograph drifts during observations (with the FP spectrum on the reference fiber during on-sky 
      observations) at a precision of a few 0.1~\ms.  
\end{itemize}
FP exposures are also used in conjunction with HC exposures to refine the wavelength solution over the whole spectral range, especially in the reddest 
orders where the U/Ne lamp features very few lines (Hobson et al.\ 2020, submitted).  
During tests and daily calibrations, light from calibration lamps is also injected into the 2 
science fibers via the calibration wheel of the Cassegrain unit, so that all 3 spectral channels can be calibrated together in a consistent and 
homogeneous way.  Typical calibration and science frames are shown in Fig.~\ref{fig:cal}.  As the FP spectrum shifts by 1~\ms\ for internal 
pressure and temperature changes of 0.006~mbar and 3.1~mK respectively, which can happen on a timescale of a few d, daily calibrations are 
essential so that all drifts can be accurately corrected for.  The calibration unit was designed and integrated at Observatoire de Haute-Provence / 
Laboratoire d'Astrophysique de Marseille \citep[OHP/LAM, France,][]{Boisse16,Perruchot18}, and the RV reference module was developed at 
Observatoire de Gen\`eve \citep[OG, Switzerland,][]{Wildi12,Cersullo17}.  

In addition to the standard calibration lamps integrated within the calibration unit and delivered with the instrument, SPIRou was recently equipped with 
a Laser Frequency Comb (LFC) engineered by Menlo Systems (MS, Germany).  The SPIRou LFC features a repetition 
rate of 13.0~GHz and an offset of 5.31~GHz, with lines covering a spectral range of 1.0-2.2~\mic, i.e., about 10\% and 20\% smaller in log scale than the 
nominal and overall SPIRou domains respectively (see Table~\ref{tab:spi}).  This limitation reflects the extreme challenge of manufacturing a 
tapered Photonic-Crystal Fiber (PCF) capable of generating the desired spectral broadening from the initial laser comb anchored at 1553~nm.  
Up to now, MS succeeded in obtaining a hybrid LFC with a blue and a red arm, each arm feeding the multi-mode output LFC fiber with its own single mode 
fiber;  whereas the red arm is equipped with an active flattening unit incorporating a spatial light modulator (SLM) supposed to trim all lines at a given 
intensity (within a few dB), the blue arm only includes a filter-based passive flattening scheme (at a level of 13~dBs).  As a result of 
time-variable fringing patterns in the SLM, line intensities in the red arm of the LFC are found to vary by up to 20~dB over the red range (and up to 
10~dB on regions of only a few lines).  
Work is ongoing at MS to manufacture a new tapered PCF that can generate a spectral broadening matching the full range of SPIRou in one 
shot, and in upgrading the flattening unit and its SLM so that the LFC can achieve nominal performance in terms of homogeneity and stability of line 
intensities.  An example LFC frame is shown in Fig.~\ref{fig:cal} (bottom left panel), where the residual fringe pattern is clearly visible.  
{\emr Once upgraded, the LFC will also be ideal for monitoring the SPIRou instrumental profile over time.  }

\subsection{Installation \& operation at the CFHT}
\label{sec:ope}

SPIRou was first integrated and tested at OMP/IRAP until 2017 December, then shipped to the CFHT.  The spectrograph was installed in the Coud\'e room 
at the 3rd floor of the telescope building in 2018 February, whereas the spectrograph control electronics, the compressors and the calibration unit 
were installed in the adjacent room where the Coud\'e pipe (coming from the 5th floor) emerges.  The Cassegrain unit is mounted 
at the f/8 Cassegrain focus of the telescope, and connected to the spectrograph and calibration unit by the fiber link rooted through the Coud\'e 
pipe (with a 15\degC temperature difference between the two).  
SPIRou underwent extensive on-site testing throughout 2018, both in the lab and on the sky, 
until successfully passing the final acceptance review in 2019 January.  

Instrument control was designed to operate in a CFHT-like environment, with a master process (the Director) controlling a number of slave 
processes (Agents), each in charge of a SPIRou subsystem (such as the Cassegrain unit, the spectrograph temperature control, the calibration 
unit).  The most challenging aspects were the control of the TTM on the Cassegrain side (to ensure optimal behaviour up to 50~Hz), and the temperature 
control on the spectrograph side (to achieve the best possible thermal stability of the optical components).  More details can be found in 
\citet{Barrick12} and in \citet{Barrick18}.  All passive and active elements / probes are constantly monitored, with data on timescales ranging from 1~hr 
to 1~wk being accessible from the web on a `status server'.  

Agitators were added to all fiber cables linking the different instrument modules (calibration and Cassegrain units, spectrograph, RV reference, LFC), 
to ensure optimal scrambling of the modal noise, that can impact observations in the nIR, especially in the reddest orders.  
Significant progress was achieved since SPIRou was installed at the CFHT, though there is still room for improvements, 
in particular with the LFC whose nearly monochromatic spectral lines are extremely sensitive to modal noise.  

Shortly after the installation of SPIRou, the CFHT suffered a 7$^{\rm th}$ magnitude earthquake, following which SPIRou became more sensitive 
to vibrations, e.g., induced by earthquakes or even by the activation of the telescope hydraulics, and causing the RV in both science and reference 
channels to jump by tens of \ms.  Fortunately, relative RVs between the science and reference channel are unaffected by these events.  
We suspect that this problem reflects a crack in one of the glue joints that bond the optical components to their mechanical supports, most likely for 
one of the prisms given the preferred direction of the observed spectral shifts.  SPIRou also suffered two breakdowns of its cooling system up to now 
(in 2019 August and 2019 December), 
each time due to a failure of a cold-head motor whose fix required to warm-up the whole instrument;  now equipped with more robust components, and 
with the planned maintenance of the whole cooling system every second year, cold heads should no longer be a source of issues in the future.  

Considerable efforts were also invested into cooling down the Coud\'e room where SPIRou is installed.  With an average room temperature that jumped 
from about 12 to 20\degC following the installation of SPIRou, the thermal background caused by the hermetic feedthroughs reached a high point of 
about 40~\phpspA\ at 2.35~\mic.  Heat extraction was implemented (especially in the adjacent room hosting the spectrograph compressors) to bring the Coud\'e 
room back to a temperature of about 15\degC, and further action is undergoing to further decrease it to about 10\degC\ \citep[e.g.,][]{Look18}.  Active 
cooling is also being implemented on the hermetic feedthroughs to attempt bringing their temperature down to 5\degC.  

A hardware upgrade is planned for early 2021, aiming at reducing SPIRou's sensitivity to vibrations, at changing the pupil slicer {\emr (now including 
a stress-free mount of the octagonal fibers)}  to enhance 
the near-field scrambling efficiency, and at improving the cooling performance on the hermetic feedthroughs in order to further decrease 
their (currently dominant) contribution to the thermal background (see Secs.~\ref{sec:fib} and \ref{sec:perf})  

Science observations, including the SLS, started in 2019 February at a rate of about 130 nights per year (mostly during bright time), with 
SPIRou being operated in Queue Service Observing (QSO) mode from the start.  
{\emq On-sky science SPIRou data either consist of individual exposures collected in a fixed rhomb configuration, or of (circular or 
linear, i.e., Stokes $V$, $Q$ or $U$) polarization sequences of 4 subexposures in pre-defined rhomb orientations.}     
In addition to the science programmes being carried out, 
SPIRou nightly collects exposures on a few telluric and RV standard stars, as well as on the sky.  
SPIRou spectra of calibration lamps are also recorded twice a day, at the beginning and at the end of each SPIRou observing night.  
An average SPIRou night is found to last 7~hr of science observing time (and read-out), once pointing overheads, night calibration time and losses to 
weather are discounted.

\subsection{Data reduction pipeline}

The SPIRou reduction pipeline, called APERO (standing for A PipelinE to Reduce Observations, Cook et al.\ 2020, in prep), works in several steps.  
The first one consists in analyzing calibration files collected at the beginning and end of any given observing night, to derive the geometrical elements 
(shape of orders from FF frames, shape of slit from FP frames, position of lines in HC and FP frames) thanks to which all frames from this observing 
night can be straightened into a reference coordinate system associated with a set of reference calibration files.  In a second step, stellar spectra on 
the science channel (and their reference FP spectra in the reference channel whenever relevant) are extracted from each straightened science frame using 
optimal extraction \citep{Horne86}, and wavelength calibrated using a master pixel-to-wavelength conversion formula also derived from the set of 
reference calibrations (using lines from both HC and FP frames, see Sec.~\ref{sec:cal}, as in, e.g., Hobson et al.\ 2020, submitted).  
Finally, stellar spectra are corrected from telluric lines using a PCA approach applied to a data base of SPIRou 
spectra of telluric standards collected in a wide range of atmospheric conditions \citep[with several such standards observed every single SPIRou night, 
further expanding the data base,][Artigau et al.\ 2020, in prep]{Artigau14};  spectra are also automatically corrected for the Barycentric Earth Radial 
Velocity (BERV).  

Processed (telluric uncorrected and corrected) spectra are saved in several formats, with both 2D FITS files containing the extracted unnormalized spectra 
of all orders (e2ds files, which can be flattened later-on using a similar s2d file from a FF frame), and 1D FITS files containing flattened spectra with 
all orders merged together and re-binned on a wavelength grid featuring a constant velocity (or wavelength) step.  At the CFHT, an automatic 
trigger ensures that each new recorded SPIRou frame, either calibration or science ones, are processed by APERO as they are being collected;  in a 
second step, full reprocessing of all data from the night is carried out to obtain the final reduction products.  {\emq Whenever individual exposures 
are part of a polarization sequence, APERO also computes a polarization spectrum from the extracted spectra of individual exposures 
\citep[e.g.,][]{Moutou20, Martioli20}.}   

In the case of polarization sequences (i.e., most SLS data), consisting of series of 4 consecutive sub-exposures collected in different 
positions of the Fresnel rhombs \citep[to remove systematics to first order,][]{Donati97b}, APERO produces the same kind of data output, along with 
the corresponding polarization spectra (for the selected polarization state / Stokes parameter) and a null polarization spectrum \citep[with the 4 
subexposures processed in a different way so that polarization signatures cancel out,][]{Donati97b}.  

In addition to the extracted SPIRou spectra, APERO generates several by-products, e.g., the cross-correlation function (CCF) of 
the spectrum with a line mask matching best the spectral type of the observed target, from which preliminary RV estimates are obtained.  
(In the near future, APERO will incorporate a larger set of line masks than the current limited sample of 4).  
Zeeman signatures will also be derived using Least-Squares Deconvolution \citep[LSD][]{Donati97b} as a by product in the case of polarization 
sequences.  Additional work may be needed from the PIs, in particular regarding telluric correction, to further optimize the derivation of precise 
RVs and accurate Zeeman signatures.  

Currently in version 0.6.100, APERO is available from GitHub, and is still undergoing regular updates to improve reduction performances, in 
particular regarding telluric correction and RV precision \citep[][Cook et al.\ 2020, in prep]{Artigau12}.  
Alternative pipelines are also being developed in parallel \citep[one being based on the ESPaDOnS pipeline {\tt Libre-ESpRIT},][]{Donati97b}, to 
compare with APERO results and ultimately ensure that reduction recipes are optimized so that SPIRou data can be exploited at their best.  

\begin{table*}
\caption[]{Summary of the main characteristics and performances of SPIRou} 
\center{
\begin{tabular}{cc}
\hline 
Nominal spectral range in a single exposure (w/ no gaps)   & 0.98--2.35~\mic\ in 46 orders (\#78 to \#33)  \\
Overall spectral range (w/ a 2-nm gap between 2.4371 and 2.4391~\mic) & 0.95--2.50~\mic\ in 50 orders (\#80 to \#31)  \\
Spectral resolving power estimated from calibration lines  & 64$\pm$1k                  \\
Spectral resolving power estimated from stellar spectra    & 70$\pm$3k                  \\
Velocity bin size of detector pixel                        & 2.28~\kms            \\
On-sky diameter of circular instrument aperture            & 1.29\arcsec         \\
Tip-tilt / guiding precision on instrument aperture (RMS)  & 0.01\arcsec\ up to H$\simeq$10   \\
\hline 
Throughput                                                 & 3.5\%\ / 7\%\ / 10\%\ / 12\%\ in YJHK bands \\
SNR performances                                           & peak S/N$\simeq$110 per pixel  at H$\simeq$8 in 300~s for a {\emr M4 dwarf} \\
Polarimetric performances                                  & circular \& linear, sensitivity 10~ppm, crosstalk $\simeq$1\% \\
Radial-velocity stability (RMS on standard stars)          & $\simeq$2~\ms\ on a timescale of a few weeks, goal 1~\ms   \\
Equivalent H magnitude of thermal background at 2.35~\mic  & 8.6, goal 9.5 \\ 
\hline
\end{tabular}}
\label{tab:spi}
\end{table*}

\begin{figure*}
%%% \mbox{\hspace{-2mm}\includegraphics[angle=-90,scale=0.25]{fig/spirou_dom.ps}
%%% \hspace{2mm}\includegraphics[angle=-90,scale=0.25]{fig/spirou_tel.ps} 
%%% \hspace{2mm}\includegraphics[angle=-90,scale=0.25]{fig/spirou_res.ps}} 
\mbox{\hspace{-2mm}\includegraphics[scale=0.25,bb=40 80 700 560]{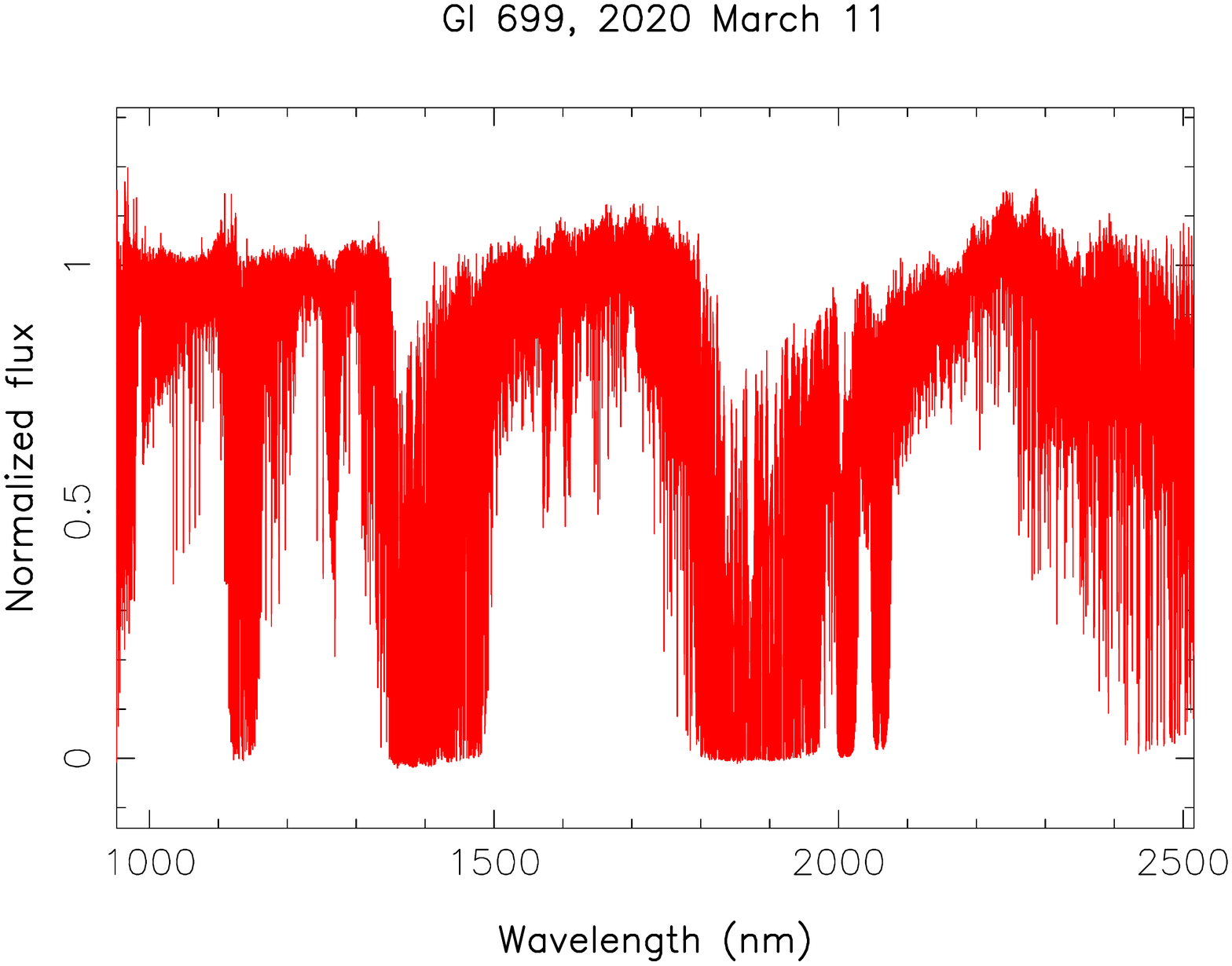}
       \hspace{2mm}\includegraphics[scale=0.25,bb=40 80 700 560]{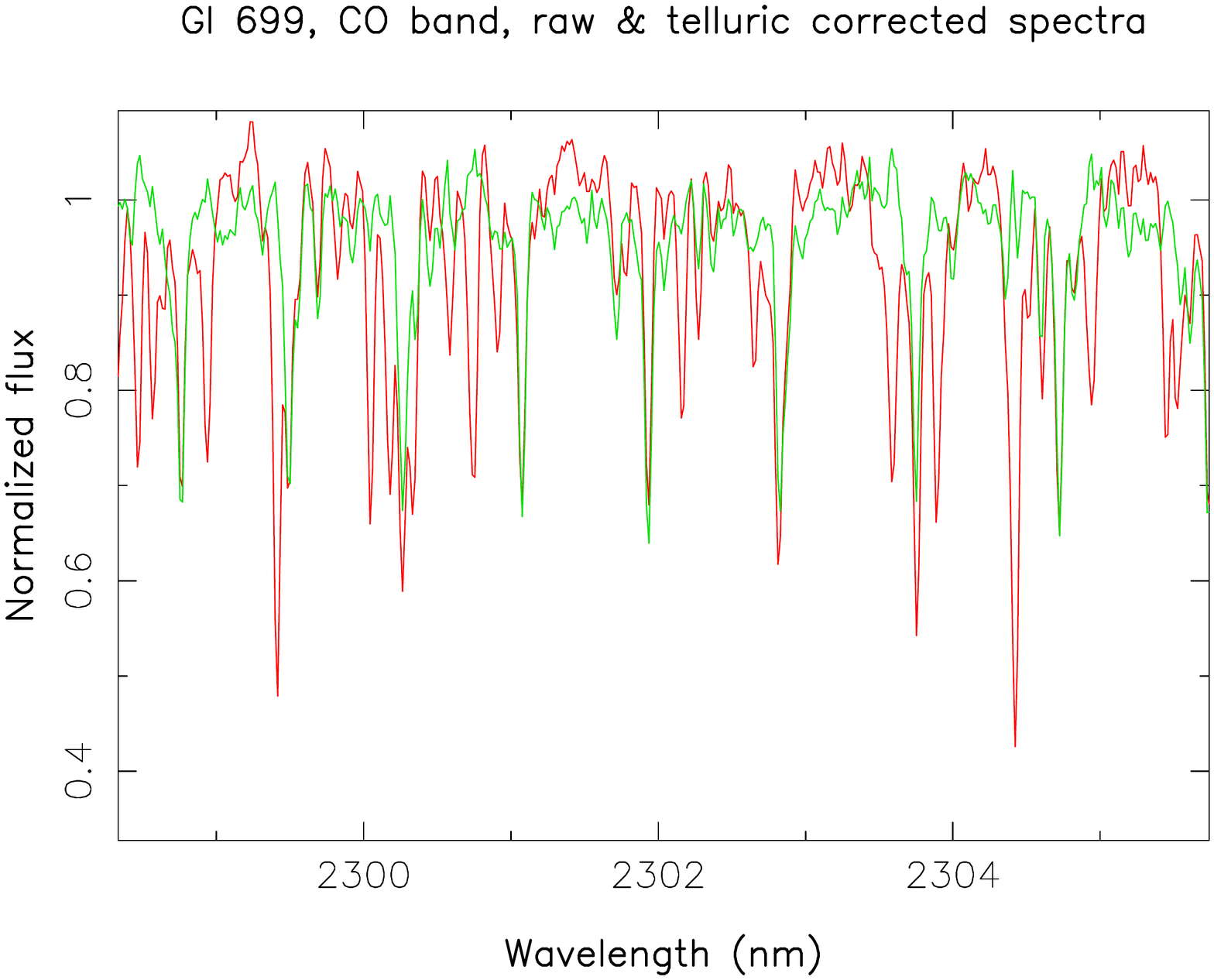} 
       \hspace{2mm}\includegraphics[scale=0.25,bb=40 80 700 560]{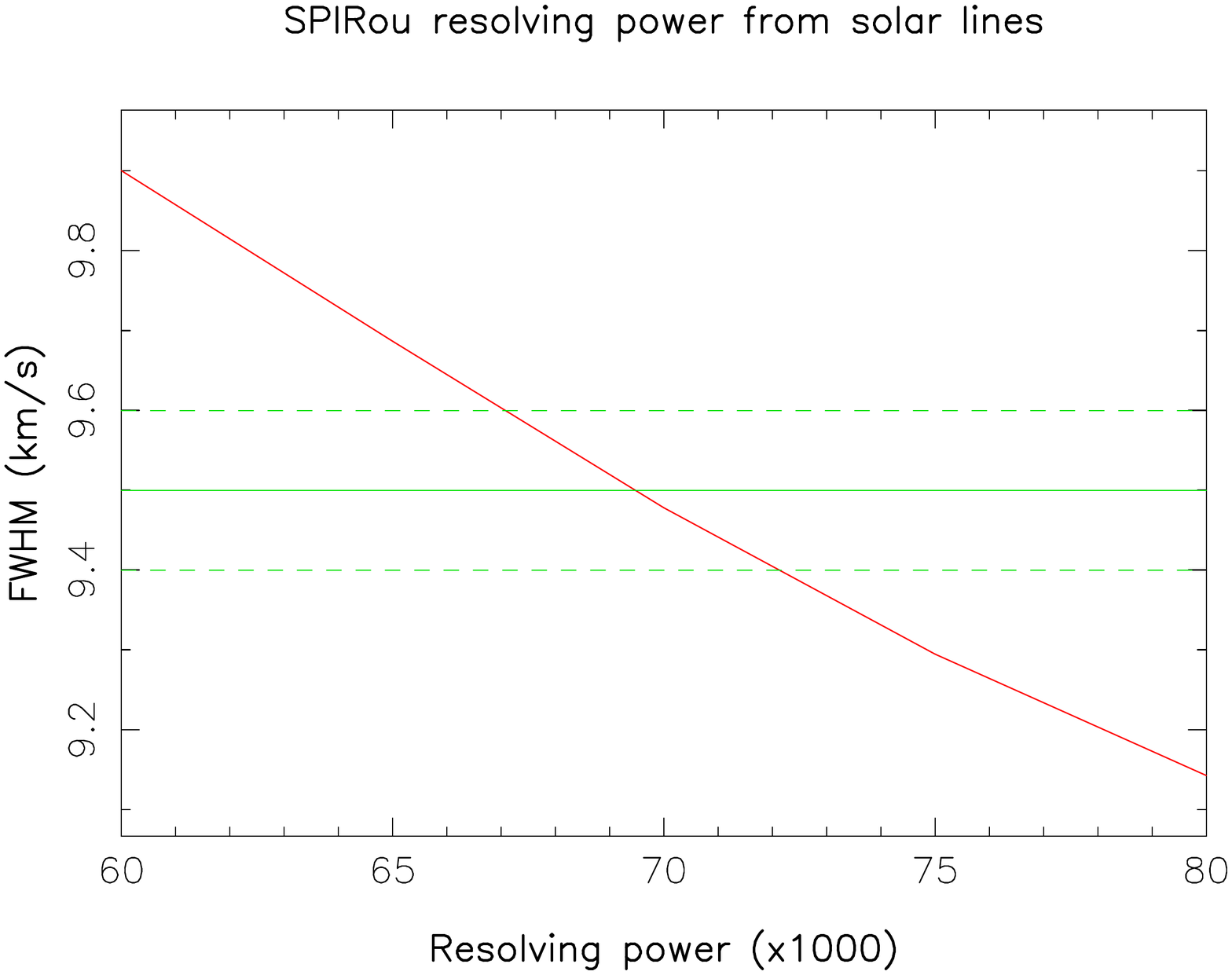}} 
\caption{Left: SPIRou spectrum of the M dwarf Gl~699 over the full spectral domain. 
Middle: Small spectral region of Gl~699 in the $^{12}$CO(2,0) band, before (red) and after (green) telluric correction.  
Right: FWHM of the LSD profiles of solar lines for various resolving powers.  The green lines indicate the value and error bars estimated from SPIRou data, 
whereas the red line traces the prediction computed from a very high-resolution nIR spectrum of the Sun. } 
\label{fig:dom}
\end{figure*}

\subsection{The SPIRou project team} 

Including design, assembly, integration, tests and installation at the CFHT,  SPIRou took about a decade to come to life, with the initial CFHT decision 
to build SPIRou dating from 2010, the Final Design Review occurring in late 2014, the pre-shipping and final acceptance reviews passed in 2017 December and 
2019 January respectively.  Over this decade, the SPIRou project team involved about 50 engineers and scientists from 10 institutes 
around the world, managed by OMP/IRAP hosting the Project Manager (PM), the Principal Investigator (PI), the deputy PM 
and the System Engineer.  
IPAG hosted the SPIRou Project Scientist (PS), UdeM provided the deputy PI and the deputy PS whereas CFHT allocated the Observatory Scientists.  
In each of the main partner sites (IRAP/OMP, OHP-LAM and IPAG in France; CFHT in Hawaii; NRC-H, UdeM and UL in Canada; 
OG in Switzerland), a local PM was in charge of the handled subsystem;  additional contributions were provided by the Institute of Astronomy \& Astrophysics 
of Academia Sinica (ASIAA, Taiwan), the Laborat\'orio Nacional de Astrof\' \i sica (LNA, Brazil) and the Instituto de Astrof\'isica e Ci\^encias do Espa\c{c}o 
(IA, Portugal).

\section{In-lab and on-sky performances}
\label{sec:perf}

In this section, we summarize the performances of SPIRou regarding the main design specifications, which we recall in Table~\ref{tab:spi}.  

\subsection{Spectral domain \& resolving power}

Ranging all the way from 0.95 to 2.50~\mic\ with only a small gap (between 2.4371 and 2.4391~\mic), the overall spectral domain of SPIRou is 
$\simeq$10\%\ larger (in log scale) than the nominal window on which the optical design was optimized.  We show in Fig.~\ref{fig:dom} (left panel) 
a SPIRou spectrum of Gl~699 secured on 2020 March 11, where the Y, J, H and K bands lie between regions featuring strong telluric absorption 
(centred at 1.14, 1.40 and 1.90~\mic).  

The resolving power of SPIRou as measured from U lines in HC frames is equal to 64$\pm$1k, in agreement with design predictions.  
Reducing the pupil in front of the spectrograph camera to a 75~mm circular aperture (with the help of the Hartmann mask) degrades the resolution 
(by about 10\% in the red), demonstrating that the spectrograph is indeed diffraction limited and that optical aberrations have a small impact on 
the FWHM of the instrument profile.  

Estimating the resolving power of SPIRou from the Gaussian broadening impact of the instrument on stellar spectra requires one to have in 
hand a very high-resolution reference spectrum of the star to be observed with SPIRou, so that the profiles of spectral lines can be 
compared.  As the most obvious reference for this test is the Sun for which very high-resolution nIR spectra are available in 
the literature \citep[e.g.,][]{Reiners16}, we observed the spectrum of the Sun (reflected off the Moon) with SPIRou, computed a LSD 
profile of all lines showing no (or very small) telluric contamination, 
and measured its FWHM on a velocity interval of $\pm$16~\kms\ ($\pm$7~pix) about the line center;  repeating the exact same operation on the 
reference spectrum of the Sun smeared at different resolving powers (ranging from 60k to 80k), we find that the Gaussian broadening impact of 
SPIRou corresponds to a resolving power of 70$\pm$3k, in agreement with design predictions (see Fig.~\ref{fig:dom} right panel).  

\begin{figure*}
%%% \mbox{\hspace{-2mm}\includegraphics[angle=-90,scale=0.25]{fig/spirou_thru.ps}
%%% \hspace{2mm}\includegraphics[angle=-90,scale=0.25]{fig/spirou_gl699.ps}
%%% \hspace{2mm}\includegraphics[angle=-90,scale=0.25]{fig/spirou_gj1151.ps}} 
\mbox{\hspace{-2mm}\includegraphics[scale=0.25,bb=40 80 700 560]{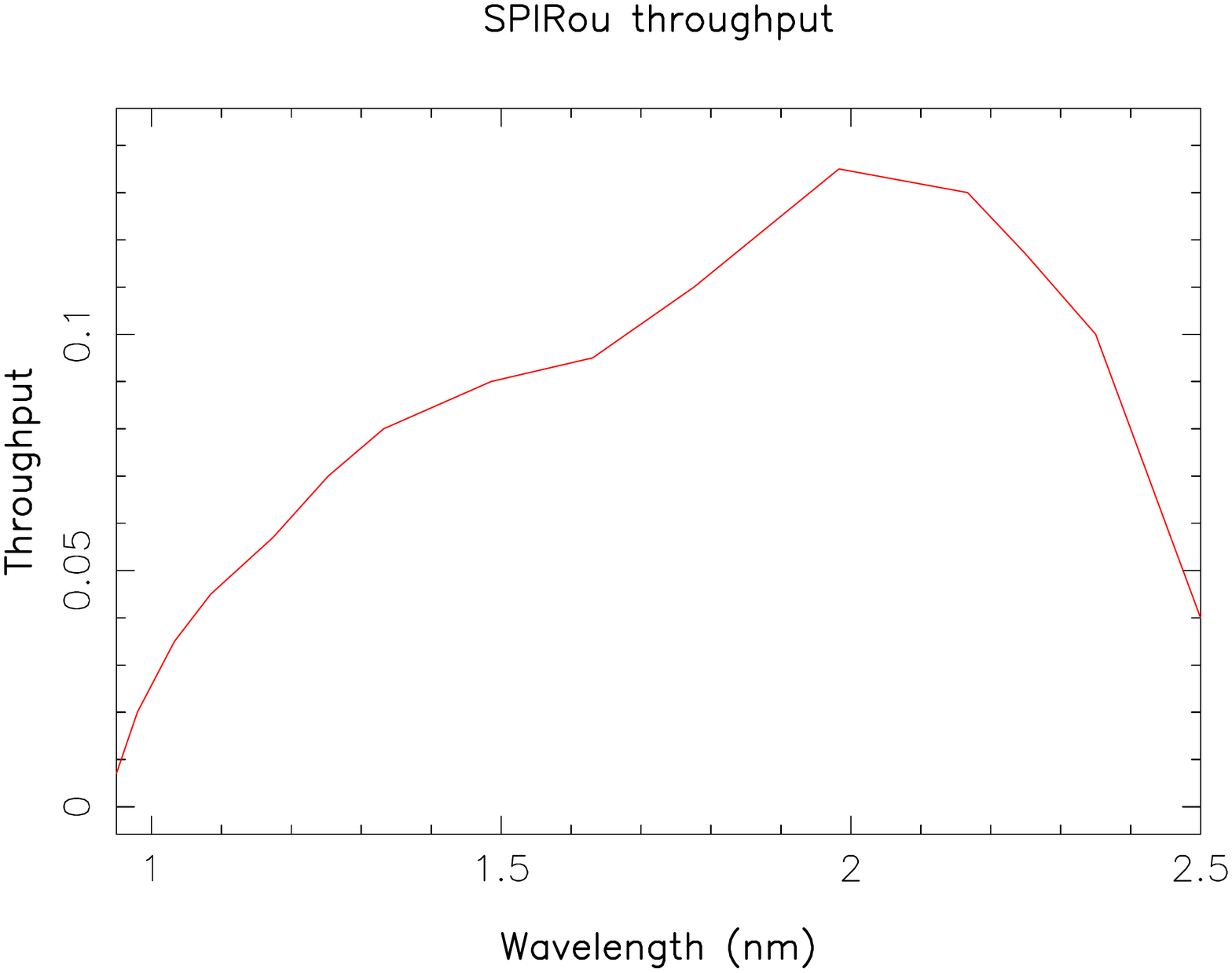}
       \hspace{2mm}\includegraphics[scale=0.25,bb=40 80 700 560]{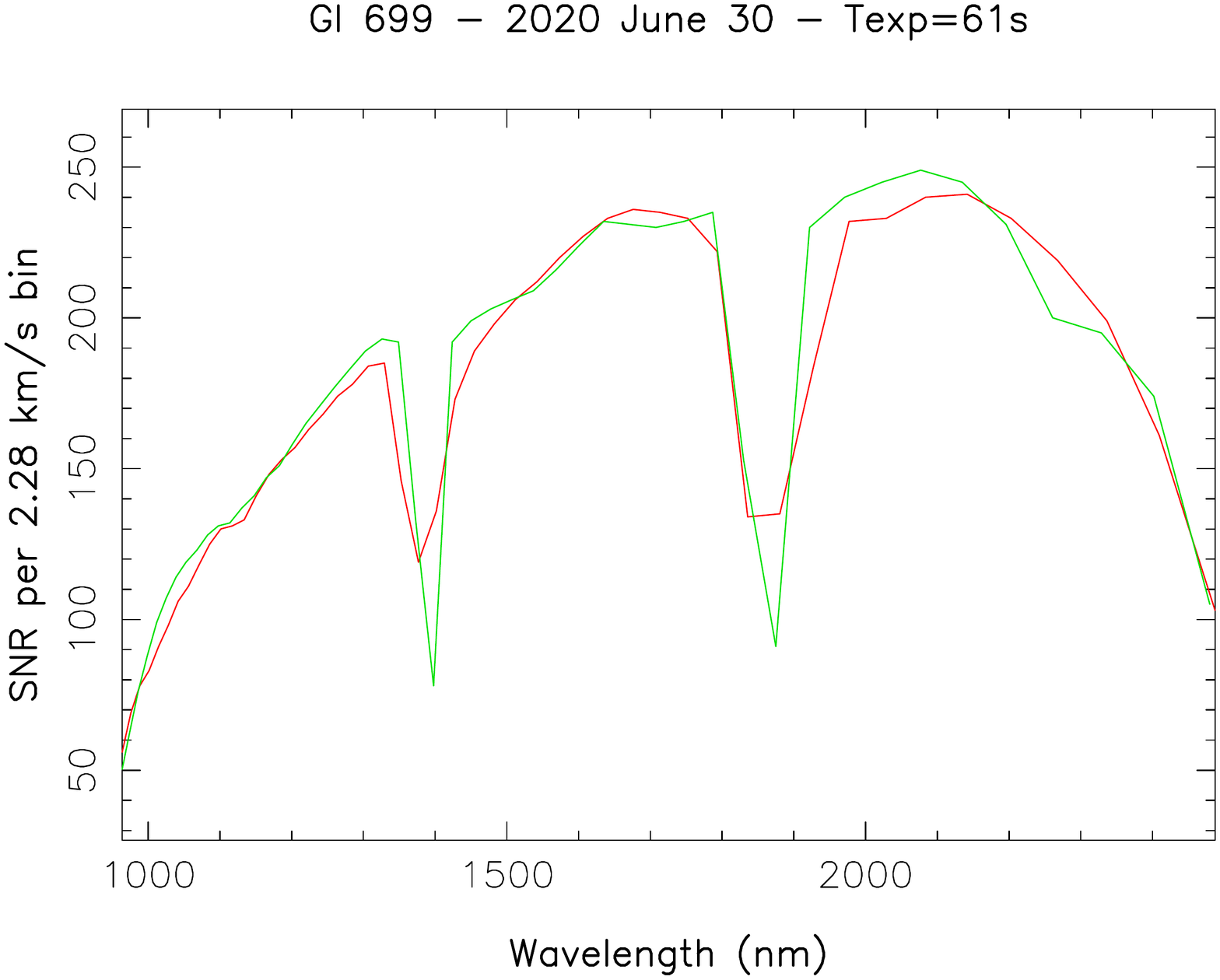}
       \hspace{2mm}\includegraphics[scale=0.25,bb=40 80 700 560]{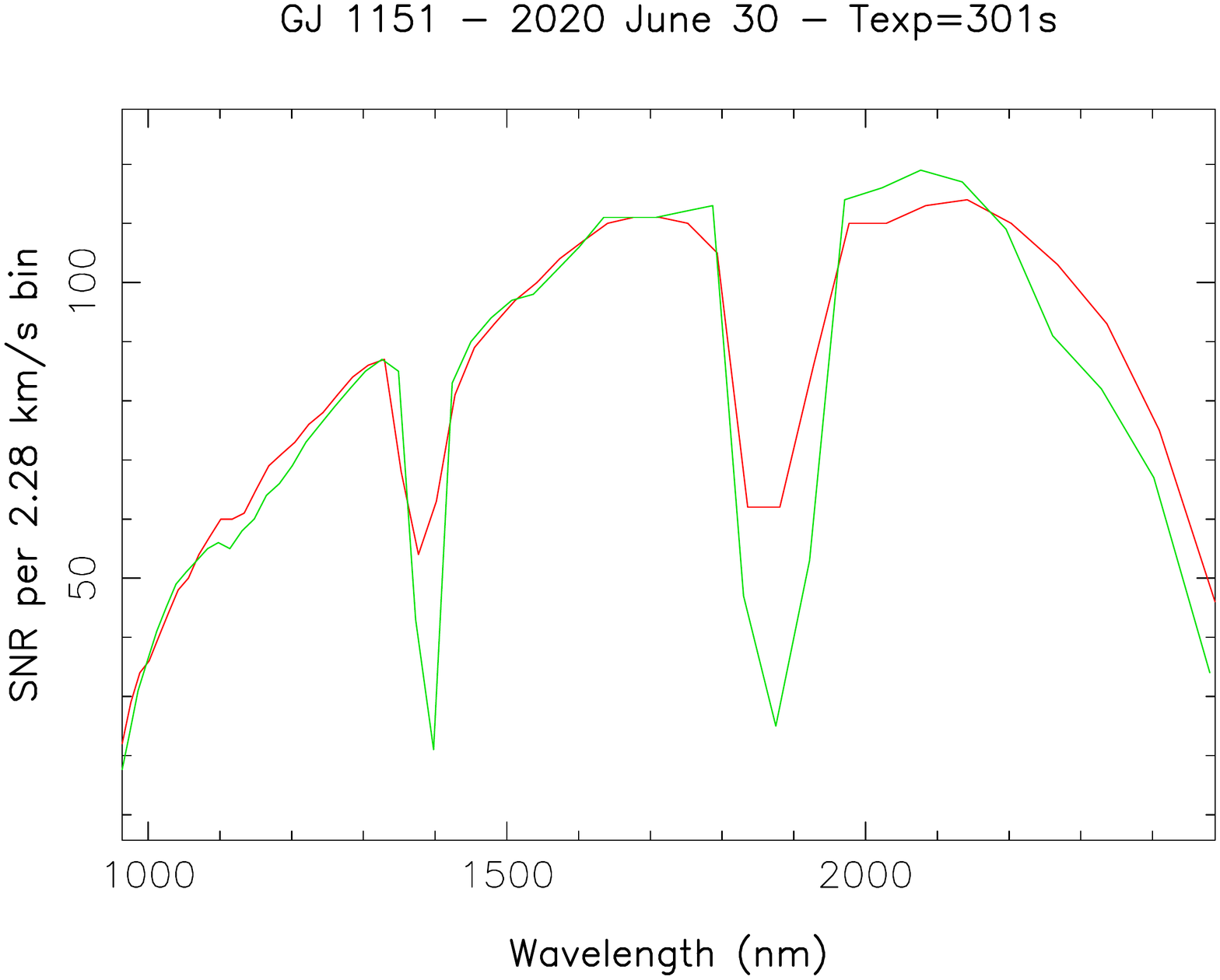}} 
\caption{SPIRou overall throughput (injection and atmospheric losses excluded, left panel) and SNR curves for 61~s and 301~s exposures on the 
M dwarfs Gl~699 (H=4.83) and GJ~1151 (H=7.95) typical of SLS-PS targets (middle and right panels respectively) over the whole spectral domain (except 
in regions where telluric lines are too dense for the stellar continuum to show up, i.e., around 1.40 and 1.85~\mic, see Fig.~\ref{fig:dom} left panel).  
The green lines are the observations (secured on 2020 June 30 in good weather conditions) and the red lines the ETC predictions (not taking into 
account thermal background).  } 
\label{fig:thru}
\end{figure*}

\begin{figure*}
%%% \mbox{\hspace{-2mm}\includegraphics[angle=-90,scale=0.25]{fig/spirou_rvfp.ps}
%%% \hspace{2mm}\includegraphics[angle=-90,scale=0.25]{fig/spirou_rvgl514.ps}
%%% \hspace{2mm}\includegraphics[angle=-90,scale=0.25]{fig/spirou_rvgl699.ps}} 
\mbox{\hspace{-2mm}\includegraphics[scale=0.25,bb=40 80 700 560]{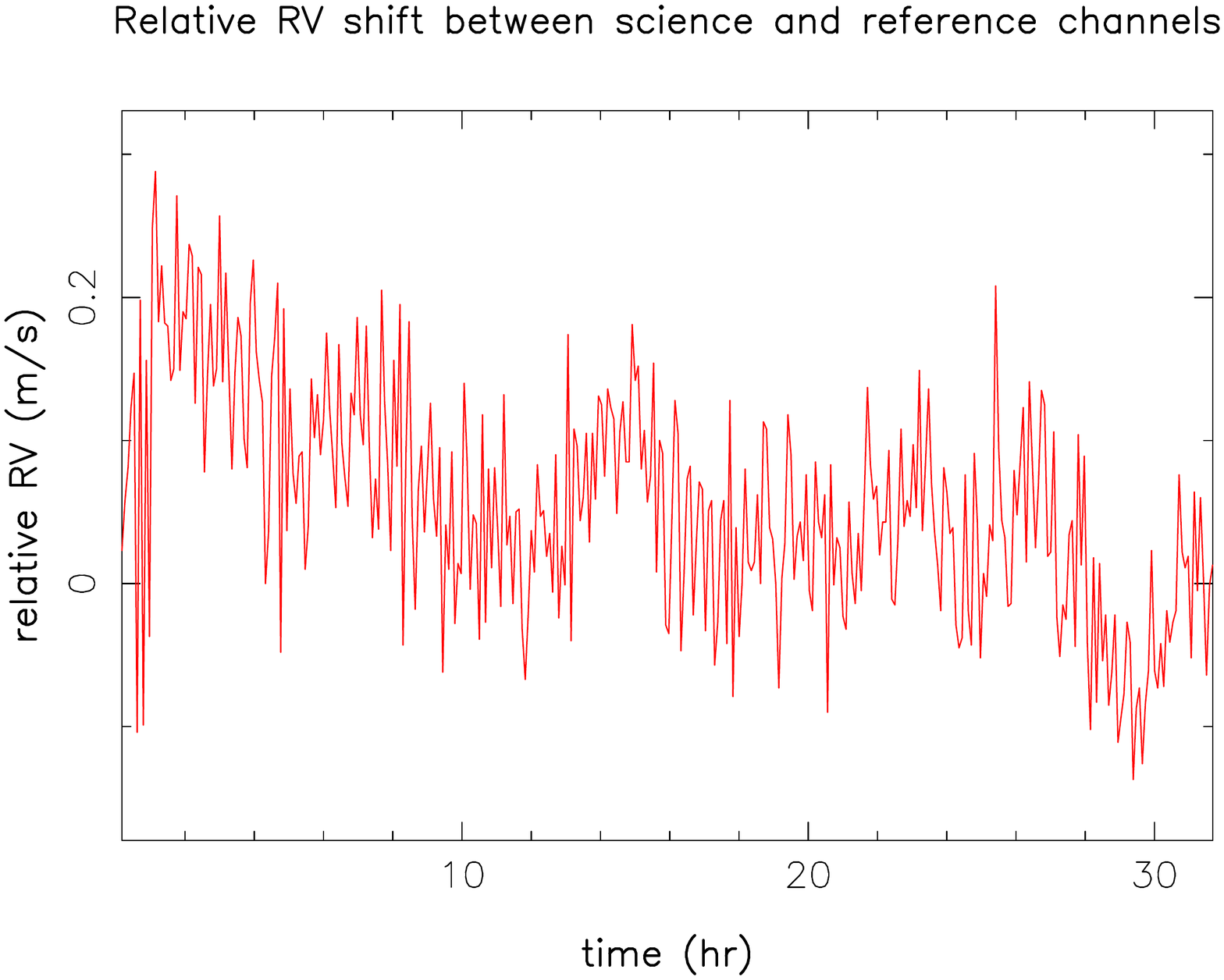}
       \hspace{2mm}\includegraphics[scale=0.25,bb=40 80 700 560]{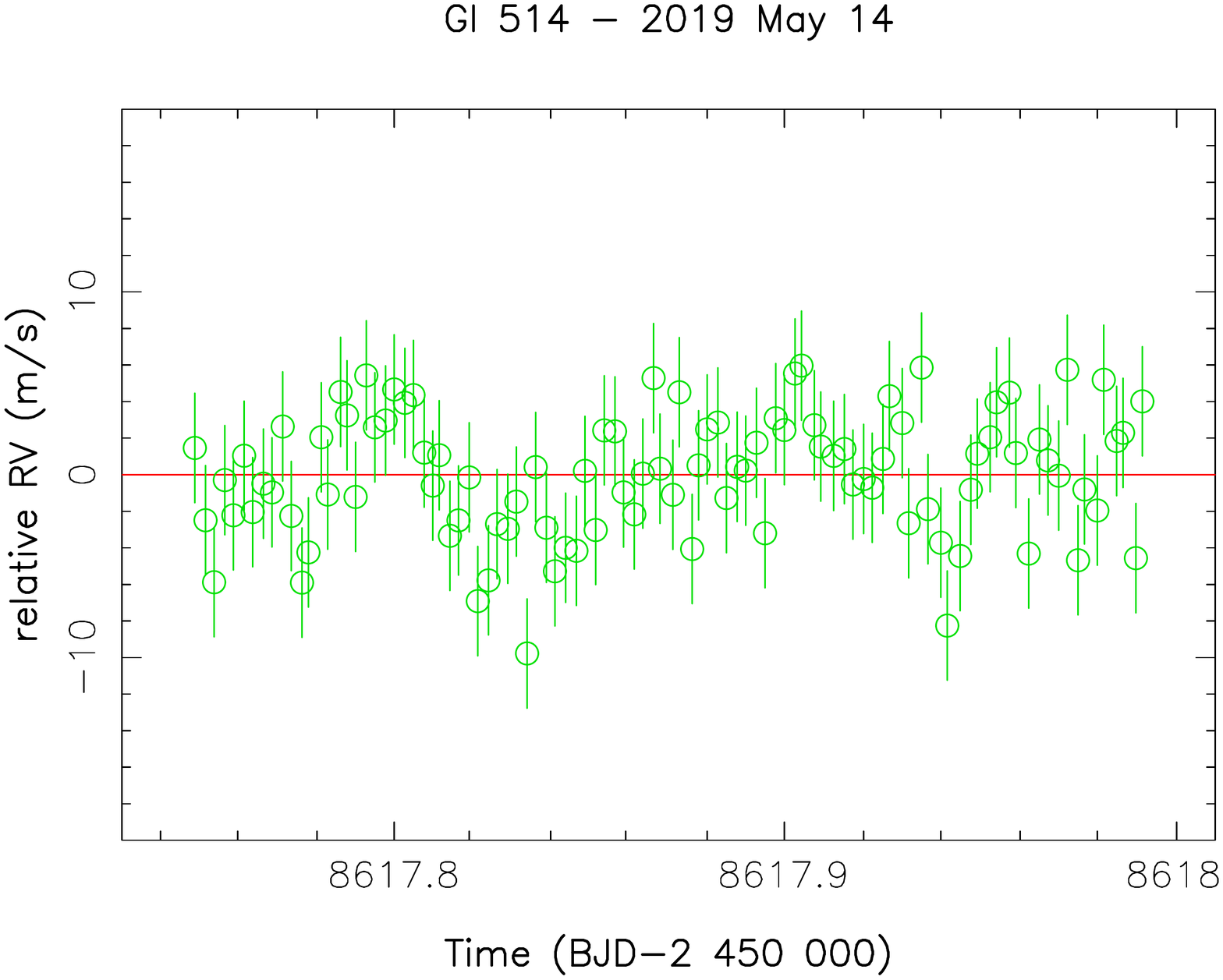}
       \hspace{2mm}\includegraphics[scale=0.25,bb=40 80 700 560]{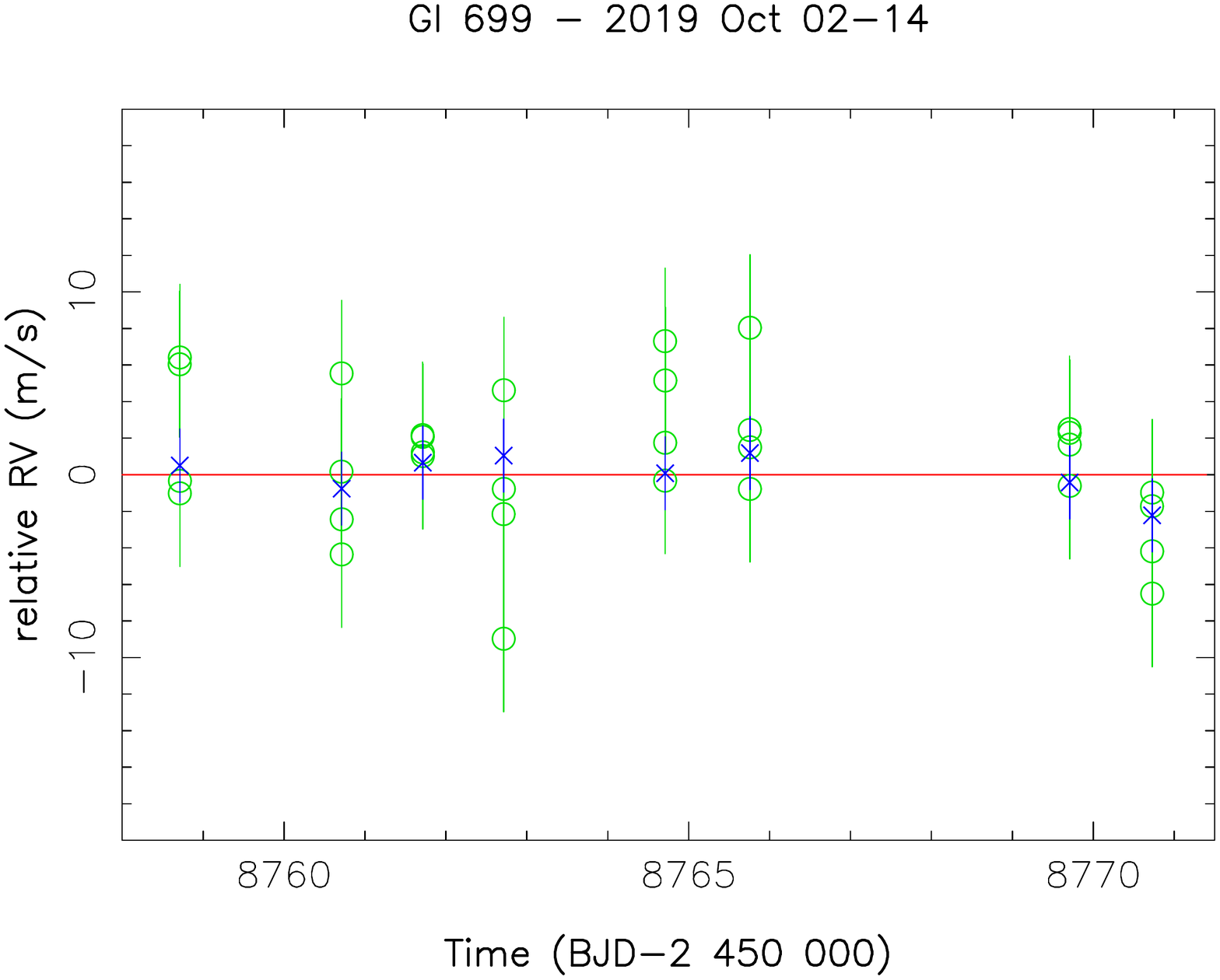}} 
\caption{\emr Monitoring the relative RV between the science and reference channels, through a 32~hr test run with the FP RV reference injected 
in both channels (left panel), a 6-hr science run on the M2 dwarf Gl~514 (100$\times$ 180-s exposures with fixed rhomb configuration, middle panel) 
and a 13~n run on the M4 dwarf Gl~699 (8 visits of 4$\times$61-s polarimetric sequences, right panel).  {\emq Green circles and blue crosses 
respectively indicate RVs on individual stellar exposures and on polarimetric sequences, with associated error bars set to 4 and 2~\ms.}   
The corresponding dispersions on relative RVs are respectively equal to 0.07~\ms, 3.4~\ms\ and 1.0~\ms\ (considering polarimetric sequences in 
the case of Gl~699), while the estimated photon noises are 0.03~\ms, 2~\ms\ and 0.5~\ms. } 
\label{fig:rvpr}
\end{figure*}

\subsection{Throughput \& thermal background}

The overall throughput of SPIRou (injection and atmospheric losses excluded) peaks slightly above $\simeq$10\%\ in the K band (see Fig~\ref{fig:thru} 
left panel), dropping progressively 
towards the blue side of the domain as a result of ZnSe absorption (from the polarimeter rhombs and the spectrograph prisms), and falling sharply 
towards the red mostly due to the detector cutoff (at 2.4~\mic).  This leads to predicted SNR curves with the SPIRou Exposure Time Calculator 
like those shown in Fig~\ref{fig:thru} (middle and right panels) in the case of the mid to late M dwarfs Gl~699 and GJ~1151 for exposure times 
of 61 and 301~s respectively, in reasonable agreement (where telluric lines are not too strong and dense) with observations secured in 
good weather conditions.  Injection losses depend mostly on seeing, increasing from 10\% to 50\% when seeing degrades from 0.6 to 1.2\arcsec.  

Thermal background sums up to a total of 90~\phpspA\ at 2.35~\mic\ for the SPIRou science channel, when adding the contributions of the polarimeter, 
fibers and hermetic feedthroughs (for polarimeter and feedthrough temperatures of about 2 and 15\degC\ respectively).  This corresponds to the 
stellar flux of a mid-M dwarf of magnitude H$\simeq$8.6 at 2.35~\mic.  The impact of the thermal background on SNR is negligible for bright stars 
like Gl~699, but starts to show up in the reddest orders for stars with H$\simeq$8 (like GJ~1151, see Fig.~\ref{fig:thru} right panel where the 
achieved SNR is below the prediction redder than 2.25~\mic, i.e., for orders \#34 to \#31) and becomes 
dominant for targets fainter than H$\simeq$9.  By cooling down the hermetic feedthroughs down to $\simeq$5\degC\ (see Sec.~\ref{sec:ope}), we hope 
to improve the situation and lower the thermal background to a stellar equivalent H magnitude of about 9.5.

\subsection{RV precision \& telluric correction}

RV precision is estimated by measuring the relative RVs between the science and reference channel and computing the dispersion on these measurements.  
Various contributors participate to the RV precision budget, in particular the instrument intrinsic stability, the light injection stability (of both 
near and far fields) and the pollution from telluric lines (that are variable with time in both position and strength with respect to stellar lines).  

The first term of this budget can be easily estimated with in-lab tests, e.g., by injecting the light from our RV-reference FP etalon in both the 
science and the reference channels simultaneously.  The left panel of Fig.~\ref{fig:rvpr} shows the result of such a test (routinely done at the CFHT 
to monitor the instrument behaviour), where SPIRou exhibited a RMS stability of 0.07~\ms\ in relative RV over a timescale of 32~hr.  
When rotating the rhombs, this budget used to increase to 0.40~\ms\ as a result of the residual beam deviation (of a few \arcsec, i.e., a few \mic\ 
at the fiber level).  With the latest rhomb combination mounted in SPIRou in late 2020 June, no more RV jitter is detected so that the overall 
instrumental contribution to the RV budget is 0.1--0.2~\ms\ RMS.

\begin{figure*}
%%% \mbox{\hspace{-2mm}\includegraphics[angle=-90,scale=0.25]{fig/spirou_polsun.ps}
%%% \hspace{2mm}\includegraphics[angle=-90,scale=0.25]{fig/spirou_polad.ps}
%%% \hspace{2mm}\includegraphics[angle=-90,scale=0.25]{fig/spirou_polge.ps}} 
\mbox{\hspace{-2mm}\includegraphics[scale=0.25,bb=40 80 700 560]{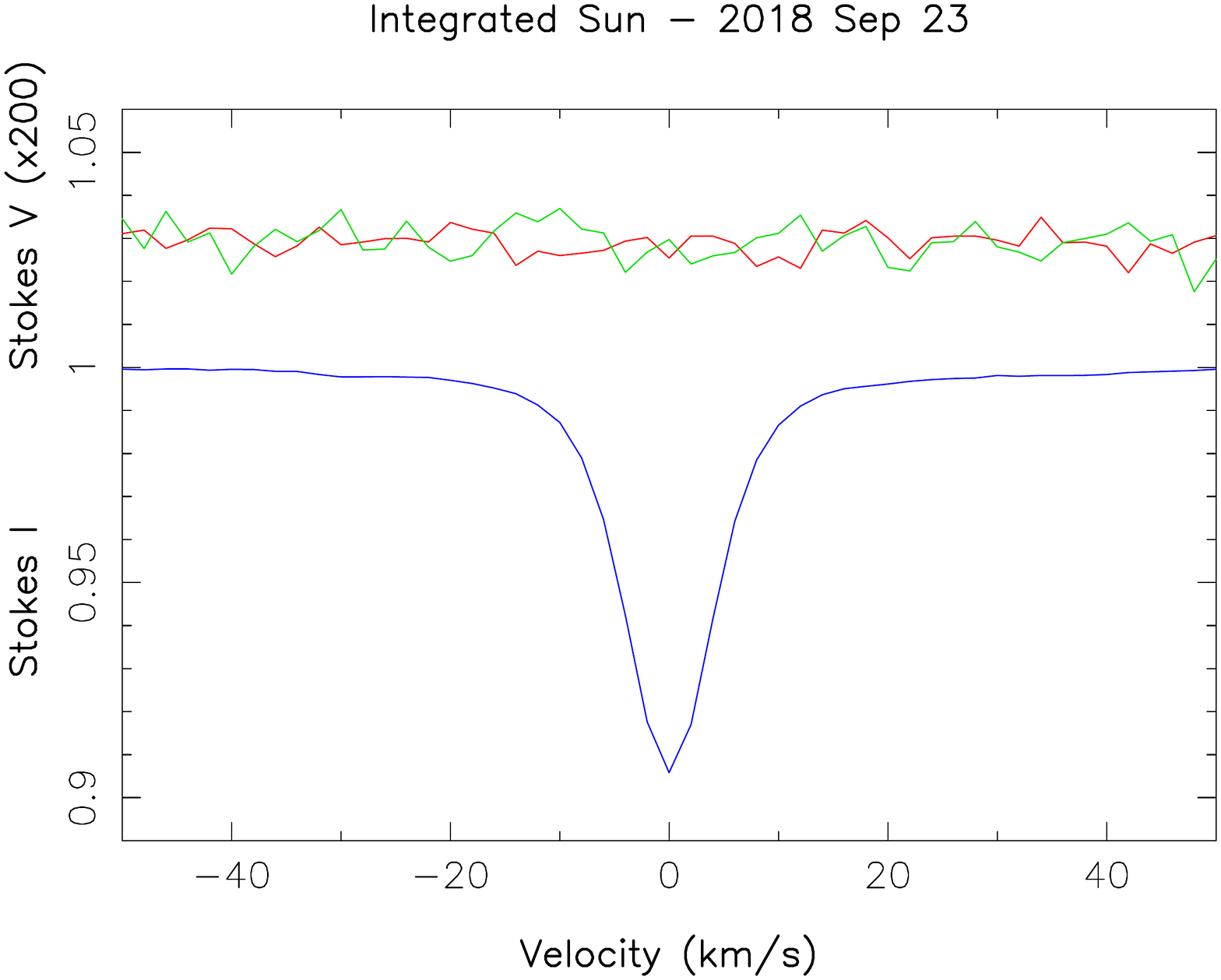}
       \hspace{2mm}\includegraphics[scale=0.25,bb=40 80 700 560]{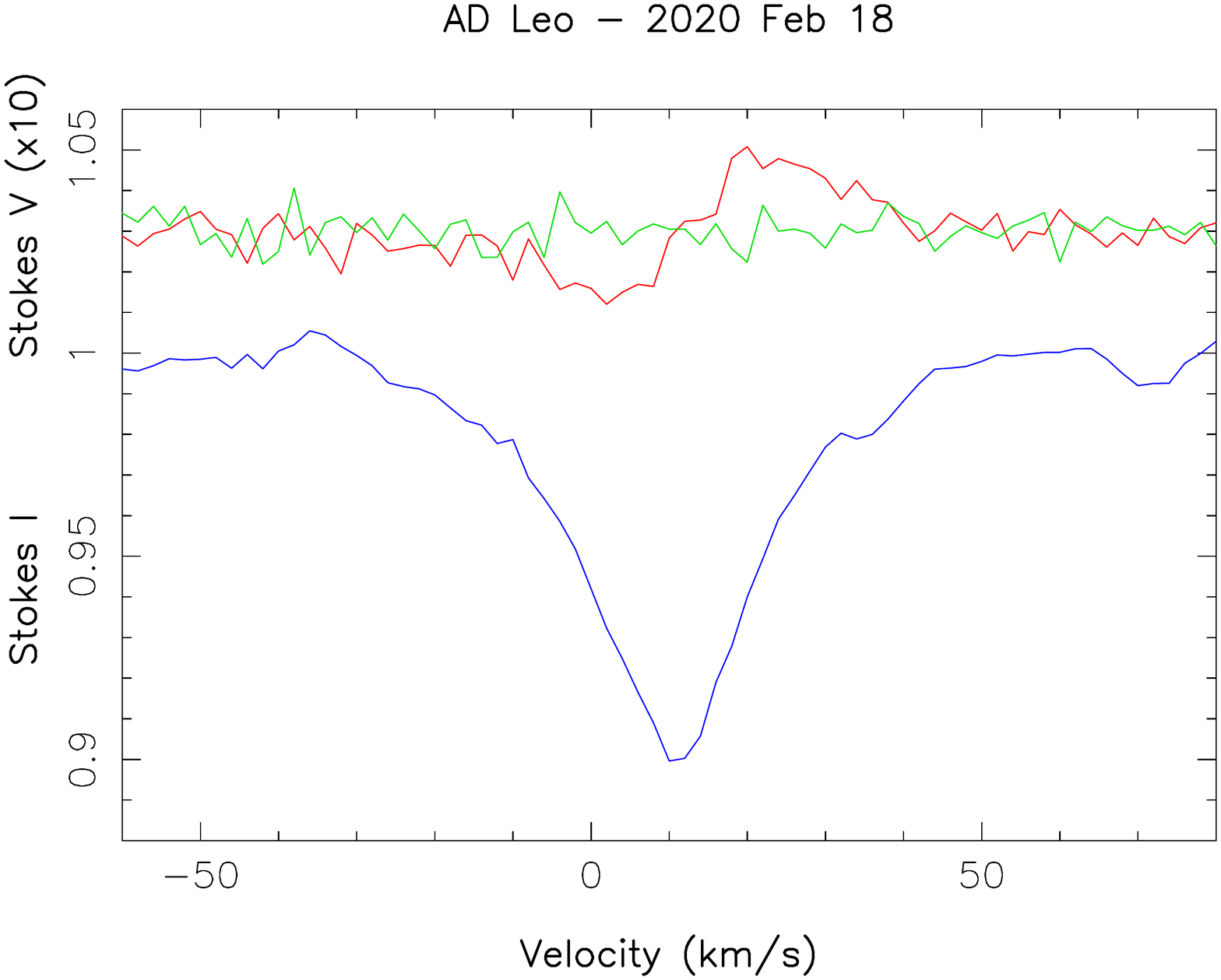}
       \hspace{2mm}\includegraphics[scale=0.25,bb=40 80 700 560]{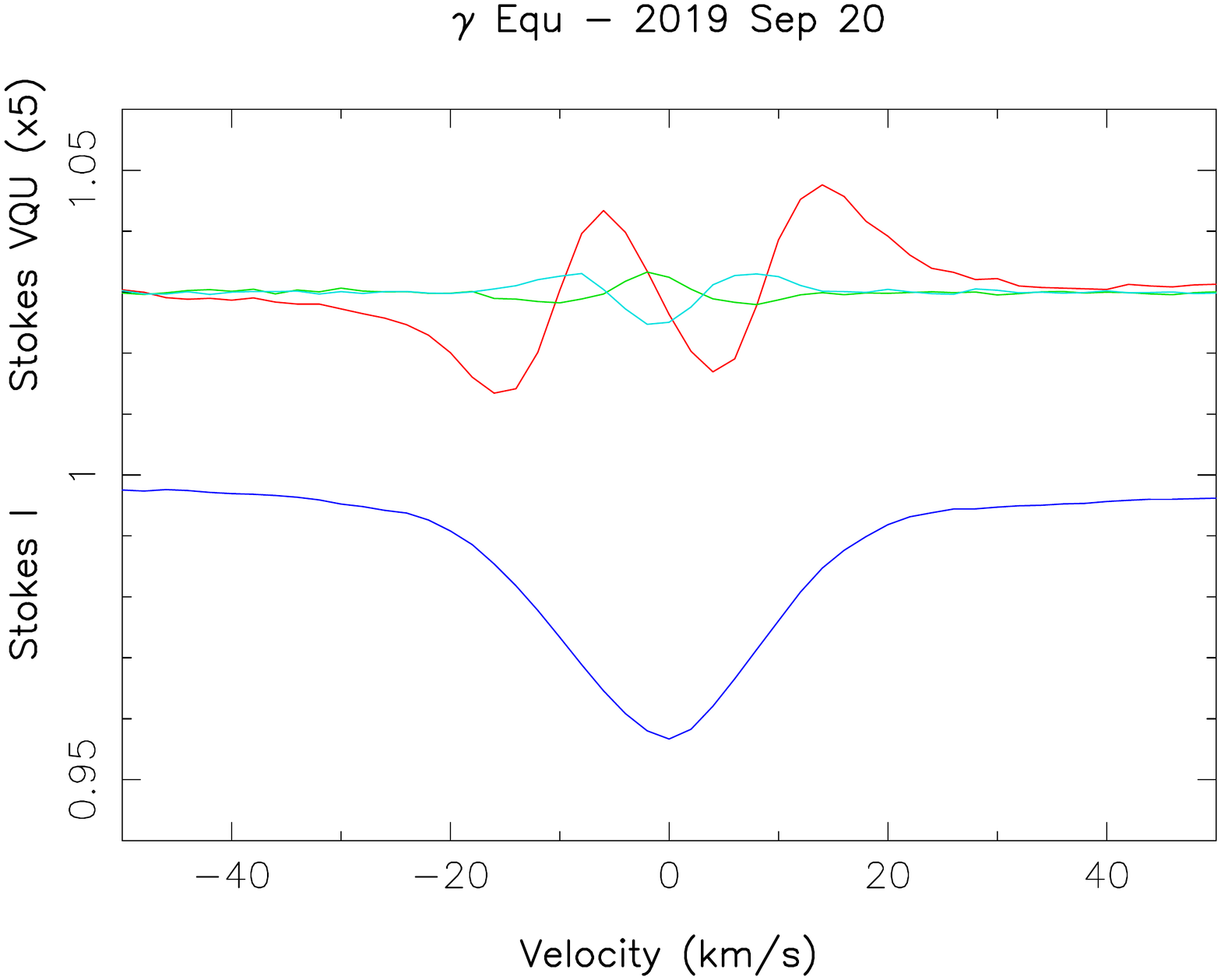}} 
\caption{Polarimetric sensitivity of SPIRou at a precision of 10~ppm, as estimated from the observation of the integrated Sun (left panel). 
Observing the magnetic stars AD~Leo and $\gamma$~Equ with SPIRou yields magnetic detections consistent with expectations in both circular and 
linear polarization (middle and right panels), and demonstrates that crosstalk between both polarization states is nominal.  
Stokes $I$ and $V$ LSD profiles are shown as blue and red lines respectively, whereas Stokes $Q$ and $U$ LSD profiles (of $\gamma$~Equ) are shown 
in green and cyan.  All polarized LSD profiles are shifted upwards (by 1.03) and expanded (by a different factor for each star) for display 
purposes. In the left and middle panel, the null polarization LSD profile is also shown (in green, with the same shift and expansion factor as those 
of the Stokes $V$ LSD profiles) to emphasize that no spurious polarization signal is present.  }  
\label{fig:pol}
\end{figure*}

Similar experiments were also carried out with the SPIRou LFC used as a RV reference.  A relative RV precision of 
$\simeq$0.2~\ms\ RMS was achieved up to now, which is encouraging, but still below the extreme RV precision (of a few \cms\ RMS) a LFC has the 
potential to deliver \citep[e.g.,][]{Probst20}.  This limitation reflects various technical problems encountered so far.  For instance, the RV jitter 
in red orders is often 2 to 3$\times$ larger than that in blue orders, likely indicating the presence of residual modal noise in the fibers despite 
the implementation of fiber agitators and scrambling devices at various points along the link between the LFC and the SPIRou calibration unit;  modal 
noise is indeed notoriously difficult to filter from LFC spectra, prone to this problem as a result of the monochromatic 
nature of individual lines.    
Moreover, flux variations accompanied by time-dependent RV shifts between the red and blue LFC orders are observed at times, which may be caused by 
injection fluctuations and differences between the two arms in the current hybrid LFC setup (see Sec.~\ref{sec:cal});  a maintenance visit from MS 
is needed at this stage to further improve the overall LFC stability so that it can progressively replace the HC and become the ultimate absolute RV 
reference it was designed to be.  

The second term of the RV precision budget (i.e., impact of light injection) is more tricky to evaluate through lab tests, requiring a device called 
`artificial star' (AS) to inject light within the instrument in a way that mimics the behaviour of real stars, flickering under the 
effect of atmospheric turbulence and variable weather.  Whereas near-field perturbations can be simulated with the AS, and 
mostly corrected for by the TTM to an accuracy of 0.01\arcsec\ RMS \citep[i.e., 1.4~\mic\ RMS at the level of the instrument aperture, 
e.g.,][]{Barrick18}, far-field perturbations are much more tricky to generate in a realistic way, and can only be estimated on the sky with real 
stars, along with the impact of spectral pollution from telluric lines (the third main term in the RV precision budget).  

The middle panel of Fig.~\ref{fig:rvpr} shows one such experiment carried out with SPIRou in 2019 May 14, where the bright inactive M2 dwarf Gl~514 
was observed continuously for 6~hr (100$\times$ 180-s exposures with fixed rhomb configuration) to study the relative RV jitter of individual exposures (with 
SNRs per pixel in H ranging from 220 to 280) that results from weather fluctuations and atmospheric variability.  The inferred relative RV precision is 
3.4~\ms\ RMS over the 6~hr run, and improves to $\simeq$2.5~\ms\ on shorter timescales, with a median dispersion of 2.6~\ms\ on groups of 4 consecutive 
exposures, typical to individual exposures of polarimetric sequences and $\simeq$1.5$\times$ larger than the expected photon-noise per exposure 
(1.5--2.0~\ms\ depending on SNR).  We note the presence of correlated noise (e.g., around JD=2\,458\,617.8), whose origin is unclear as no correlation 
shows up with guiding parameters or atmospheric conditions, as would be expected if induced by residual near- and far-field perturbations not 
fully filtered out by the fiber link, or imperfect telluric correction (see, e.g., Fig.~\ref{fig:dom} middle panel).  

Over the longer term, observations of inactive M dwarfs, such as Gl~699, are regularly being carried out as polarimetric sequences of 4 individual 
exposures, like in the 13-night SPIRou run of 2019 October during which 8 sequences of 4$\times$61-s individual exposures (each with 
a median SNR per pixel of 180 in H) were collected.  The relative RV dispersion of these observations is 1.0~\ms\ RMS 
for polarimetric sequences (see Fig.~\ref{fig:rvpr} right panel), whereas the dispersion of RV measurements from individual exposures within polarimetric 
sequences ranges between 0.5 and 4.8~\ms\ (with a median value of 3.1~\ms);  if weighting the relative RVs by the inverse variance within each polarization 
sequence, the RV precision on this series of data points reaches 0.7~\ms.  
We note in particular that no systematic pattern is observed between RVs of individual exposures within 
polarimetric sequences, in agreement with lab tests indicating that rhomb rotation is not degrading RV precision beyond a few 0.1~\ms\ RMS.  

Altogether, these observations indicate that the relative RV precision of SPIRou is currently of order 2~\ms\ on a timescale of a few weeks for spectra 
of mid-M dwarfs with SNRs of $\simeq$200, and is expected 
to regularly improve as upgrades are implemented, both on hardware (e.g., scrambling of the octagonal fibers feeding the pupil slicer, throughput and RV impact 
of rhombs) and on software (e.g., telluric correction).  In particular, detector persistence is found to be an issue on the faintest targets 
(H$>$10), with exposures taken earlier in the night (especially on bright stars) having in some cases a significant impact on the RV estimate;  
similarly, FP spectra recorded in the reference channel may affect the spectra of faint targets in the science channels as a result of pixel crosstalk.  
{\emr Whereas the latter effect is now reliably corrected for within APERO, the former is currently under investigation so that the achievable RV precision 
is better quantified for the faintest targets.}

\subsection{Polarimetric sensitivity \& crosstalk}

By observing the integrated Sun through its reflection on the Moon and carrying out a 4$\times$60-s circular polarization sequence, a ultra-high SNR 
spectrum (SNR$\simeq$2000 per pixel in H) was secured, from which Stokes $I$ and $V$ LSD profiles were computed (following telluric correction), 
yielding a noise level as low as 10~ppm per pixel in the Stokes $V$ LSD profile (see Fig.~\ref{fig:pol} left panel).  
As in \citet{Donati97b}, this test, showing that no spurious polarization is detected in conjunction with the narrow spectral lines of the Sun, 
demonstrates that SPIRou has the potential to reliably recover Zeeman signatures with a polarimetric sensitivity of at least 10~ppm, provided that 
enough photons are collected from the observed star.  In the case of the Sun, no magnetic field is detected down to a precision of 0.2~G.  

SPIRou is also able to detect polarized Zeeman signatures in the line profiles of known magnetic stars, such as the active M dwarf AD~Leo and the 
chemically-peculiar F star $\gamma$~Equ (see Fig.~\ref{fig:pol} middle and right panels).  In the case of AD~Leo, the Stokes $V$ LSD signature we 
detect in a 4$\times$61-s polarization sequence (with SNR=360 per pixel in H) yields a longitudinal magnetic field of $-135\pm10$~G, significantly 
smaller than the one measured (and mapped) a decade ago with ESPaDOnS observations \citep[e.g.,][]{Morin08b} and consistent with recent findings that 
the magnetic field of this star has been regularly decreasing with time over the last decade \citep{Lavail18}.   
We note in particular that the Stokes $I$ and $V$ profiles, spreading over a velocity range of at least 50~\kms, are significantly broader and 
shallower than their optical counterparts \citep[spanning no more than 40~\kms,][]{Morin08b, Lavail18} as a result of the larger Zeeman broadening 
at nIR wavelengths.  

SPIRou also very clearly detects circular and linear polarization Zeeman signatures in the lines of $\gamma$~Equ, which can be used to accurately 
estimate SPIRou's crosstalk level between polarization states {\emr (through measurements carried out at various Bonnette angles)}.  We find that 
crosstalk from Stokes $V$ to Stokes $Q$ and Stokes $U$ polarizations was respectively equal to 1.2\%\ and 1.1\%\ at the time of our observations, 
in agreement with the specification ($<$2\%).

\section{Overview of first results} 
\label{sec:pano}

In this penultimate section, we present a quick overview of the very first results obtained with SPIRou from data secured during commissionning and 
the beginning of the SLS.  We stress that the goal of this paper is not to carry out detailed analyses on any single data set discussed below 
\citep[to be presented in future papers from members of the SLS team, e.g.,][]{Moutou20, Martioli20}, but rather to highlight the potential of SPIRou 
with a number of examples taken from most science topics that SPIRou will tackle (in particular those of the SLS, see Sec.~\ref{sec:scig}).

\subsection{Detecting \& characterizing exoplanets of M dwarfs}

\begin{figure}
\includegraphics[scale=0.34,bb=20 45 650 560]{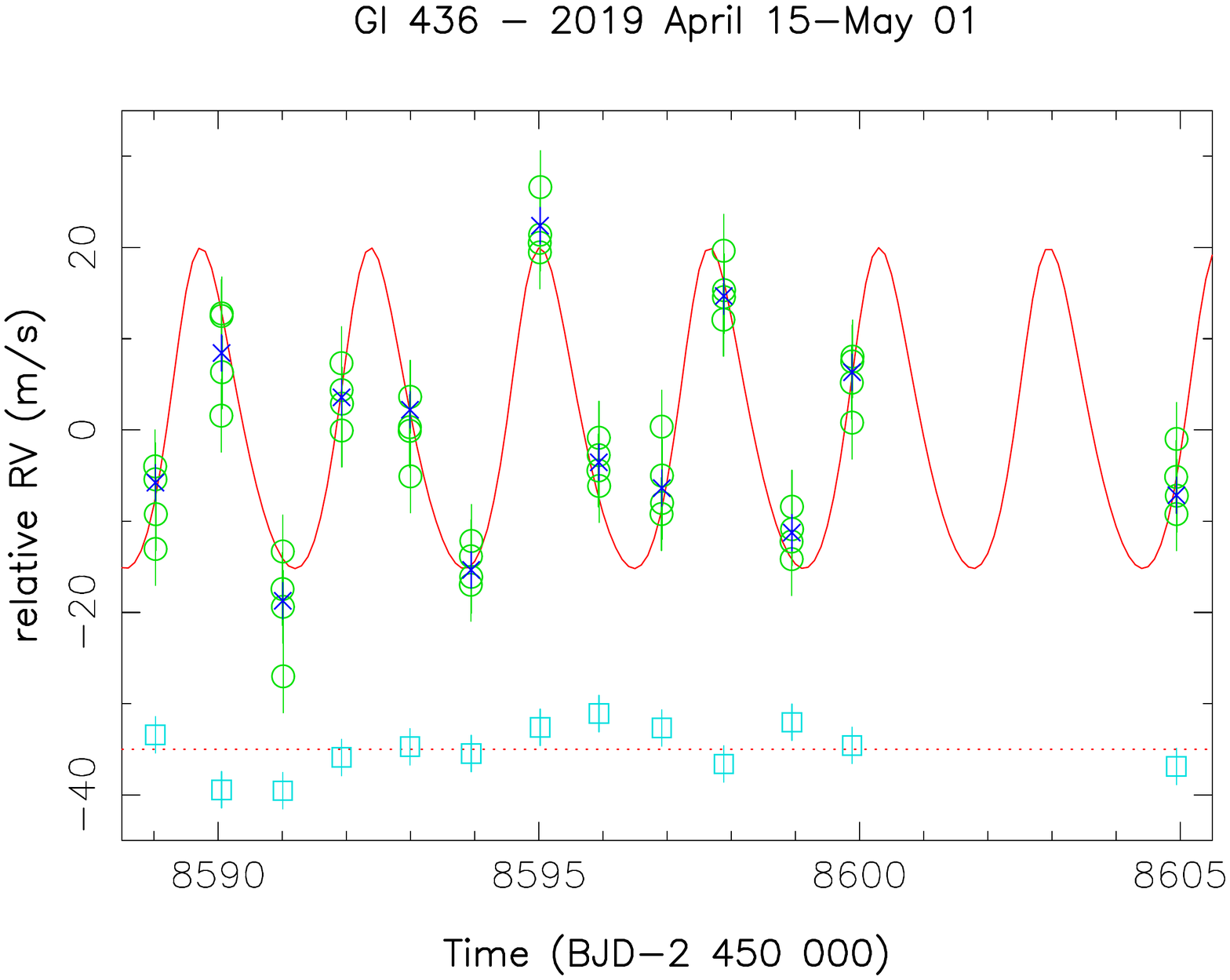}
\vspace{2mm}\\
\includegraphics[scale=0.34,bb=20 45 650 560]{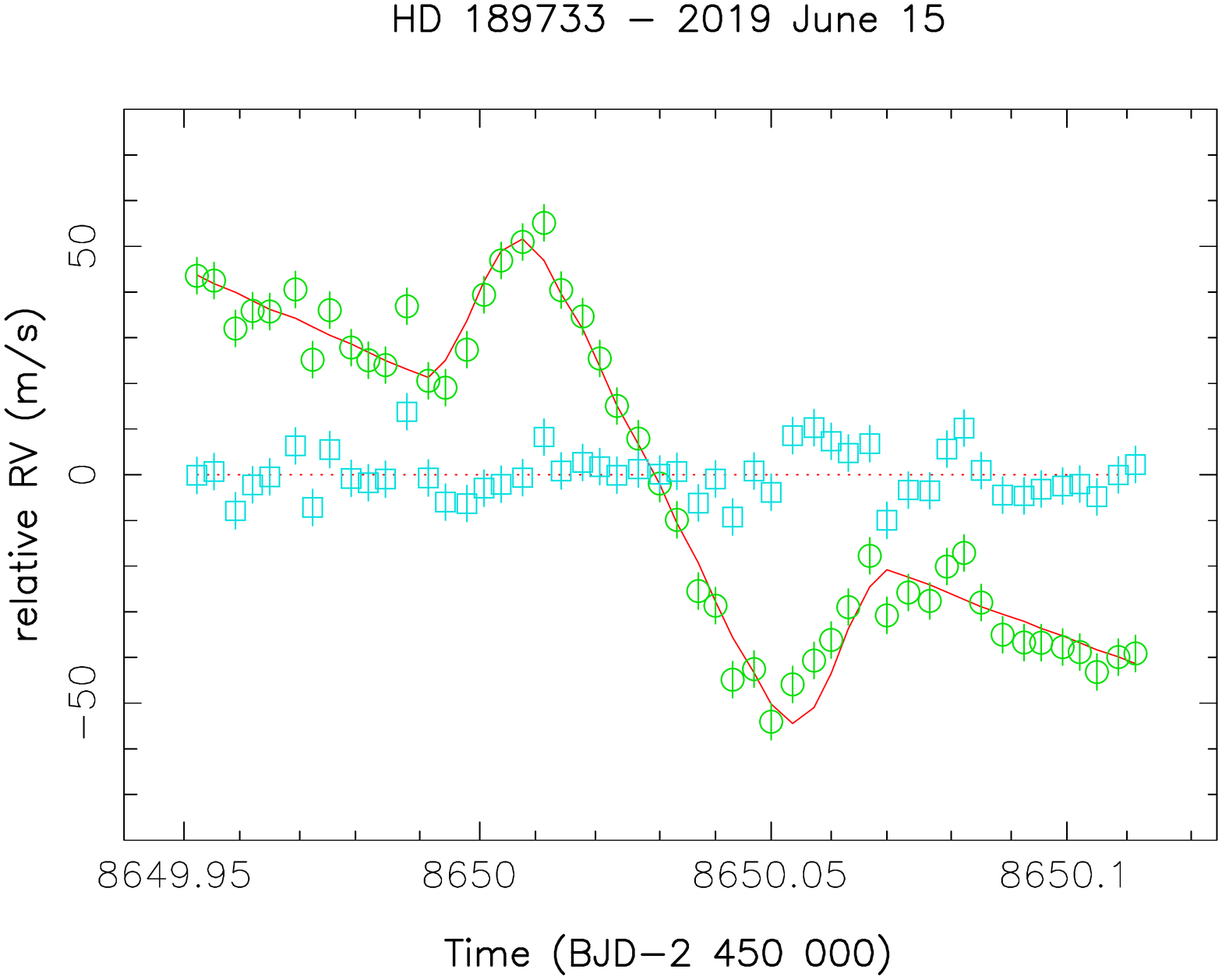}
\caption{RV measurements for the planet-hosting M3 dwarf Gl~436 (top panel) and for the K2 dwarf HD~189733 (bottom).  
The RMS dispersion with respect to the published ephemeris of Gl~436 \citep{Lanotte14} (red line) is equal to 2.6~\ms\ for the RVs derived from  
polarimetric sequences (blue crosses, with residuals, shown as cyan squares, shifted by $-$35~\ms\ for clarity).  The median dispersion of RVs measured 
from individual exposures within polarization sequences (green circles) is equal to 2.9~\ms, and the RMS dispersion of the main RV measurements reaches 
2.3~\ms\ when weighting each RV point by the inverse variance within each polarization sequence.  
In the case of HD~189733, {\emr observed with a fixed rhomb configuration}, the RMS dispersion of the residuals (cyan squares) with respect to the 
Rossiter-McLaughlin signature model (red line) of \citet{Moutou20} is equal to 5.3~\ms, and 4.3~\ms\ on the out-of-transit points. 
{\emq As in Fig.~\ref{fig:rvpr}, error bars are set to 4~\ms\ for the points corresponding to individual exposures (green circles in both panels and cyan 
squares in bottom panel), and to 2~\ms\ for those associated with polarization exposures (blue crosses and cyan squares in top panel).} }  
\label{fig:rvs}
\end{figure}

We start this panorama with a small sample of 13 observations of the M3 dwarf Gl~436, known to host a transiting warm Neptune on 
an inclined, eccentric orbit \citep{Lanotte14}.  This data set was obtained with SPIRou during a 16-n run in the second half of 2019 April 
(see Fig.~\ref{fig:rvs}).  
The measured RVs from the 13 4$\times$245-s polarization sequences (with a median SNR of 300 per pixel in H) collected on Gl~436 are found to be in 
good agreement with the published ephemeris, with an RMS dispersion {\emr about the predicted RV curve} of 2.6~\ms\ if assuming equal weights for 
all points, and 2.3~\ms\ with a weight equal to the inverse variance of RVs within each polarization sequence;  the internal dispersion of RVs between 
individual exposures ranges from 1.9 to 5.0~\ms\ from night to night, with a median of 2.9~\ms, similar to the internal RV precision obtained within 
polarization sequences of the inactive dwarf Gl~699 (see Sec.~\ref{sec:perf}).  {\emr The dispersion of RV residuals we report for this star is 
consistent with the RV precision 
of $\simeq$2~\ms\ RMS found on other quiet M dwarfs (see Fig.~\ref{fig:rvpr} middle and right panels), and larger than both the expected photon noise 
level (of 0.5~\ms) and of the RV precision (of 1~\ms) SPIRou aims at, suggesting that there is still room for improvements}.  
No magnetic field is detected at the surface of this (relatively inactive) dwarf, down to a precision of about 3~G.  A lot more results of this type 
are expected in the framework of the SLS-PS and SLS-TF (see Sec.~\ref{sec:scig}), thanks to which new planets of nearby M dwarfs will be discovered or 
confirmed, and characterized {\emr (e.g., AU~Mic~b, Klein et al., 2020a, submitted)}.  

\begin{figure}
\includegraphics[scale=0.34,bb=30 45 650 560]{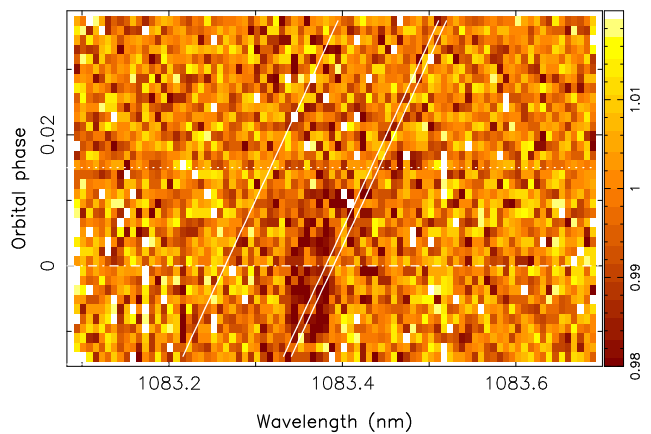}
\caption{SPIRou spectra of HD~189733 around the 1083~nm \hei\ line during the transit of the hJ, after normalizing all spectra by the median 
out-of-transit spectrum.  The slanted full lines trace the velocity curve of the planet for each of the \hei\ triplet 
component, whereas the horizontal dash-dot and dotted lines respectively depict the orbital phases of mid transit and egress.  
A patch of increased \hei\ absorption is detected at mid transit in association with the 2 main (red) components of the \hei\ 1083~nm triplet 
(about an order of magnitude deeper than the third blue component).  }   
\label{fig:tran}
\end{figure}

\begin{figure}
\includegraphics[scale=0.34,bb=20 45 650 560]{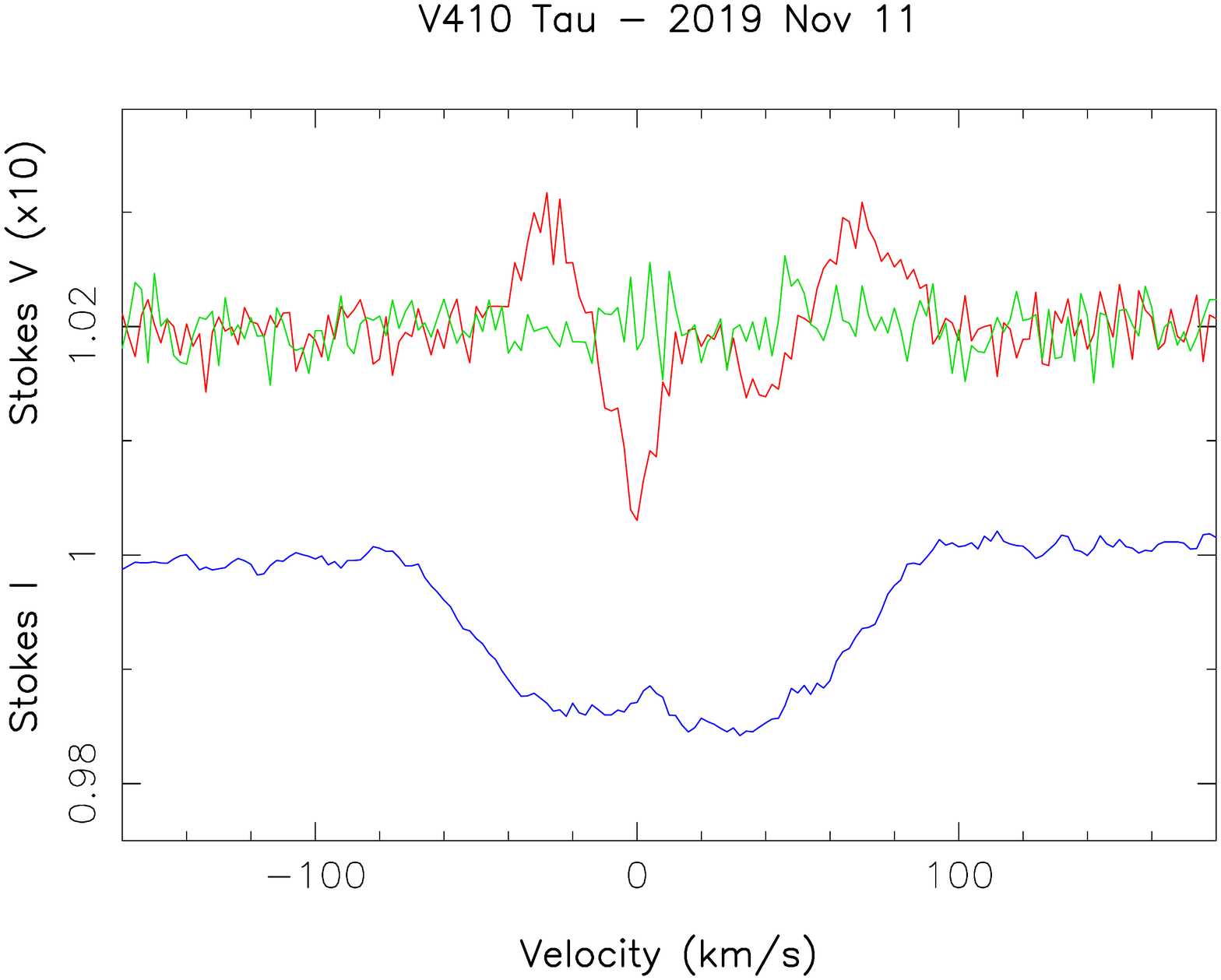}
\vspace{3mm}\\
%%% \mbox{\hspace{-2mm}\includegraphics[angle=-90,scale=0.58]{fig/spirou_stiv410tau.ps}\hspace{2mm}\includegraphics[angle=-90,scale=0.58]{fig/spirou_stvv410tau.ps}} 
\mbox{\includegraphics[scale=0.57,bb=120 95 320 590]{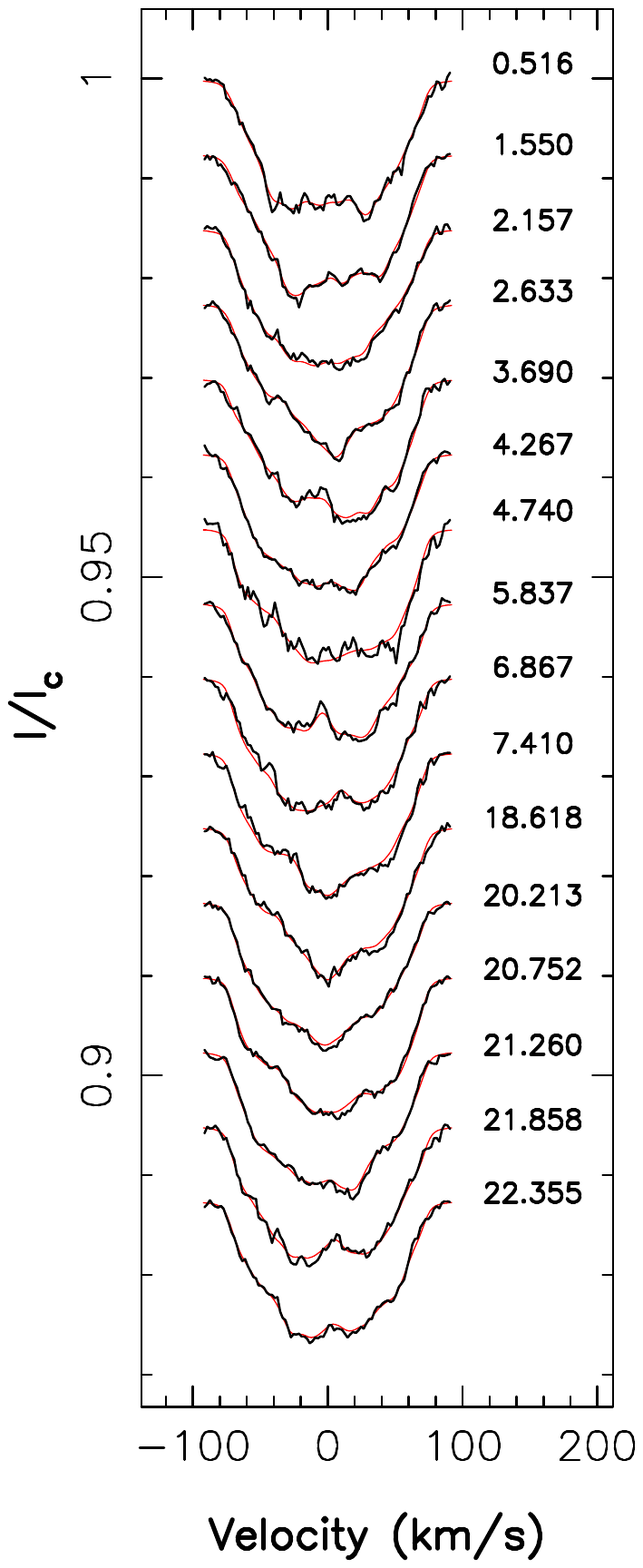}\hspace{2mm}\includegraphics[scale=0.57,bb=120 95 320 590]{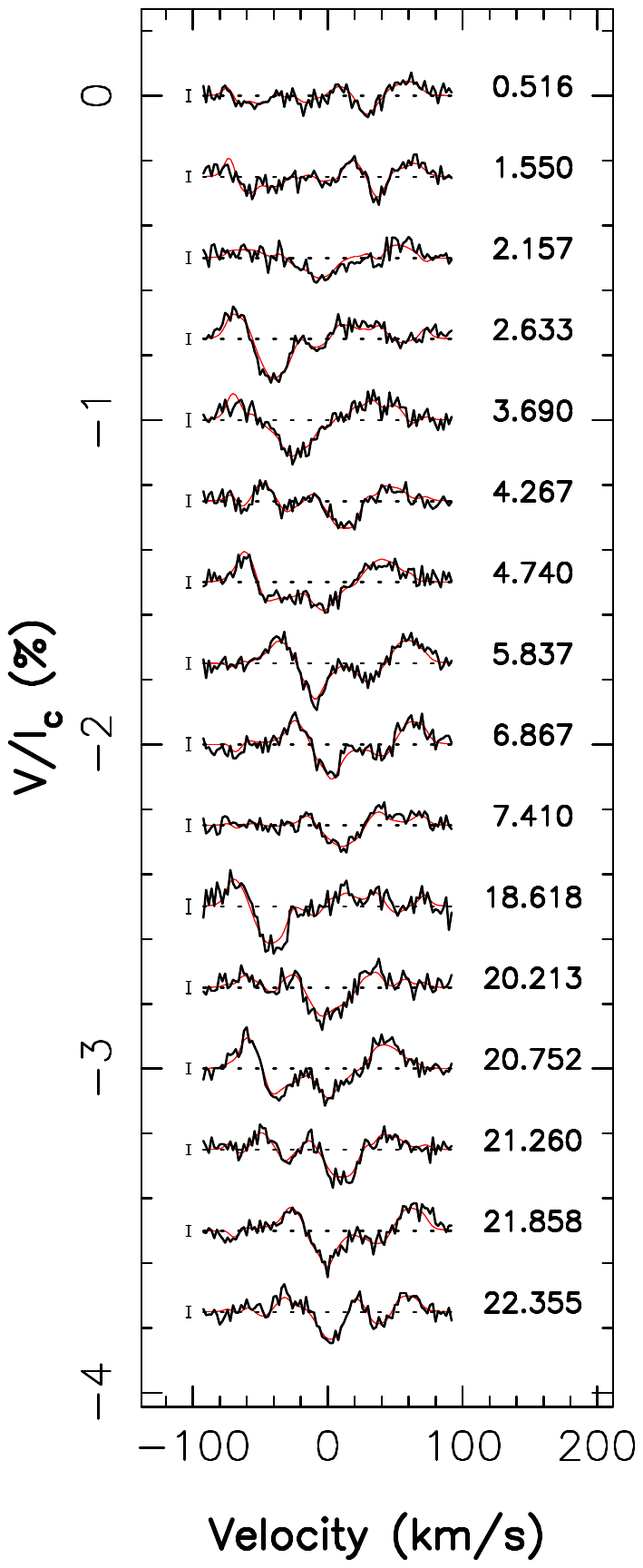}} 
\caption{LSD Stokes $I$ and $V$ profiles of the wTTS V410~Tau on 2019 Nov~11 (top panel, blue and red curves respectively, with the null polarization 
profile also shown in green) showing obvious Zeeman detections in conjunction with photospheric 
lines probing the presence of strong magnetic fields at the surface of the young PMS star, and full set of LSD Stokes $I$ and $V$ profiles collected 
with SPIRou in 2019 November \& December (bottom panels, thick black line) along with the ZDI fit (thin red line) to these observations (with rotational 
phases indicated next to each LSD profile, and a $\pm$1$\sigma$ error bar shown for each Zeeman signature).  } 
\label{fig:tts}
\end{figure}

\begin{figure*}
\includegraphics[scale=0.65,bb=0 150 800 300]{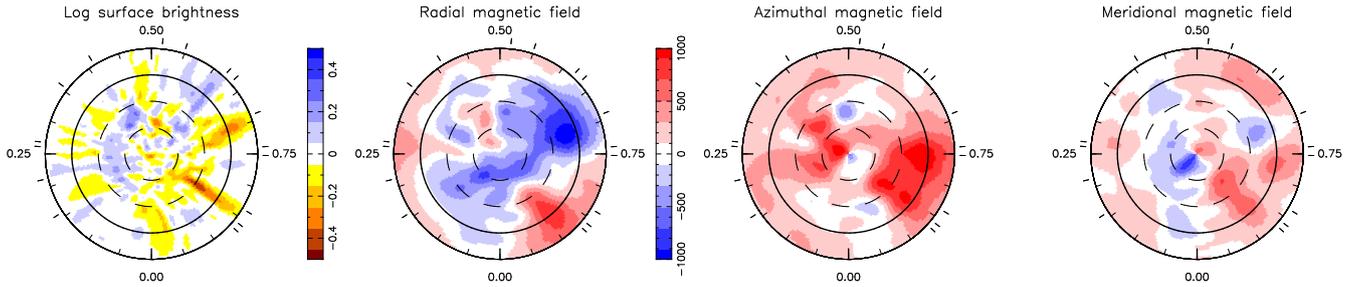}
\caption{Preliminary ZDI map of the logarithmic relative surface brightness (left panel), and the radial, meridional and azimuthal magnetic field (second 
to fourth panels) of V410~Tau, reconstructed from the data set shown in the bottom panels of Fig.~\ref{fig:tts}.  The star is shown in a flattened polar 
projection, with the equator depicted as a bold circle and the 30\degr\ and 60\degr\ latitude parallels as dashed lines.  Ticks around the star mark the 
rotational phases of our observations. }
\label{fig:mapv}
\end{figure*}

\begin{figure}
\includegraphics[scale=0.34,bb=20 45 650 560]{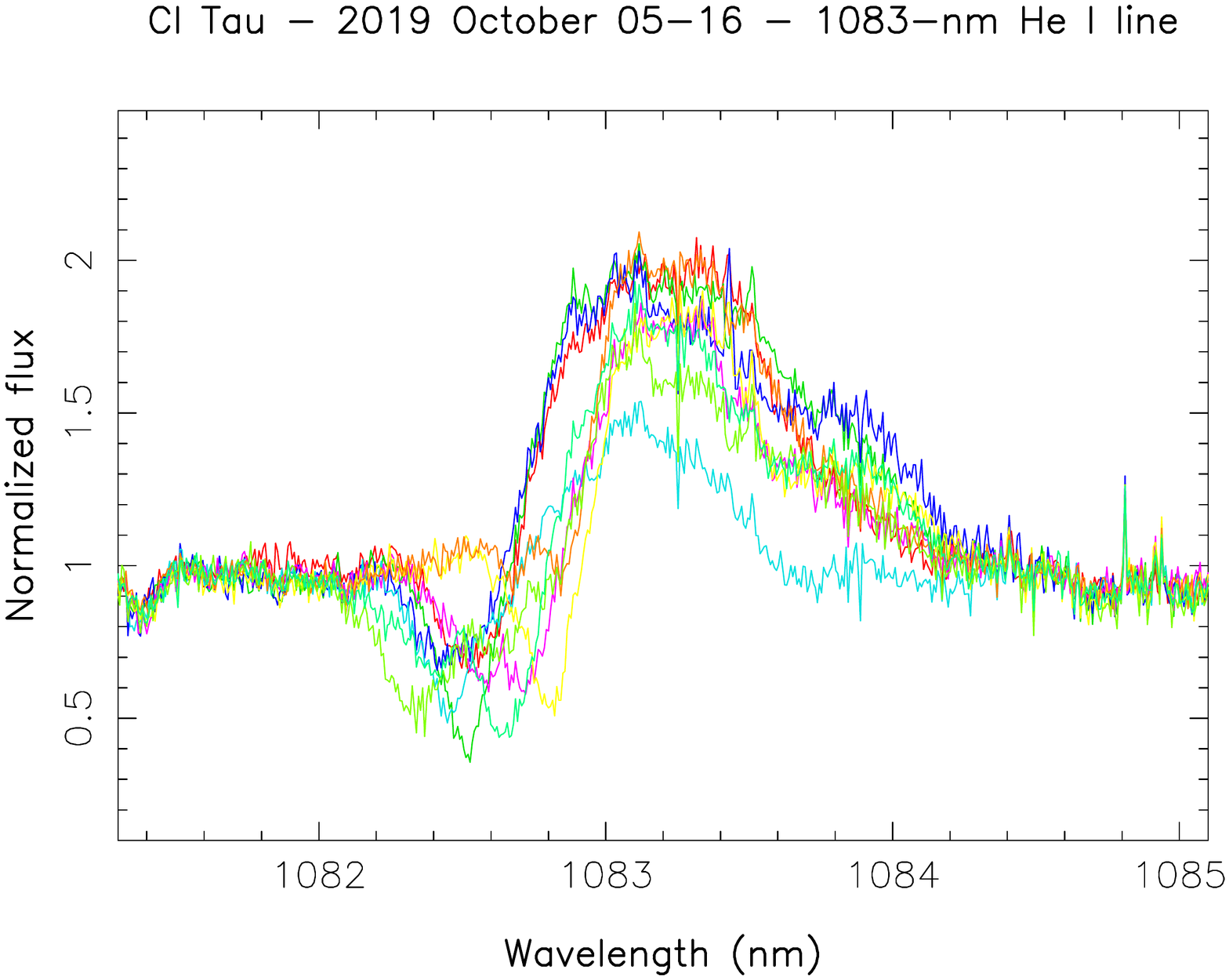}
\caption{Temporal evolution of the 1083-nm \hei\ line of the cTTS
CI~Tau over the 2019 October 02-15 SPIRou run showing variability of both blue absorption and red emission components probing accretion
and ejection processes taking place in the close circumstellar environment. } 
\label{fig:hei}
\end{figure}

\subsection{Characterizing transiting exoplanets \& their atmospheres}

As part of the SLS, one full transit of the hJ of the K2 star HD~189733 was monitored with SPIRou on 2019 June 15, with 50$\times$250-s 
exposures {\emr in a fixed rhomb configuration} and a median SNR of 230 per pixel in H.  
The Rossiter-McLaughlin effect is easily detected and consistent with the published literature, whereas the RV precision throughout the transit 
is found to be 5.3~\ms\ RMS (4.3~\ms\ RMS on the out-of-transit points, see Fig.~\ref{fig:rvs} bottom panel), larger than the photon noise level (of 
1.8~\ms\ on average) {\emr and in line with the activity jitter of this (moderately active) star}.  
A detailed analysis of these transit data, of additional data from another 
transit (2018 September 22), as well as spectropolarimetric data (with detected magnetic fields of a few G) collected over several 
rotation cycles of HD~189733 and orbital cycles of its hJ, are presented in \citet{Moutou20}, whereas a similar study on the young 
M dwarf AU~Mic was recently completed \citep{Martioli20}.  

SPIRou is also a powerful instrument for investigating exoplanet atmospheres through transmission or reflection spectroscopy \citep{Brogi12, Brogi16}, 
thanks to the extra-wide spectral domain recordable in a single exposure and the associated multiplex gain \citep{Brogi18, Brogi19}, {\emr making it at least 
as (if not more) efficient for such applications than CRIRES on the VLT, even in its updated version CRIRES+}.  For instance, the 36$\times$250-s spectra 
of HD~189733 collected with SPIRou on 2018 September~22 reveal increased \hei\ absorption during the transit of its hJ (see Fig.~\ref{fig:tran}) with 
a signature similar to those detected with Carmenes \citep{Salz18}, whereas the other transit monitored with SPIRou (on 2019 June 15) shows a much weaker 
absorption signal (if any).  {\emq We also note that the detected signature does not closely follow the planet trajectory (depicted by a slanted line on 
Fig.~\ref{fig:tran}), as for the second transit reported in \citet{Salz18};  this suggests that the detected \hei\ signature may at least partly be 
attributable to stellar activity, e.g., when chromospheric regions are masked by the transiting planet.}   
This detection of \hei\ absorption and its temporal variability illustrates the potential of such 
observations to investigate extended atmospheres of exoplanets and their possible evolution on timescales of a few months.  Looking for molecular 
species such as \hdo\ and \cod\ is also promising, for diagnosing elemental abundances and ratios such as [C/O] bringing key constraints for theoretical 
models.  Studies from SPIRou data on these topics are currently being carried out for a number of transiting exoplanets (Darveau-Bernier et al.\ 2020; 
Boucher et al.\ 2020; Klein et al.\ 2020b, in prep).

\subsection{Magnetic fields \& accretion / ejection processes of TTSs}

Looking at TTSs in the context of the SLS-MP, SPIRou routinely detects polarized Zeeman signatures in their spectral lines, probing the presence of 
strong magnetic fields at their surfaces.  For instance, large Zeeman signatures are detected in conjunction with atomic lines in the spectrum of 
the very young wTTS V410~Tau (see Fig.~\ref{fig:tts} top panel for the Stokes $I$ and $V$ profiles of a typical 4$\times$300-s polarization sequence 
with a SNR of 200 per pixel in H, and bottom panel for the full data set collected on this star in 2019 November and December), with multiple 
lobes of opposite signs revealing a complex magnetic field at the surface of the star.  The corresponding longitudinal fields are 
found to vary from 40 to $-$240~G with typical error bars of 30~G, i.e., similar to those reported from ESPaDOnS spectra \citep{Yu19} but with a 
smaller error bar in half the exposure time.  It demonstrates the gain brought by observing PMS stars at nIR wavelengths and thereby the improved 
sensitivity of SPIRou for investigating magnetic fields of young stars.  

The preliminary brightness and magnetic maps derived from this first SPIRou data set on V410~Tau (see lower panel of Fig.~\ref{fig:tts} and 
Fig.~\ref{fig:mapv}) are similar to those obtained from optical data \citep{Yu19}, though with less contrasted surface brightness inhomogeneities 
(as expected for cool features of low-mass stars observed in the nIR); 
in particular, the cool spot complex repeatedly reconstructed close to the pole from optical data, is much less conspicuous (if even present) in this 
new map from SPIRou data.  The inferred magnetic topology is also quite consistent with the findings of \citet{Yu19}, with a 
(mainly non-axisymmetric) poloidal component enclosing 65\%\ of the total magnetic energy and a dipole component with a polar field strength of 400~G 
(tilted at 20$\degr$ to the rotation axis).  A detailed analysis of these data is being carried out, to be published soon (Finociety et al.\ 2020, in prep).  

The nIR domain also features a number of interesting new proxies for studying accretion and ejection processes taking place in the circumstellar 
environments of cTTSs, which are key for addressing the main goals of the SLS regarding star / planet formation (see Sec.~\ref{sec:scig}).  
More specifically, these proxies are being used to monitor the inner regions of their accretion discs and the magnetospheric gaps the intense 
magnetic fields of these young stars are able to carve at disc center, where complex time-variable patterns of inflows and outflows are occurring.  
We show in Fig.~\ref{fig:hei} (bottom panel) the example of the 1083-nm \hei\ line of the well-known cTTS CI~Tau, whose accretion disc is thought 
to be the location of active planet formation \citep{Clarke18}.  This line is known to be a powerful tracer for investigating both outflows and 
inflows through its blue-shifted absorption and red-shifted emission components respectively \citep{Edwards09}.  In the case of CI~Tau, the 
variability of both components is obvious, e.g., with the blue-shifted absorption moving in position and strength with time and even 
disappearing entirely at some epochs.  By making it possible to connect the main characteristics of this variability with a detailed modeling of 
the magnetospheric topologies of cTTSs \citep[as in, e.g.,][]{Donati20}, SLS-MP studies will bring new insight for our understanding of 
magnetospheric accretion / ejection processes of cTTSs, especially for younger, lower-mass and more strongly accreting PMS stars than previously 
possible.

\subsection{Stellar atmospheres, activity \& Earth's atmosphere}

\begin{figure}
\includegraphics[scale=0.34,bb=20 45 650 560]{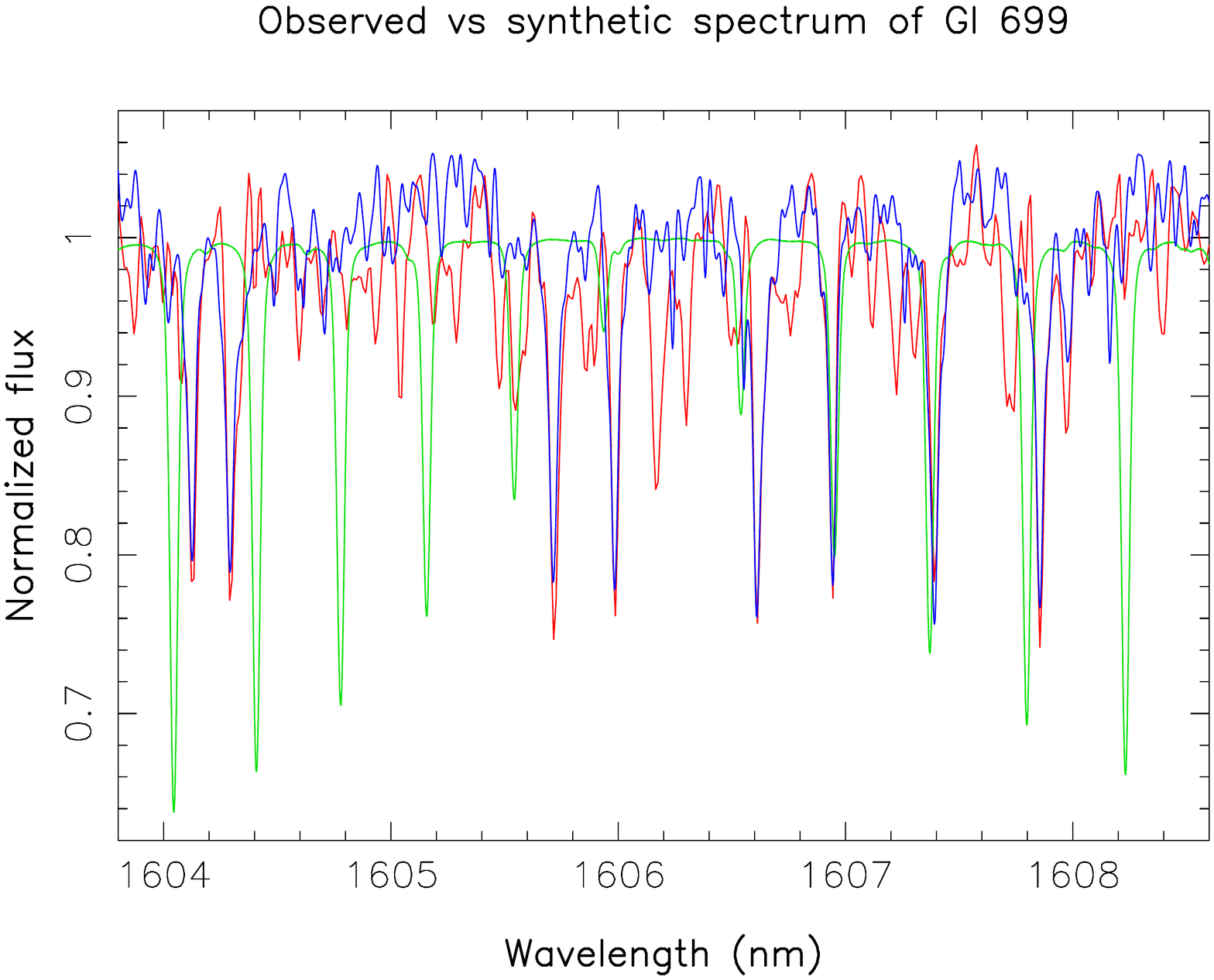}
\vspace{3mm}\\
\includegraphics[scale=0.34,bb=20 45 650 560]{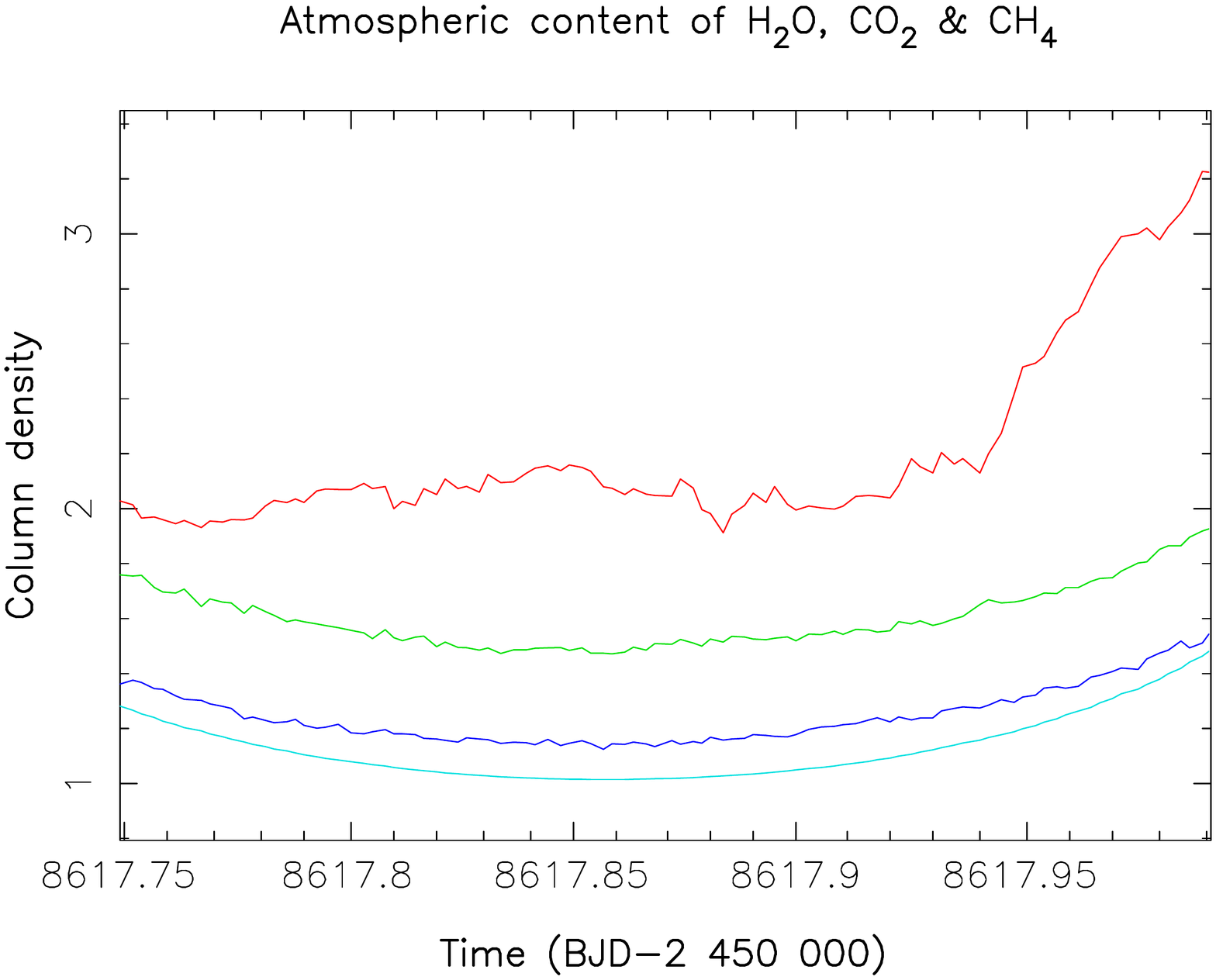}
\caption{Top panel: Small portion of the (telluric corrected) SPIRou spectrum of the M4 dwarf Gl~699 (red) compared with the (broadened) 
synthetic high-resolution spectrum in the Phoenix database of \citet{Husser13} that matches best the atmospheric parameters of Gl~699 (blue), 
along with a TAPAS spectrum for average atmospheric conditions at the CFHT (green), {\emr found to provide a very accurate description of the 
true telluric features over most of the SPIRou domain including the one shown here}.   Whereas the main lines in this spectral window are in 
reasonable agreement, significant spectral content in the SPIRou observation is absent from the model.  Bottom panel: \hdo\ (red), 
\cod\ (green) and \chq\ (blue) atmospheric content as a function of time throughout the 6-hr SPIRou monitoring of Gl~514 (see Fig.~\ref{fig:rvpr} middle 
panel), as derived with a TAPAS model of atmospheric transmission \citep{Bertaux14} fitted to the SPIRou data.  {\emr Airmass is also shown (cyan).} }  
\label{fig:modatm}
\end{figure}

As pointed out in Sec.~\ref{sec:scig}, there are many more science goals that SPIRou can investigate beyond those on which the SLS is focusing.  
An obvious one, on which the SLS is also contributing to as a legacy output, is the study of the atmospheres of M dwarfs from their high-resolution 
nIR spectra, which already motivated some attention and triggered new developments with the Phoenix model atmospheres \citep{Rajpurohit13, Allard13, 
Rajpurohit18a, Rajpurohit18b}.  Thanks to the extended domain and the high spectral resolution, SPIRou spectra are particularly adapted for such 
studies;  a straightforward comparison of the closest synthetic Phoenix spectrum \citep[extracted from the database of][]{Husser13} matching the 
atmospheric parameters of Gl~699 (\teff=3200~K, \logg=5.0, [Fe/H]=$-0.5$, e.g., \citealt{Mann15}) and broadened to the spectral resolution of SPIRou 
shows that, although the main lines of the observed spectrum agree with model predictions, obvious discrepancies remain and abundant spectral content 
is absent from the model (see Fig.~\ref{fig:modatm}).  Ongoing studies from SPIRou data are being carried out in this field, including within the SLS.  

Studies of stellar activity, complementing those on precise RVs (to improve the efficiency at filtering RV curves) and magnetometry of late-type dwarfs
(to study how activity relates to and fuels from the magnetic field that produces it), will also benefit from SPIRou and the new activity diagnostics 
that are available in the nIR domain \citep[e.g.,][]{Shofer19}, in particular \hei\ (see Figs.~\ref{fig:tran} and \ref{fig:hei}), 
\pbe, \pga\ and \bga.  Measuring surface magnetic fluxes from small-scale fields (generating little to no polarization signatures as a result of their 
tangled structures) thanks to the stronger Zeeman splitting at nIR wavelengths will be quite valuable in this respect, as this proxy was shown to be 
the one that correlates best with RV variations in the case of the Sun \citep{Haywood16}.  
Papers based on (or using) SPIRou data are being worked on, with studies on, e.g., the activity and magnetic field of 
$\epsilon$~Eri as seen by Narval, SPIRou and K2 (Petit et al.\ 2020, in prep).  

Last but not least, SPIRou is an ideal tool for studying column densities of specific molecules in the Earth's atmosphere, in particular \hdo, 
\cod, \chq\ and \od\ that exhibit a rich absorption spectrum in the SPIRou domain, and their evolution with time on both short- and long-term.  
An example of such modeling is given in Fig.~\ref{fig:modatm} (bottom panel), showing the atmospheric content of \hdo, \cod\ and \chq\ as a function 
of time over the 6~hr monitoring of Gl~514 with SPIRou on 2019 May 14 (see Fig.~\ref{fig:rvpr} middle panel), and derived by adjusting TAPAS 
atmospheric transmission models \citep{Bertaux14} to SPIRou data.  This example illustrates that, whereas \chq\ and \cod\  
are strongly correlated with airmass (being evenly distributed in the atmosphere), \hdo\ is much more variable and subject to (more rapidly 
varying) weather conditions.  This kind of analysis can be used not only to quantitatively optimize in almost real time ongoing SPIRou observations given 
current weather conditions, but also to significantly contribute, in the context of global warming, to the long-term monitoring of the Earth's 
evolving atmosphere in the best temperate astronomical site worldwide.

\section{Summary and perspectives}
\label{sec:conc}

In this paper, we first present the wide variety of science goals for which SPIRou was constructed;  we focus in particular on the two main ones, namely 
the quest for planetary systems around nearby M dwarfs and the study of magnetized star / planet formation, on which the SLS concentrates 
through a dedicated 300-night Large Programme at the CFHT.  We then detail the main modules the instrument consists of, starting with the Cassegrain 
unit and its achromatic polarimeter mounted on the telescope, the fiber link conveying the stellar light to the CFHT Coud\'e room, where the cryogenic 
bench-mounted cross-dispersed high-resolution spectrograph is mounted along with the calibration unit providing the instrument with spectra of 
reference calibration lamps.  We also provide basic information on the automatic reduction pipeline APERO, including its telluric correction process.  

Next, we outline the current instrument performances as estimated from both lab tests and on-sky observations.  SPIRou yields stellar spectra covering 
0.95-2.50~\mic\ in a single exposure, with a sampling rate of 2.28~\kms\ per pixel and a spectral resolving power of 70k (in terms of the 
broadening impact on stellar spectra).  
Throughput rises from 3\% in the Y band, to 7\%, 10\% and 12\% in the J, H and K band respectively, making it capable of collecting spectra of M 
dwarfs with peak SNRs of 110 per pixel at H$\simeq$8 in a 5-min exposure in good weather conditions.  Thermal background at 2.35~\mic\ is currently 
equivalent to the flux of a H$\simeq$8.6 mid-M dwarf, with plans to reduce it in the near future.  RV precision is currently found to be $\simeq$2~\ms\ 
RMS on a timescale of at least a few weeks, reaching at times 1~\ms\ RMS on bright stars in good weather, and is expected to further improve with 
forthcoming hardware and software (APERO) upgrades.  
Polarimetric performances are similar to those of ESPaDOnS, with a sensitivity of at least 10~ppm and a crosstalk level from circular to linear 
polarization of order 1\%. 

We finally present a quick overview of a few basic results obtained so far with SPIRou, which are (or will soon be) described in detail in dedicated 
papers.  We show in particular that SPIRou is able to detect the RV impact of a known exoplanet (Gl~436b), the Rossiter McLaughlin effect of a transiting 
hJ (HD~189733b), the \hei\ absorption from the atmosphere of a transiting hJ (again HD~189733b), and the Zeeman signatures from surface magnetic fields of 
a M dwarf (AD~Leo) and a TTS (V410~Tau).  We also illustrate how SPIRou contributes to studies of stellar atmospheres and of the Earth's atmosphere.  

At this time, SPIRou has been carrying out science programmes, including the SLS, at the CFHT since the beginning of 2019.  SPIRou is already the 
most popular CFHT instrument, capitalizing the largest number of PI proposals since it has been offered to the whole 
community.  SPIRou programmes were allocated a total of 209 nights on the CFHT in the first 3 semesters of science operation (2019a to 2020a). 
Data collection has however progressed more slowly than expected due to a number of unfortunate issues (from poor weather to technical failures 
at the CFHT, TMT protests in Hawaii, and lately the COVID-19 pandemic) that altogether caused the loss of almost half the allocated SPIRou time so far.  
SPIRou being mounted in bright time periods only also comes as a limitation, in particular for the observation of transiting planets with orbital 
periods longer than a week.  The situation will hopefully improve in the coming semesters, with more SPIRou time allocated to ensure faster 
progression and better completion of ongoing programmes.  

Over the longer term, SPIRou is expected to be a key contributor in coordinated programmes with most major facilities to come, in particular the JWST from 
2021, the ELTs from 2025, then PLATO from 2026 and ARIEL from 2028.  More specifically, SPIRou velocimetric and spectropolarimetric observations will be 
instrumental for identifying the most promising nearby Earth-like HZ planets to be scrutinized by the JWST and the ELTs for detailed atmospheric 
characterization, and for providing consistent modelings of the star / disc / planet interactions taking place at the heart of forming star / planet 
systems.  Later on, SPIRou will be needed for confirming and characterizing planetary systems around M stars to be unveiled by PLATO, and for monitoring 
and modeling the activity of stars to be investigated with ARIEL.  Altogether, it requires SPIRou to progressively ramp up in duty cycle so that it can 
achieve very dense temporal monitoring for the large stellar samples to be observed in coordination with the major facilities of the decade.  
Coupling SPIRou with ESPaDOnS so that both instruments can observe the same targets at the same time and provide high-resolution spectra 
of the observed stars all the way from 0.39 to 2.50~\mic\ in a single exposure for any given polarization state, would further boost the overall 
science return and make the CFHT the only observatory worldwide offering such capabilities.

\section*{Acknowledgements}
\small 
This paper is dedicated to the memory of Les Saddlemyer from NRC-H, who managed the development of the SPIRou dewar, and to Laurent Par\`es from OMP/IRAP 
in charge of the optical design and tests of the Cassegrain unit, who sadly passed away in 2017 January and 2020 February, respectively.

We thank all partners for funding SPIRou, whose construction cost (including reviews and travels) 
reached $\simeq$5~M\euro, namely the IDEX initiative at UFTMP, UPS, the DIM-ACAV programme in R\'egion Ile de France, the MIDEX initiative 
at AMU, the Labex@OSUG2020 programme, UGA, INSU/CNRS, CFI, CFHT, LNA, DIAS, ASIAA and IA.  
We are grateful for in-kind manpower allocated to SPIRou by OMP/IRAP, OHP/LAM, IPAG, CFHT, NRC-H, UdeM, UL, OG, LNA and 
ASIAA, amounting to about 75~FTEs including installation and ongoing upgrades.  We also thank an anonymous referee for valuable comments that improved 
the manuscript.  

JFD and TPR acknowledge funding from the European Research Council under the H2020 research \& innovation programme (grants \#740651 NewWorlds 
and \#743029 EASY).  We also acknowledge funding from the French ``Agence Nationale pour la Recherche'' (grants ANR-18-CE31-0019 and ANR-15-IDEX-02). 
This work was supported by the Funda\c c\~ao para a Ci\~encia e a Tecnologia and FEDER through COMPETE2020 (UID/FIS/04434/2019, UIDB/04434/2020, 
UIDP/04434/2020, PTDC/FIS-AST/32113/2017, POCI-01-0145-FEDER-032113, PTDC/FIS-AST/28953/2017, POCI-01-0145-FEDER-028953).  SYW acknowledges funding from ASIAA.  

Our study is based on data obtained at the CFHT, operated by the CNRC (Canada), INSU/CNRS (France) and the University of Hawaii.  
The authors wish to recognise and acknowledge the very significant cultural role and reverence that the summit of Maunakea has always had within the 
indigenous Hawaiian community. We are most fortunate to have the opportunity to conduct observations from this mountain. 
This work also benefited from the TAPAS service through the ETHER center at URL {\tt http://ether.ipsl.jussieu.fr/tapas/}, the SIMBAD / 
CDS database at URL {\tt http://simbad.u-strasbg.fr/simbad} and the ADS system at URL {\tt https://ui.adsabs.harvard.edu}.

\section*{Data availability}  Most data underlying this article are part of the SLS, and will be publicly available from the Canadian Astronomy Data 
Center 1~yr following the completion of the SLS.  Data on RV and telluric standards are already public.  

\bibliography{spirou}
\bibliographystyle{mnras}

\vspace{5mm}\small
\noindent $^1$ Univ.\ de Toulouse, CNRS, IRAP, 14 av.\ Belin, 31400 Toulouse, France \\ 
$^2$ Univ.\ de Toulouse, CNRS, OMP, 14 av.\ Belin, 31400 Toulouse, France \\ 
$^3$ Observatoire du Mont-M\'egantic, D\'epartement de physique de l'Universit\'e de Montr\'eal, iREx, Montr\'eal, Canada \\ 
$^4$ Univ.\ Grenoble Alpes, CNRS, IPAG, 38000 Grenoble, France \\ 
$^5$ Canada-France-Hawaii Telescope, Kamuela, Hawaii, USA \\ 
$^6$ Institut d'astrophysique de Paris, UMR7095 CNRS, Universit\'e Pierre \& Marie Curie, 98bis boulevard Arago, 75014 Paris, France \\ 
$^7$ Obs.\ Astronomique de l'Universit\'e de Gen\`eve, Gen\`eve, Switzerland \\ 
$^8$ Aix Marseille University, CNRS, LAM, 13388 Marseille, France \\ 
$^9$ Hertzberg Institute of Astrophysics, National Research Council of Canada, Victoria, Canada \\ 
$^{10}$ Observatoire de Haute-Provence, CNRS, Aix Marseille University, Institut Pyth\'eas, 04870 St Michel l'Observatoire, France \\ 
$^{11}$ Centre of Optics, Photonics and Lasers,  Universit\'e Laval, Quebec, Canada \\ 
$^{12}$ Laborat\'orio Nacional de Astrof\' \i sica, Itajuba, MG, Brazil \\ 
$^{13}$ Institute of Astronomy \& Astrophysics Academia Sinica, Taipei, Taiwan \\ 
$^{14}$ European Southern Observatory, Vitacura, Santiago, Chile \\ 
$^{15}$ Instituto de Astrof\'isica e Ci\^encias do Espa\c{c}o \& Faculdade de Ci\^encias, Universidade do Porto, Porto, Portugal \\ 
$^{16}$ Dublin Institute for Advanced Studies, Dublin, Ireland \\ 
$^{17}$ LUPM, Universit\'e de Montpellier, CNRS, Place E.~Bataillon, 34095 Montpellier, France \\ 
$^{18}$ Departamento de Fisica - ICEx - UFMG, Belo Horizonte, MG, Brazil \\ 
$^{19}$ Plan\'etarium Rio Tinto Alcan, iREx, Montr\'eal, Canada \\ 
$^{20}$ Max-Planck-Institut f\"ur Quantenoptik, Garching, Germany; Menlo Systems GmbH, Martinsried, Germany

\bsp	% typesetting comment
\label{lastpage}
\end{document}